\documentclass[aps,prb,a4paper,twocolumn,10pt,floatfix,showpacs,showkeys,amsmath,amssymb,nobibnotes,unsortedaddress,superscriptaddress]{revtex4-2}

\setcitestyle{super}
\usepackage{graphicx}
\usepackage[export]{adjustbox}
\usepackage[dvipsnames]{xcolor}
\usepackage[utf8]{inputenc}
\usepackage[T1]{fontenc}
\usepackage{amsmath}
\usepackage{xr}
\usepackage{hyperref}
\usepackage{upgreek}
\usepackage{textgreek}
\usepackage[percent]{overpic}
\usepackage{xcolor}
\usepackage{physics}
\usepackage{verbatim}
\usepackage[explicit]{titlesec}
\usepackage{lipsum}
\usepackage{xspace}
\usepackage{lineno}
\usepackage{float}
\usepackage{diagbox}
\usepackage{titletoc}
\usepackage{tocloft}
\usepackage{makecell}
\usepackage{booktabs}
\usepackage{threeparttable}

\graphicspath{{figures/}}
\renewcommand{\figurename}{Figure}
\renewcommand{\thefigure}{\arabic{figure}}
\renewcommand{\tablename}{Table}
\renewcommand{\thetable}{\arabic{table}}
\renewcommand{\thesection}{\arabic{section}}

\titleformat{\section}{\large\bfseries\filcenter}{\thesection.}{1em}{#1}

\newcommand{\jgen}{J_\mathrm{gen}}
\newcommand{\jrec}{J_\mathrm{rec}}

\newcommand{\jsc}{J_\mathrm{sc}}

\newcommand{\Voc}{V_\mathrm{oc}}

\newcommand{\FF}{F\!F}
\newcommand{\pFF}{pF\!F}

\newcommand{\nid}{n_\mathrm{id}}
\newcommand{\nsig}{n_\mathrm{\sigma}}

\newcommand{\napp}{n_\mathrm{app}}
\newcommand{\Eu}{E_{U}}
\newcommand{\Eg}{E_{g}}

\newcommand{\mueff}{\mu_\mathrm{eff}}
\newcommand{\sigmaeff}{\sigma_\mathrm{eff}}
\newcommand{\sigmaoc}{\sigma_\mathrm{oc}}

\newcommand{\etadiss}{\eta_\mathrm{diss}}
\newcommand{\etacol}{\eta_\mathrm{col}}

\newcommand{\etaexc}{\eta_\mathrm{exc}}
\newcommand{\tpl}{t^{-3/2}}
\newcommand{\bl}{\left(}

\newcommand{\br}{\right)}

\newcommand{\E}{\,\cdot\, 10^}
\newcommand{\kT}{k_\mathrm{B} T}
\newcommand{\der}{\mathrm{d}}

\begin{document}

\title{Rethinking Charge Transport and Recombination in Donor-diluted Organic Solar Cells}

\author{Chen Wang}
\affiliation{Institut für Physik, Technische Universität Chemnitz, 09126 Chemnitz, Germany}

\author{Christopher Wöpke}
\affiliation{Institut für Physik, Technische Universität Chemnitz, 09126 Chemnitz, Germany}

\author{Toni Seiler}
\affiliation{Institut für Physik, Technische Universität Chemnitz, 09126 Chemnitz, Germany}

\author{Jared Faisst}
\affiliation{Institute of Physics, University of Freiburg, 79104 Freiburg, Germany}
\affiliation{Fraunhofer Institute for Solar Energy Systems ISE, 79110 Freiburg, Germany}

\author{Mathias List}
\affiliation{Fraunhofer Institute for Solar Energy Systems ISE, 79110 Freiburg, Germany}
\affiliation{Freiburg Materials Research Center FMF, University of Freiburg, 79104 Freiburg, Germany}

\author{Meike Kuhn}
\affiliation{Dynamics and Structure Formation -- Herzig Group, University of Bayreuth, 95447 Bayreuth, Germany}

\author{Bekcy Joseph}
\affiliation{Leibniz-Institut für Festkörper- und Werkstoffforschung Dresden (IFW), 01069 Dresden, Germany}

\author{Alexander Ehm}
\affiliation{Institut für Physik, Technische Universität Chemnitz, 09126 Chemnitz, Germany}

\author{Dietrich R.\ T.\ Zahn}
\affiliation{Institut für Physik, Technische Universität Chemnitz, 09126 Chemnitz, Germany}

\author{Yana Vaynzof}
\affiliation{Leibniz-Institut für Festkörper- und Werkstoffforschung Dresden (IFW), 01069 Dresden, Germany}

\author{Eva M. Herzig}
\affiliation{Dynamics and Structure Formation -- Herzig Group, University of Bayreuth, 95447 Bayreuth, Germany}

\author{Roderick C.\ I.\ Mackenzie}
\affiliation{Department of Engineering, Durham University, Lower Mount Joy, South Road, Durham DH1 3LE, United Kingdom}

\author{Uli Würfel}
\affiliation{Fraunhofer Institute for Solar Energy Systems ISE, 79110 Freiburg, Germany}
\affiliation{Freiburg Materials Research Center FMF, University of Freiburg, 79104 Freiburg, Germany}

\author{Maria Saladina}
\email[Corresponding author. Email: ]{maria.saladina@physik.tu-chemnitz.de}
\affiliation{Institut für Physik, Technische Universität Chemnitz, 09126 Chemnitz, Germany}

\author{Carsten Deibel}
\email[Corresponding author. Email: ]{deibel@physik.tu-chemnitz.de}
\affiliation{Institut für Physik, Technische Universität Chemnitz, 09126 Chemnitz, Germany}

\begin{abstract}
We systematically investigate PM6:Y12 bulk-heterojunction solar cells with donor fractions ranging from 1\% to 45\%, linking morphology, charge transport, and recombination to device performance. Complementary structural and spectroscopic methods reveal that a percolating PM6 network forms even at below 5\% donor content, with lamellar stacking and vertical composition gradients that do not hinder the charge extraction. The reduction of the effective active layer conductivity towards low donor fractions obeys a three-dimensional percolation model, indicating that charge transport is governed by network topology rather without a pronounced percolation threshold. A transition from nongeminate Langevin recombination to a dispersive Smoluchowski-type loss occurs below 5\% donor fraction. The latter regime is also nongeminate, i.e., pertains to recombination of the total charge carrier density. Correspondingly, we observe that the Langevin reduction in the higher donor fractions -- mostly dominated by redissociation of electron--hole pairs after encounter -- changes towards low donor fractions: in these cases, the nongeminate loss rate exceeds the prediction of the Langevin model. This regime coincides with increasing transport resistance due to topology-limited hole conduction, leading to reduced fill factors despite a high retained charge-generation efficiency. Our results demonstrate that strong donor dilution preserves photogeneration if a continuous donor network is maintained, and unveil how topology-controlled transport and non-Langevin recombination jointly define the performance limits of donor-diluted organic solar blends.
\end{abstract}


\maketitle

\section{Introduction}

Organic solar cells based on non-fullerene acceptors (NFAs) have reached power conversion efficiencies exceeding 20\%, driven by improved absorption, reduced energetic losses, and efficient charge generation.\cite{fu2025two,jiang2024non,li2025non} Beyond optimized donor--acceptor ratios, recent studies have shown that surprisingly high efficiencies can be retained upon strong donor dilution, in some cases down to only a few percent polymer content.\cite{dolan2024enhanced,sharma2025elucidating,wu2025carrier} Donor dilution is attractive for semi-transparent applications, where lowering the polymer fraction improves optical transmission without necessarily compromising device performance.\cite{sharma2024semitransparent} In general, it allows for studying the impact of material connectivity without changing the material system.

Several experimental studies have explored such donor-diluted NFA systems.\cite{schopp2022unraveling,sharma2025elucidating,tang2022molecular} From a structural perspective, Ortner et al.\ reported that the polymer donor D18 forms an interconnected, fibrillar network down to about 2\% donor content, below which donor connectivity collapses and acceptor aggregation dominates the morphology.\cite{ortner2025donor} In line with this, photogeneration has been reported to remain efficient down to around 10--15\% donor-fraction range since donor--acceptor interfaces remain accessible for excitons.\cite{sharma2024semitransparent,wang2022quasi,yao2021efficient} At very low donor contents, increased photoluminescence and reduced photocurrent have been attributed to incomplete exciton harvesting when acceptor domains exceed the exciton diffusion length.\cite{haffner2025high,ortner2025donor}

Reports on nongeminate recombination in diluted systems are less uniform. While some studies report nearly composition-independent recombination down to moderate donor dilution,\cite{sharma2024semitransparent} others find that the effective nongeminate recombination coefficient decreases systematically with increasing donor fraction, indicating that recombination kinetics are sensitive to how charge carriers are spatially distributed rather than to charge carrier density alone.\cite{haffner2025high} 
Despite these differences, effective mobilities have often been reported to vary only weakly with donor fraction, attributed to ambipolar charge transport in NFAs.\cite{ortner2025donor,sharma2024semitransparent,yao2021efficient,wang2022quasi} In this context, fill-factor losses at very low donor contents have been explained by disrupted long-range connectivity and extraction limitations rather than decrease in mobility,\cite{ortner2025donor,yao2021efficient,sharma2024semitransparent} or attributed to composition-dependent nongeminate recombination.\cite{haffner2025high}

Taken together, these studies indicate that donor dilution modifies morphology, transport, and recombination, but they do not yet provide a unified picture of their interplay in such systems.
In particular, three key questions remain open. First, how does nanoscale and mesoscale morphology evolve when the donor fraction is reduced far below conventional bulk heterojunction composition? Second, how does reduced donor connectivity affect charge transport and recombination, especially given the strong mobility imbalance typically present in diluted polymer:NFA blends? Third, can recombination still be described within a Langevin framework in highly diluted systems, or does a fundamentally different, spatially controlled recombination regime emerge?

In this work, we address these questions using the PM6:Y12 system, varying the donor fraction from 1\% to 45\%. By combining complementary optical, electrical, and structural characterization techniques, we establish a consistent picture linking nanomorphology, charge transport, and recombination. We show that donor dilution primarily modifies mesoscale connectivity that governs macroscopic charge transport, and that recombination crosses over from a Langevin-type behavior to dispersive, Smoluchowski-type kinetics at low donor fractions. Together, these results clarify how morphology and domain connectivity control performance in strongly donor-diluted NFA solar cells and provide a unified framework for interpreting transport and recombination beyond the conventional optimized bulk heterojunction concept.

\section{Results and Discussion}


First, we give an overview over the impact of the donor fraction change from 1\% to 45\% in terms of the solar cell parameters and the internal quantum efficiency. Then, we discuss the charge transport and find that both mobility and conductivity are limited particularly in the lower donor fractions. We can describe the donor mobility in accordance with a simple percolation model. Subsequently, we consider recombination and show that the mobility imbalance plays an important role for the higher donor fractions with respect to reduced Langevin-type recombination. However, for lower donor fractions we propose that our findings are best described by Smoluchowski-type recombination based on a diffusion-limited process of spatially distributed mobile charge carriers forming and recombination as charge-transfer (CT) states. Finally, we consider the charge collection by combining recombination and charge transport. We show that the fill factor of all donor fractions is limited by transport resistance, which is due to the imbalance and generally low values of electron and hole mobilities. Using our recently proposed figure-of-merit $\beta$ that quantifies the voltage loss due to transport resistance, all the fill factors can be well predicted. We discuss all findings in the context of the nanomorphology and vertical phase segregation.

\subsection{PV performance}

\begin{figure}[b]
    \centering
    \includegraphics[width=0.9\linewidth]{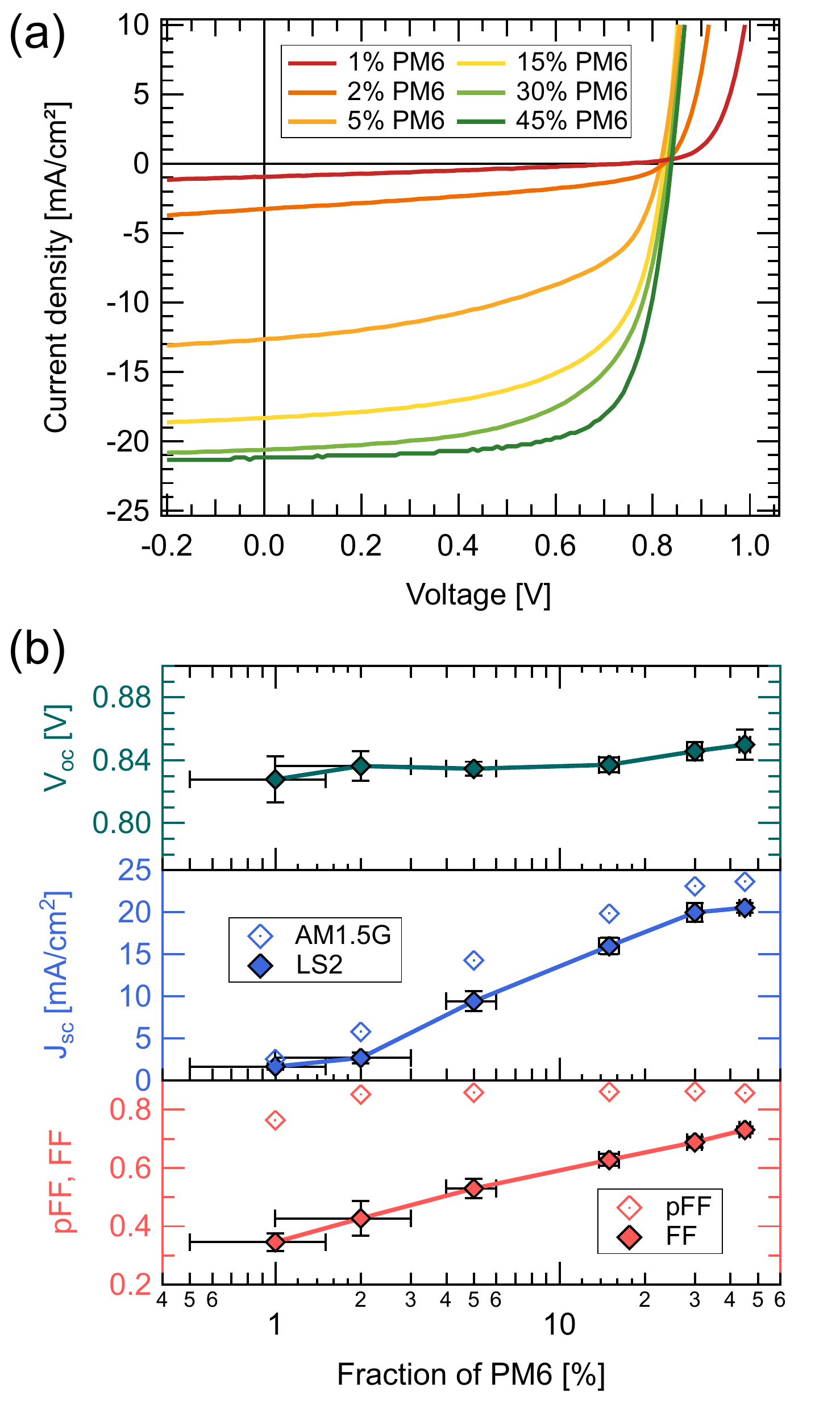}
    \caption{(a) JV characteristics of PM6:Y12 solar cells with varying PM6 fraction, under 1~sun (i.e., 100~mW\,cm$^{-2}$ under AM1.5G spectrum) equivalent illumination. (b) Statistical $\Voc$, $\jsc$, and $\FF$ as a function of PM6 content measured using the LS2 solar simulator. Calibrated $\jsc$ values from AM1.5G spectrum are shown for reference. With decreasing donor fraction, the difference between $\pFF$ and $\FF$ increases, showing growing contribution of transport resistance.}
    \label{fig:PV_performance}
\end{figure}

The conjugated polymer donor PM6 and the non-fullerene small molecule acceptor Y12 are selected as the donor--acceptor combination. These materials are commonly used in organic solar cells (OSCs) and have demonstrated reasonable efficiencies of $\approx 16\%$.\cite{ma2020achieving,kramdi2025blade} Solar cell devices were fabricated using an ITO/ZnO/active layer/PEDOT:F/Ag structure. The photoactive layer was spin-cast from chloroform solutions, with a wide variation of donor--acceptor weight ratios from highly donor-diluted (1\%) to optimized (45\%) ratio.

Figure~\ref{fig:PV_performance}(a) shows representative current density--voltage (JV) characteristics of the resulting solar cells under 1~sun equivalent illumination (100~mW\,cm$^{-2}$, AM1.5G), corresponding parameters are listed in Table~\ref{SI_table:solar_cell_parameters}. The open‐circuit voltage ($\Voc$) remains nearly constant across all investigated ratios. In contrast, both the short‐circuit current density ($\jsc$) and fill factor ($\FF$) increase significantly with higher PM6 fraction, particularly beyond 5\%. The best efficiency is achieved at 45\% PM6 fraction, delivering a power conversion efficiency (PCE) of 13.3\% measured with our solar simulator (LS2). After correcting for the spectral mismatch between the LS2 and the AM1.5G spectrum (Figure~\ref{SI_fig:EQEJV}(a) and~(b), Supporting Information), the PCE increases to 15\%. Slightly higher PCE can be expected via additional optimisation of interfacial layers or employing regular architecture. This value is confirmed by external quantum efficiency (EQE) measurements and corresponding integrated $\jsc$ (Figure~\ref{SI_fig:EQE_intJsc}).

The solar cell parameters extracted from 20 independently fabricated cells are shown in Figure~\ref{fig:PV_performance}(b). From the ideal diode equation, $\Voc \approx \nid \kT/e \ln(\jsc/J_0)$, where $\nid$ is the recombination ideality factor, $\kT/e$ the thermal voltage, and $J_0$ the dark saturation current density. Given that $\jsc$ is changing by roughly a factor of 15 from 1\% to 45\% donor fraction, we would have expected -- if the other parameters in the diode equation remained unchanged -- a corresponding increase of $\Voc$ by around 70~meV. We make surface recombination at around 1 sun responsible for limiting the $\Voc$ in the higher donor fraction devices,\cite{kirchartz2013differences} where ideality factors below unity are observed (see Figure~\ref{SI_fig:sunsvoc_nid}, Supporting Information). The $\FF$ of all photovoltaic devices shown are limited by transport resistance, i.e., recombination losses during imperfect charge collection.\cite{neher2016new,saladina2024transport} The pseudo fill factors ($\pFF$) -- the fill factors that the devices would have, if they were just limited by recombination and not the additional transport resistance loss -- is calculated from the $\nid$ and $\Voc$ using the Green equation.\cite{green_accuracy_1982} The narrowing $\pFF$ and $\FF$ gap suggests a reduced transport resistance loss as the PM6 fraction increases. 

The intuitive expectation is that reducing the donor fraction decreases the donor--acceptor interfacial area and lowers nonradiative losses, as reported previously for fullerene-based systems.\cite{vandewal2014increased} We measured sensitive EQE spectra (see Figure~\ref{SI_fig:sensEQE} and Figure~\ref{SI_fig:energy_gap}, Supporting Information) and analyze the $\Voc$ loss of solar cells based on the reciprocity relation.\cite{rau2007reciprocity} However, Table~\ref{SI_table:voc_losses} shows that the nonradiative open-circuit voltage loss $\Delta V_{\mathrm{oc,nonrad}}$ in PM6:Y12 blends remains roughly constant (approximately 0.22-0.25~V) over the full range of PM6 fractions, despite small variations in the optical bandgap and the radiative limit. Possible explanations are that (i) the interface density relevant for recombination is almost independent of composition, and/or (ii) reductions in interface density are compensated by increased recombination rate. Nanomorphology results presented in the following section reveal reduced PM6 order at low donor content, which enhances charge recombination and renders the latter contribution likely dominant.

\subsection{Nanomorphology and vertical phase segregation, optical properties and exciton dynamics}

\begin{figure*}[t]
    \centering
    \includegraphics[width=0.9\linewidth]{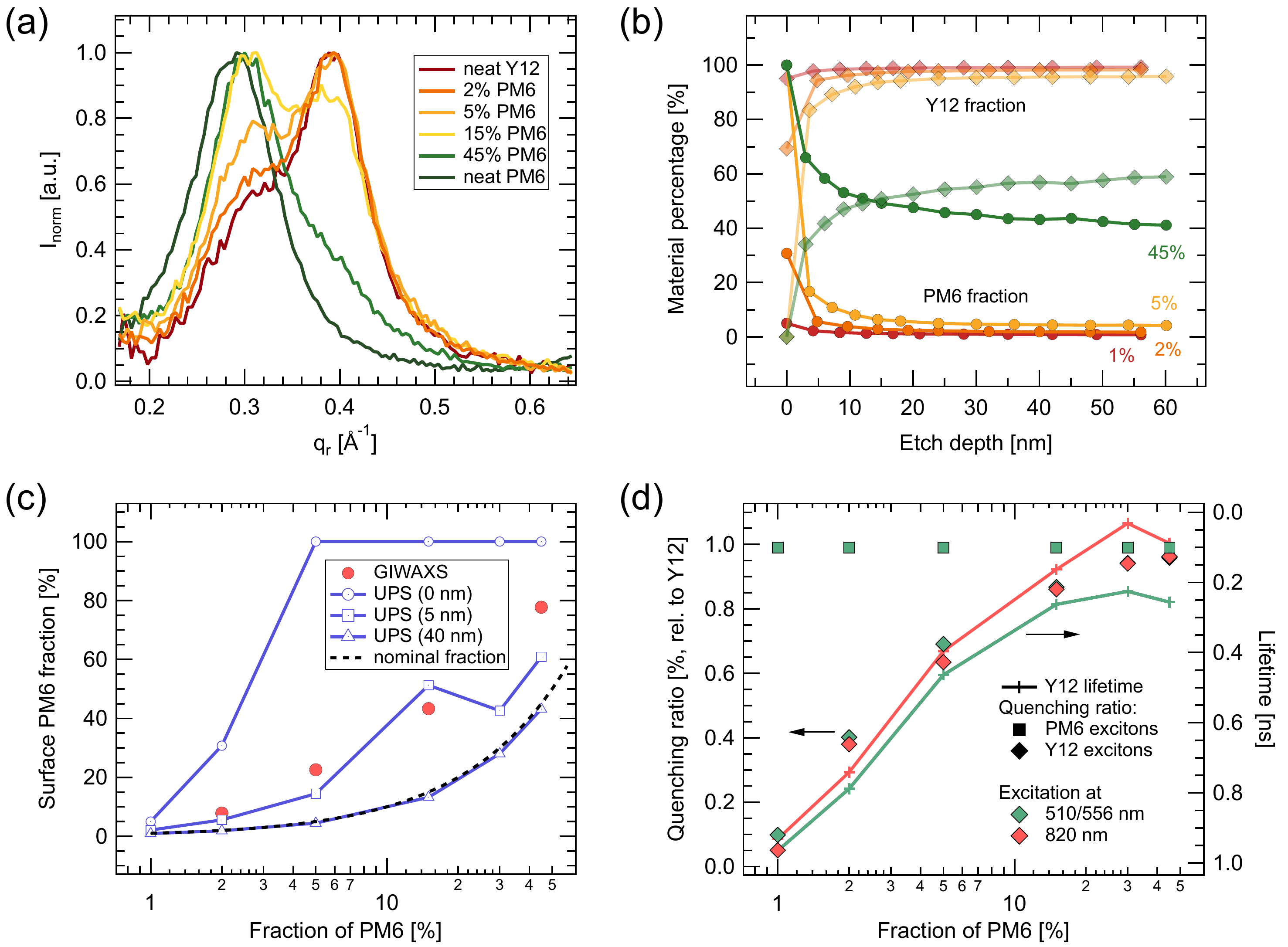}
    \caption{(a) Horizontal line cuts (q$_r$) of the 2D GIWAXS data shown in Figure~\ref{SI_fig:GIWAXS:2D} for PM6:Y12 blend films with varying PM6 content, measured at an incidence angle of 0.18°. The curves are background-corrected and min–max normalized for clarity. (b) Vertical composition profiles of various donor--acceptor ratios derived from UPS depth profiles. 
    (c) The fraction of PM6 at various cross-sections of UPS depth profile in (b) as a function of input PM6 content. Bulk UPS at 40~nm from the surface aligns well with the expected ratio. At 5~nm, it corresponds to the relative ordered material fraction obtained from GIWAXS line cuts. (d) PL quenching ratio for PM6 and Y12 excitons, and Y12 exciton lifetime in PM6:Y12 blend films as a function of PM6 content.}
    \label{fig:nano}
\end{figure*}

To investigate the nanomorphology of PM6:Y12 blends and their evolution with donor content, we employed grazing-incidence wide-angle X-ray scattering (GIWAXS) and resonant soft X-ray scattering (RSoXS). We also investigated the optical properties by means of spectral ellipsometry, as well as the singlet exciton dynamics and quenching yields.

\subsubsection{Nanomorphology}

GIWAXS reveals a systematic change in molecular ordering with increasing PM6 fraction. The analysis focuses on the low-$q_r$ region, where PM6 and Y12 exhibit distinct horizontal diffraction features (see Figure~\ref{SI_fig:GIWAXS:2D}). Details can be found in the Supporting Information, Section~\ref{SI_sec:GIWAXS}. PM6 shows a lamellar stacking peak at $q_r = 0.29$~\AA$^{-1}$, while Y12 exhibits two characteristic peaks at $q_r = 0.32$ and $0.40$~\AA$^{-1}$. As shown in Figure~\ref{fig:nano}(a), the PM6-associated lamellar peak increases in intensity with donor content, while the Y12-related features are progressively suppressed. Notably, a pronounced increase in PM6-related order already occurs at 5\% PM6, indicating a morphological transition from isolated donor inclusions to a more connected and ordered PM6 network. 

Structural information on larger length scales (5--100\,nm) is provided by RSoXS (see Figure~\ref{SI_fig:GIWAXS:rsoxs}). The 2\% PM6 blend exhibits a scattering profile similar to neat Y12, whereas increasing the PM6 content to 5\% leads to the emergence of larger structural features, which persist in the 15\% PM6 sample. The 45\% PM6 blend shows a scattering profile qualitatively similar to neat PM6, indicating donor-dominated morphology on the 5–100~nm scale. Intermediate blends display a broad distribution of length scales, consistent with a heterogeneous but interconnected morphology. Owing to the energy-dependent contrast in RSoXS, a clear separation between surface-related and internal structures cannot be made.

While spectroscopic ellipsometry does not directly resolve morphology, it provides complementary insight into changes in the local electronic environment with composition (see Section~\ref{SI_sec:ellipsometry}, Supporting Information). With increasing PM6 fraction, the dielectric function in Figure~\ref{SI_fig:ellips:e1e2alpha} evolves toward that of neat PM6, indicating the emergence of bulk-like donor electronic states. At low PM6 content, however, PM6-related optical features are redshifted relative to neat PM6, showing that the PM6 electronic states are influenced by the surrounding Y12 matrix. 
Consistent with RSoXS, the ellipsometry results support a picture in which PM6 is predominantly present in interface-proximate or finely dispersed regions within a Y12-rich matrix. At higher PM6 fractions, the recovery of PM6-like optical features reflects the formation of more extended and ordered donor domains, in agreement with the GIWAXS results. More disordered PM6 at low donor content likely increases nonradiative voltage losses.\cite{liu2023understanding} This compensates for the reduced voltage losses due to lower interfacial density, leading to similar $\Delta V_{\mathrm{oc,nonrad}}$ values.

\subsubsection{Vertical phase segregation}

To assess vertical phase segregation in the PM6:Y12 blends, we performed ultraviolet photoelectron spectroscopy (UPS) depth profiling (see Section~\ref{SI_sec:XPSUPS}, Supporting Information). The compositional profile extracted from the UPS measurements shown in Figure~\ref{fig:nano}(b) confirms that, beyond the near-surface region, the bulk of the active layer exhibits the expected donor--acceptor ratios across all compositions (for all PM6 fractions, refer to Figure~\ref{SI_fig:UPSXPS}(b)). At the same time, UPS reveals a systematic enrichment of PM6 at the top surface with increasing donor fraction, which is attributed to the lower surface energy of the donor polymer.\cite{Vaynzof2011,lami2020} Notably, a threshold of donor fraction is required for this effect to become apparent. In the present system, a PM6 content of approximately 5\% marks the onset of a thin PM6-rich overlayer.

Importantly, this vertical segregation does not pose a limitation for charge extraction in the present device architecture. Electrons are extracted at the ITO/ZnO interface, while holes are collected at the top electrode. A donor-enriched region toward the top contact therefore locally facilitates hole extraction under operating conditions, thereby reducing the probability of recombination.\cite{cai2022vertically} For regular device architecture, the same vertical composition profile would likely result in higher recombination and collection losses. However, a lower surface energy of the underlying transport layer would lead to a different vertical gradient in real devices.

Figure~\ref{fig:nano}(c) shows cross-sections of UPS at different etching depths. The bulk-sensitive cross-section at 40~nm confirms that the active layer composition closely follows the nominal donor--acceptor ratio. 
The UPS depth sensitivity provides a useful reference for interpreting the GIWAXS results. GIWAXS predominantly probes the near-surface region. When this is taken into account, the ordered PM6 fraction extracted from GIWAXS is found to scale well with the local donor fraction determined by UPS at $\sim 5$\,nm probing depth. This resolves the apparent discrepancy where the ordered PM6 fraction inferred from GIWAXS exceeds the nominal bulk weight fraction (Figure~\ref{SI_fig:GIWAXS:fraction}). Rather than indicating preferential ordering beyond the available donor content, it reflects the surface-enriched PM6 composition probed by the technique.

\subsubsection{Exciton dynamics}

To probe the exciton dynamics, which also give indirect information on the nanomorphology, we performed steady-state and time-resolved photoluminescence (PL) measurements on PM6:Y12 blend films with varying donor content. PM6 was excited predominantly at 510/556~nm (with a minor contribution from Y12 due to its finite absorption), while Y12 was selectively excited at 820~nm. PL intensities were obtained by integrating the emission over the respective spectral ranges and used to determine exciton quenching ratios relative to neat materials (see Section~\ref{SI_sec:quenching}, Supporting Information).

The resulting quenching ratios and exciton lifetimes are shown in Figure~\ref{fig:nano}(d). 
PM6 excitons are nearly completely quenched, indicating efficient exciton harvesting in the donor phase for all compositions.
Despite the higher absorbance of PM6 at 510~nm, a substantial fraction of excitons is still generated in Y12, particularly at low PM6 content. The consistently strong quenching of PM6 excitons therefore implies faster decay, which could in principle imply a higher exciton diffusivity -- we deem this unlikely -- or smaller effective domain sizes or mixed donor--acceptor domains. Notably, PM6 excitons are quenched efficiently despite the expected additional energy transfer step prior to charge separation.\cite{karuthedath2021intrinsic} This indicates that donor--acceptor interfaces are readily accessible and required exciton diffusion distances are sufficiently short.

In contrast, the quenching ratio of Y12 excitons increases monotonically with PM6 fraction, indicating that adding donor material increases the accessibility of donor--acceptor interfaces to Y12 excitons. Since steady-state PL in the blends originates almost exclusively from Y12, the transient PL decay reflects the effective Y12 exciton lifetime. The close correspondence between the PM6-fraction dependence of the lifetime and the quenching ratio confirms that Y12 exciton quenching is governed by interface accessibility rather than changes in intrinsic Y12 exciton properties.

Taken together, UPS and GIWAXS consistently show that vertical phase segregation is limited to a thin surface region, while the bulk composition remains homogeneous. The presence of a donor-enriched surface layer and the rapid development of PM6 ordering at low donor fractions imply that vertical segregation does not limit charge extraction, whereas lateral and mesoscale connectivity are likely to play a more decisive role in governing charge transport. The local exciton quenching demonstrates that the exciton harvesting in PM6 is highly efficient even in the case where the ratio to the acceptor is close to 45:55, which either indicates that PM6 domains could contain a small amount of acceptor material, or that the conjugated polymer features a nanofibrillar structure that is dispersed in the blend. In the likely neat Y12-domains, the PL quenching efficiency increases with higher donor fraction. The PL quenching yield and exciton dissociation play an important role in determining the internal quantum efficiency, which will be discussed in the next section.

\subsection{Internal quantum efficiency}

\begin{figure}[b]
    \centering
    \includegraphics[width=0.9\linewidth]{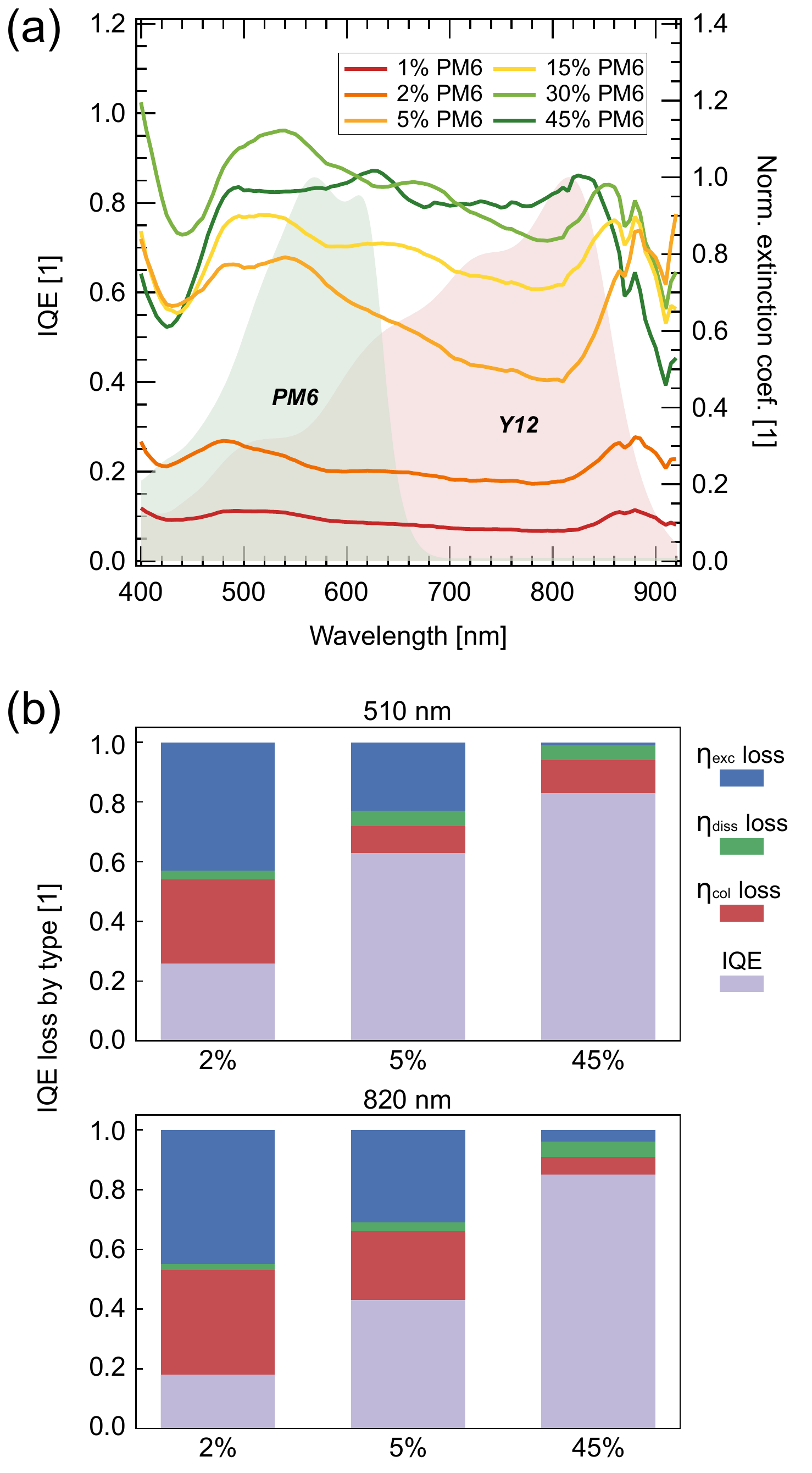}
    \caption{(a) IQE spectra of solar cells with different PM6 content. (b) The calculated contributions from individual photogeneration process steps to the IQE losses at 510~nm (top) and 820~nm (bottom).}
    \label{fig:IQE}
\end{figure}

\begin{table*}[t]
\centering
\caption{Photoelectric conversion process efficiencies and the measured IQE. Both $\etadiss$ and IQE were determined at short-circuit conditions, and the $\etacol$ is calculated by $IQE/\etaexc \cdot \etadiss$ for loss analysis.}

\small
\setlength{\tabcolsep}{4pt} 
\renewcommand{\arraystretch}{1.08}

\begin{threeparttable}
\begin{tabular*}{\textwidth}{@{\extracolsep{\fill}} lcccccccccc @{}}
\toprule
& \multicolumn{6}{c}{PM6 predominant excitation (510~nm)}
& \multicolumn{4}{c}{Y12 selective excitation (820~nm)} \\
\cmidrule(lr){2-7}\cmidrule(lr){8-11}
PM6 ratio
& $\eta_{\mathrm{exc,PM6}}$
& $\eta_{\mathrm{exc,Y12}}$
& $\eta_{\mathrm{exc}}^{\ast}$
& $\eta_{\mathrm{diss}}$
& $\eta_{\mathrm{exc}}\eta_{\mathrm{diss}}$
& \textbf{IQE}
& $\eta_{\mathrm{exc}}$
& $\eta_{\mathrm{diss}}$
& $\eta_{\mathrm{exc}}\eta_{\mathrm{diss}}$
& \textbf{IQE} \\
\midrule
2\%  & 0.99 & 0.40 & 0.43 & 0.96 & 0.40 & \textbf{0.26} & 0.38 & 0.96 & 0.36 & \textbf{0.18} \\
5\%  & 0.99 & 0.70 & 0.73 & 0.95 & 0.69 & \textbf{0.63} & 0.63 & 0.96 & 0.60 & \textbf{0.43} \\
45\% & 0.99 & 0.96 & 0.99 & 0.96 & 0.94 & \textbf{0.83} & 0.96 & 0.95 & 0.91 & \textbf{0.85} \\
\bottomrule
\end{tabular*}\label{tab:IQE}

\begin{tablenotes}[flushleft]
\footnotesize
\item[$\ast$] Weighted exciton harvesting efficiency.
\end{tablenotes}
\end{threeparttable}
\end{table*}

The internal quantum efficiency (IQE) is calculated by normalizing EQE spectra to the total absorptance ($A$) of the device, as shown in Figure~\ref{fig:IQE}(a). It allows to understand if the compositional changes result in differing contributions of donor and acceptor materials to the process from photon absorption to charge collection. The total absorptance is calculated by the matrix transfer formalism using the OghmaNano simulation\cite{mackenzie2012extracting,dattani2014general,liu2015organic,bickerdike2025unravelling} based on the experimentally determined optical refractive index and extinction coefficient (see Figure~\ref{SI_fig:oghmanano}, Supporting Information). The IQE increases across the whole spectral range as the PM6 fraction is raised to 30\%. In order to quantify the relative changes of donor and acceptor contributions, Figure~\ref{SI_fig:EQEJV}(c) and (d) presents the ratio IQE(820~nm)/IQE(510/550~nm), where 820~nm represents a region of virtually pure acceptor absorption, as compared to a dominant donor absorption in the 500~nm range. The relative contribution of Y12 to the IQE response increases with the PM6 content. This indicates that the photovoltaic conversion associated with Y12 becomes more efficient. 

To gain some insight into the processes contributing to the IQE, we consider the involved loss mechanisms. In general, IQE is defined as the product of efficiencies of three internal processes: exciton harvesting at the donor--acceptor interface ($\etaexc$), CT exciton dissociation ($\etadiss$) into the separated charge carriers, followed by the competition between recombination and charge collection ($\etacol$).\cite{nelson2003physics} The lowest efficient process corresponds to the key loss mechanism. Table~\ref{tab:IQE} summarizes the efficiencies of the discussed internal processes. In the following, we briefly outline how each efficiency was determined experimentally before discussing their implications for the overall IQE.

The $\etaexc$ was determined directly from quenching ratios. For selective Y12 excitation, $\etaexc$ equals to PL quenching ratio of Y12, while for co-excitation of donor and acceptor, we calculated an $\etaexc$ (weighted, see Table~\ref{tab:IQE}) by weighting the individual efficiencies $\eta_{\mathrm{exc},\mathrm{\scriptsize PM6}}$ and $\eta_{\mathrm{exc},\mathrm{\scriptsize Y12}}$ according to their relative exciton generation at the excitation wavelength (see Section~\ref{SI_sec:quenching}, Supporting Information, for more details). 

We studied the field-dependence of CT dissociation by the time-delayed collection field (TDCF) method, employing dominant PM6 excitation as well as selective Y12 excitation. The dissociation yield as a function of pre-bias voltage (see Figure~\ref{SI_fig:TDCF:etadiss}, Supporting Information) was fitted to determine the $\etadiss$ value at short-circuit conditions, which is used to compare with the IQE -- derived from EQE measured under short-circuit conditions -- and listed in Table~\ref{tab:IQE}.  

Using the yield values summarized in Table~\ref{tab:IQE}, we evaluate the loss contributions of each photoelectric conversion process to the IQE, as shown in Figure~\ref{fig:IQE}(b). For devices with low PM6 fraction, limited exciton harvesting at the donor--acceptor heterojunction as well as pronounced charge collection losses dominate the overall IQE reduction. In solar cells with optimized PM6 fraction, inefficient charge collection is the main photocurrent-limiting factor. These observations motivate a more detailed analysis of charge transport and recombination processes as a function of PM6 fraction.

\subsection{Charge carrier transport}

Despite the impressive performance of state-of-the-art OSCs, they are still disordered systems.\cite{heeger2010semiconducting,bartelt2013importance} Correspondingly, charge transport in disordered organic semiconductors is governed by incoherent hopping between spatially and energetically localized states,\cite{bassler1993charge,baranovskii2014theoretical} although locally -- in particular for charge photogeneration -- delocalisation can play an important role.\cite{deibel2009prl} In the following, we will report our findings of changes in the active layer conductivity and charge carrier mobilities due to the varying donor fraction. We discuss them in the context of energetic disorder as well as spatial effects. Therefore, we first describe how we determine the effective conductivities of all devices from JV measurements. Then, we determine the density of states (DOS) of the active layer, and find that the energetic disorder increases with lower donor at shallower energies, and it becomes similar deeper into the bandgap. Finally, by comparing the devices under conditions at which the effective energetic disorder is similar, we find that hole transport in lower donor fractions can be described in the framework of a simple percolation model.

\subsubsection{The effective conductivity}

We quantitatively assess charge transport in diluted donor-content devices by determining the active layer conductivity. We apply a method that we recently established and validated, which allows the direct extraction of the effective conductivity from the JV characteristics of solar cells.\cite{saladina2024transport,wang2025transport}

\begin{figure*}[t]
    \centering
    \includegraphics[width=0.9\linewidth]{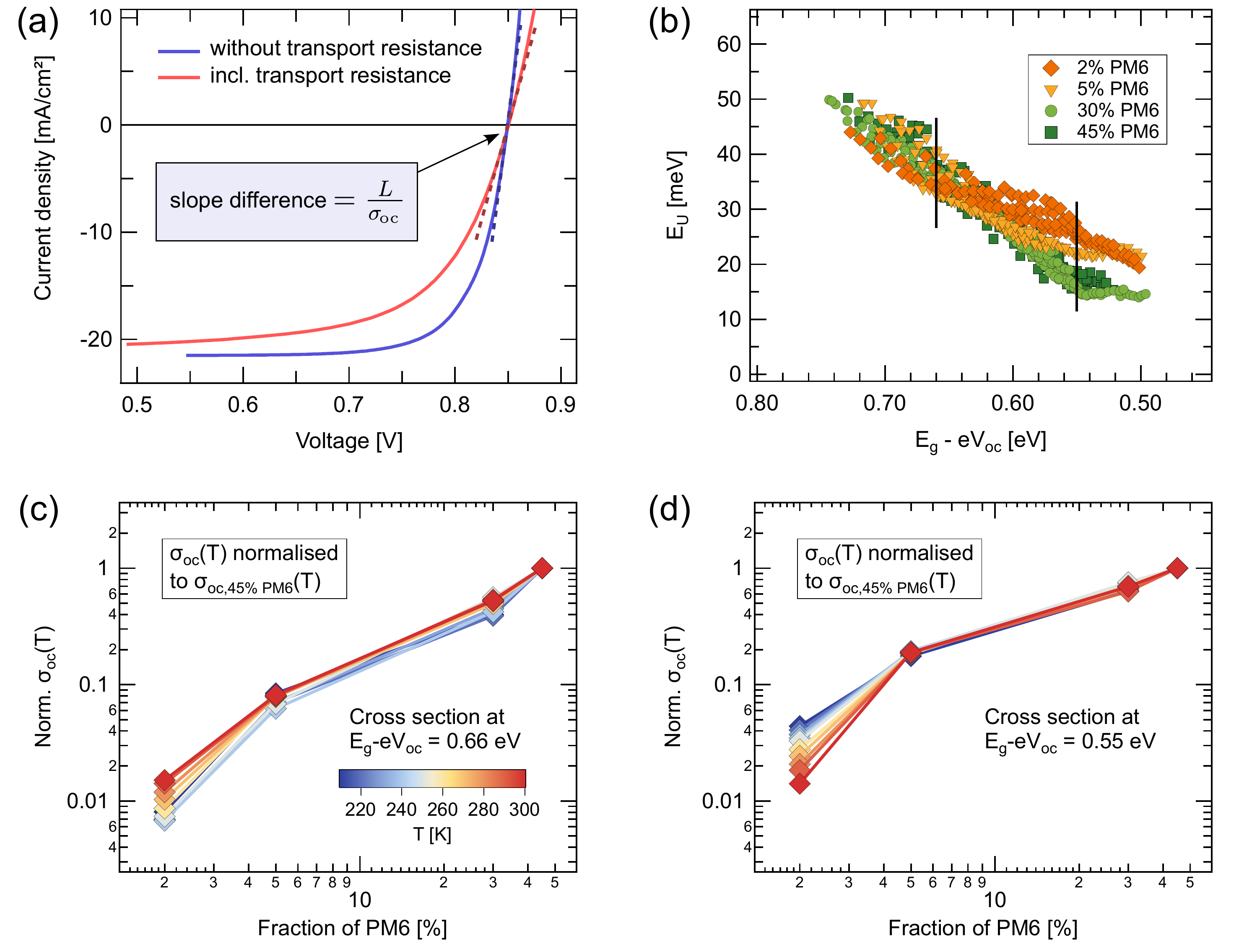}
    \caption{The effective conductivity $\sigma_{\mathrm{oc}}$ in PM6:Y12 solar cells with varying PM6 content. (a) The method of $\sigma_{\mathrm{oc}}$ extraction from the slope of JV curve at $\Voc$, separating transport and recombination contributions. (b) Energetic disorder parameter $E_\mathrm{u}$ extracted at different energetic positions within the DOS. (c) Temperature dependent $\sigma_{\mathrm{oc}}$ normalized to the value of the 45\% PM6 device, evaluated at a fixed energetic cross-section of (c) $\Eg - e\Voc = 0.66$~eV, where $\Eu$ is similar for all PM6 compositions, and (d) $\Eg - e\Voc = 0.55$~eV.}
    \label{fig:sigma}
\end{figure*}

In brief, near the open-circuit voltage, the slope of the JV curve reflects the transport resistance of the photoactive layer (see Figure~\ref{fig:sigma}a), from which the effective conductivity $\sigmaoc$ can be directly obtained without relying on separate conductivity measurements. Analytically, the effective conductivity at open circuit can be determined using
\begin{equation}\label{eq:sigmaoc}
    \frac{L}{\sigmaoc}=\left.\left(\frac{\der V}{\der J}-\frac{\nid\kT}{e\jrec}\right)\right|_{J=0} . 
\end{equation}
Here, $L$ is the thickness of the active layer. The term $\left.\mathrm{d}V/\mathrm{d}J\right|_{J=0}$ represents the differential resistance extracted from the slope of the JV curve at zero current. The second term in brackets accounts for the recombination contribution and corresponds to the inverse slope of a hypothetical JV curve with infinite conductivity, i.e., a device without transport losses (see Figure~\ref{fig:sigma}a). Here, $\jrec$ is the recombination current density at open circuit. Subtracting this recombination-related resistance yields the transport resistance of the active layer, from which $\sigmaoc$ is obtained. We point out that we correct the experimental $JV$ curve for the external series resistance before performing this evaluation,\cite{saladina2024transport} which is not shown in the above equation for clarity. It is determined by fitting $\der V/ \der J$ at large positive voltages, where transport resistance is negligible compared to the external series resistance.

Figure~\ref{SI_fig:sigma} shows the effective conductivity $\sigmaoc$ as a function of PM6 fraction for different temperatures and light intensities. As expected, the absolute conductivity increases with temperature for all compositions. Remarkably, however, the functional dependence on PM6 content remains essentially unchanged across the investigated temperature range. To draw conclusions from this finding, we will first consider the DOS of the active layer.

We investigate the DOS by extracting the characteristic energy $\Eu$, essentially the local slope of the DOS at each accessible energetic position within the bandgap. We do this by evaluating the recombination ideality factors, determined from light intensity- and temperature-dependent suns–$\Voc$ measurements,\cite{schiefer2014} interpreted within a multiple-trapping-and-release model.\cite{hofacker2017} The details are described in section~\ref{SI_sec:nid} of the Supporting Information. We introduced this approach recently,\cite{saladina2023power} and found that the DOS in PM6:Y6 blends is best described by a Gaussian shape for one carrier type and an adapted exponential shape -- indeed, a power-law distribution -- for the other. The result gives the $\Eu$, which would be constant in case of a purely exponential DOS. In the PM6:Y12 solar cells shown in Figure~\ref{fig:sigma}(b), $\Eu$ is not constant but increases towards deeper energies within the bandgap, consistent with a power-law DOS distribution. Using this information representing the DOS shape, we can now come back to the consideration of the active layer conductivity.

As the effective conductivity depends on both carrier mobility and carrier density -- and the latter determines the energetic position at which carriers predominantly reside -- we evaluate $\sigmaoc$ at a fixed energetic cross-section, and normalize the temperature dependence of the effective conductivity $\sigmaoc(T)$ of all devices to $\sigmaoc(T)$ of the 45\% PM6 device to make the similarity of the temperature dependence clearer. The conductivity values shown in Figure~\ref{fig:sigma}(c) are extracted at $\Eg - e\Voc = 0.66$~eV, where all PM6 fractions exhibit nearly identical values of $\Eu$. At shallower energies, however, the 2\% PM6 blend shows a noticeably larger $\Eu$ (cf.\ Figure~\ref{fig:sigma}b).  For the case with a different disorder, we therefore perform the energetic cross-section at $\Eg - e\Voc = 0.55$~eV, shown in Figure~\ref{fig:sigma}(d). 
In the case where energetic disorder differs for the solar cell with 2\% PM6 content, the normalized conductivity curves no longer collapse onto a single master curve, revealing a residual temperature dependence (Figure~\ref{fig:sigma}d). This deviation directly indicates the role of energetic disorder in shaping the transport behavior at very low donor concentrations. Such a residual temperature dependence is not observed in the 5\% blend or higher donor fraction at this energetic position within the bandgap. Generally, the resulting collapse of the normalized temperature-dependent curves indicates that the effective energetic disorder is essentially identical for all compositions (with the exception of a small deviation for the 2\% PM6 blend). 

We can rationalize the mostly similar thermal activation of the conductivity by qualitatively consulting a model for hopping transport. In the classical Miller–Abrahams picture, hopping is treated statistically without explicit consideration of molecular geometry,\cite{miller1960impurity,ambegaokar1971hopping} with a hopping rate
\begin{equation}\begin{split}\label{eq:th:hopping-rate}
    \nu_{ij} = \nu_0 \exp\bl -u \br, \quad u = \frac{2r_{ij}}{\alpha} + 
        \begin{cases}
            \frac{\Delta E_{ij}}{\kT}, &\quad E_i < E_j \\
            0, &\quad E_i \ge E_j \\
        \end{cases} . 
\end{split}\end{equation} 

Here, $\nu_0$ is the attempt-to-escape frequency. The exponent $u$ is the universal hopping parameter which accounts for both the spatial separation $r_{ij}$ and the energy difference between sites $\Delta E_{ij}$. The decay of wavefunction overlap with inter-site distance is characterized by the localization length $\alpha$. The upward hops in energy are thermally activated and downward hops occur solely by tunneling. As a result, upward energy hops are the rate-limiting ones in charge transport. In short, the distance dependent term is a tunneling contribution, and the temperature dependent Boltzmann term corresponds to thermal activation. For our case of devices with different donor fraction, but very similar energy landscape at given depth within the bandgap (represented by the almost constant $\Eu$ value at $\Eg - e\Voc = 0.66$~eV), the Boltzmann term contribution to the hopping transport between localised sites is comparable. That means, the main difference comes from the spatial environment with the differing connectivity of sites for different donor fractions, expressed in a varying tunneling term.

Generally, percolation can involve directional geometric effects such as relative molecular orientation and spatial arrangement,\cite{ambrosetti2010solution} as well as energetic disorder, as the density of available sites directly depends on the DOS\cite{baranovskii2014theoretical,hofacker2020critical,hofacker2021modelling} Nevertheless, as the effective disorder is comparable across all donor fractions, we feel justified in our case to use a purely geometric percolation model for the conductivity analysis at a constant temperature and constant energy depth.

\subsubsection{The percolation model}

\begin{figure}[t]
    \centering
    \includegraphics[width=0.9\linewidth]{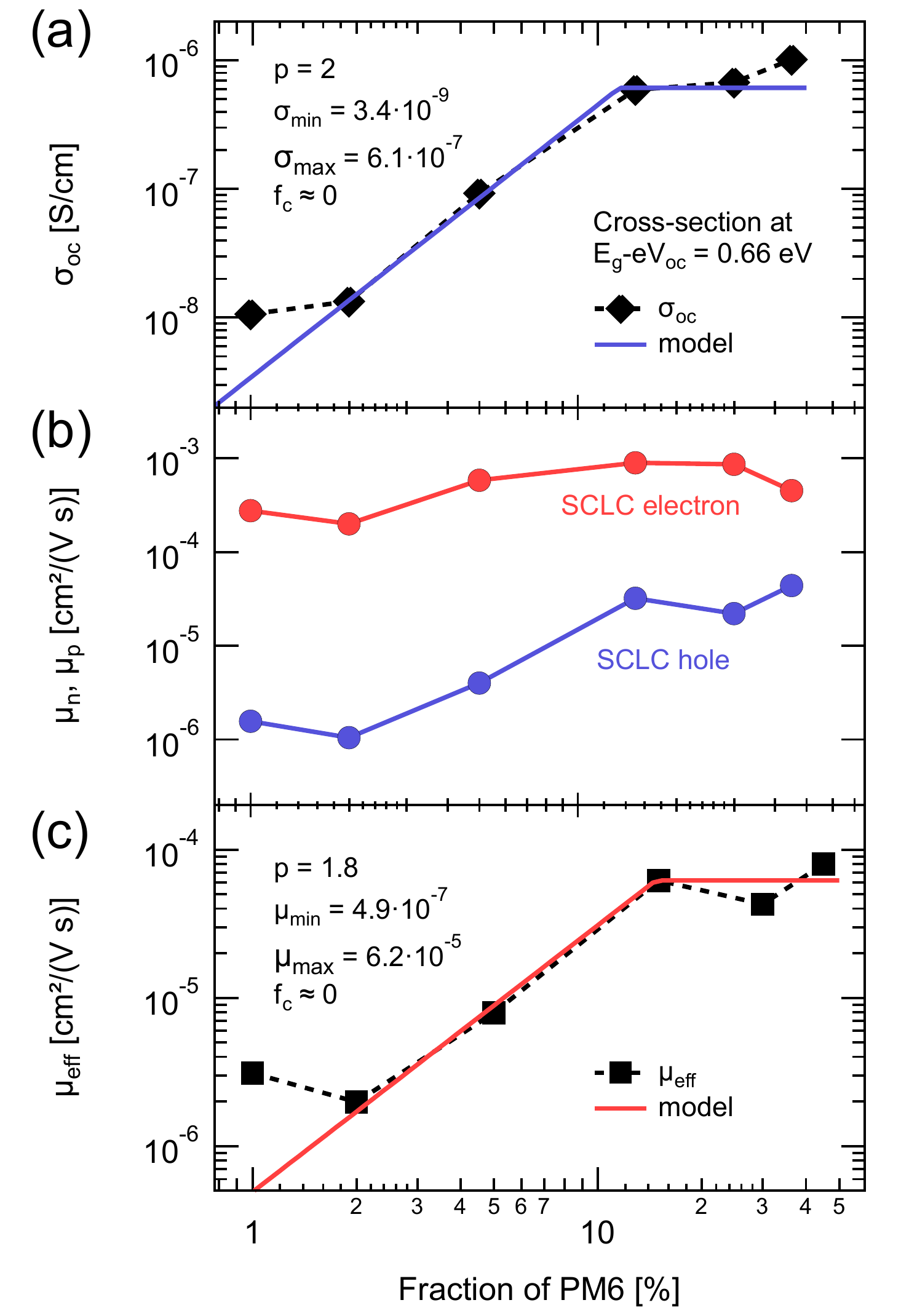}
    \caption{Percolation analysis of charge transport in PM6:Y12 blends. (a) Effective conductivity $\sigma_{\mathrm{oc}}$ at 300~K as a function of PM6 fraction, fitted with a classical percolation model, Eq.~\eqref{eq:percolation}. (b) Electron and hole mobilities extracted from SCLC measurements. (c) PM6 fraction-dependent effective mobility $\mu_{\mathrm{eff}}$, calculated as the harmonic mean of the electron and hole mobilities, fitted as in (a).} 
    \label{fig:percolation}
\end{figure}

We extracted the room-temperature values of the effective conductivity $\sigmaoc$ for our PM6:Y12 blend devices (Figure~\ref{SI_fig:sigma}, Supporting Information) at constant depth in the bandgap (at $\Eg - e\Voc = 0.66$~eV, as described in the previous section). The effective conductivity $\sigmaoc$ of the active layer shown in Figure~\ref{fig:percolation}(a) decreases strongly for lower donor fraction. In a previous work,\cite{wang2025contribution} we confirmed that the effective conductivity represents the harmonic mean of electron and hole conductivities. That means in our case, $\sigmaoc$ is dominated by the lower conductivities of holes in the PM6 donor.

The dependence of $\sigmaoc$ on the PM6 fraction can be quantitatively described using a classical percolation model:\cite{stauffer2018introduction} 
\begin{equation}\label{eq:percolation}
    \sigma = \min\bl \sigma_\mathrm{min}\cdot\bl f-f_c\br^p,\quad\sigma_\mathrm{max} \br , 
\end{equation}
where $\sigma$ is the effective conductivity, $f$ is the PM6 weight fraction, $f_c$ is the percolation threshold, $\sigma_\mathrm{min}$ is a conductivity prefactor setting the scale of the percolating network, and $\sigma_\mathrm{max}$ is the saturation conductivity of the fully connected system. We fit this model for a critical exponent of $p=2$ to $\sigmaoc$ vs.\ donor fraction $f$ at 300~K. The result is also shown in Figure~\ref{fig:percolation}(a), and confirms that the conductivity depending on $(f-f_c)^2$ follows the expectations for a three-dimensional transport network,\cite{stauffer2018introduction} demonstrating that charge transport in these blends is inherently three-dimensional. Here, percolation refers to geometrical percolation based on donor connectivity through the bulk of the active layer, rather than transport confined to specific interfaces or preferential directions. This finding implies that charge transport integrates contributions from the entire bulk morphology, consistent with the UPS depth profiling showing the absence of a strong vertical transport bottleneck even in highly diluted systems.

Geometrical percolation is closely related to the connectivity of donor domains. Similarly to our findings, Ortner et al.\cite{ortner2025donor} reported a gradual loss of donor connectivity in D18:L8-BO with decreasing donor fraction. We therefore expect that such connectivity-limited, three-dimensional charge transport can also occur in other polymer:NFA systems. The fitted percolation threshold is found to be extremely small, effectively approaching zero. Similar observations of very low percolation thresholds have been reported previously for organic nanowires and field-effect transistors.\cite{kim2012charge,goffri2006multicomponent} Such a vanishing threshold suggests that polymer chain entanglement and the intrinsic connectivity of the mixed network ensure the presence of continuous transport pathways even at very low PM6 fractions, rendering charge transport of polymer:NFA blends remarkably robust against dilution.

Together, these observations provide a consistent physical picture: while the energetic disorder increases with lower donor at shallower energies, it becomes similar deeper into the bandgap. Consequently, the variation of conductivity with PM6 content is governed primarily by connectivity of the transport network rather than by changes in the energy difference-dependent term of Miller–Abrahams hopping.

To further investigate the microscopic transport mechanism, we perform space-charge-limited current (SCLC) measurements (see Figure~\ref{fig:percolation}b). While the data for the electron-only devices can be evaluated using the Mott--Gurney equation, the hole-only devices deviate from the quadratic voltage dependence expected for a Gaussian DOS and are well described by the Mark–Helfrich model (see Supporting Information, Section~\ref{SI_sec:SCLC}).\cite{mott1948electronic,mark1962space,nicolai2011electron} Combined with the ideality factor analysis (Supporting Information, Section~\ref{SI_sec:nid}), these observations support a mixed DOS picture, in which fast electron transport is governed by a Gaussian DOS, while slow hole transport is controlled by a power-law DOS tail.\cite{saladina2023power} 

To connect the conductivity analysis with microscopic transport parameters, we convert $\sigmaoc$ into an effective charge carrier mobility $\mueff$. As recently shown, similar to the conductivities discussed above, also $\mueff$ is governed by the harmonic mean of the electron and hole mobilities.\cite{wang2025contribution} We therefore calculate $\mueff$ from the individual SCLC mobilities, $\mu_n$ and $\mu_p$, accordingly:
\begin{equation}
    \mueff = \frac{2}{\mu_n^{-1} + \mu_p^{-1}} . 
\end{equation}

Remarkably, $\mueff$ in Figure~\ref{fig:percolation}(c) follows the same percolation dependence on PM6 fraction as $\sigmaoc$, with comparable critical exponents. Here, we used $p$ as fit parameter. With a value of 1.8, it is well within the bounds of what is expected for a three-dimensional percolation model and confirms the above conclusions.

An important question concerns the role of hole transport in the Y-series acceptors, given the small HOMO-HOMO offset relative to PM6. While hole transport in Y6 has been reported previously,\cite{zhang2020delocalization} our results indicate that its contribution to macroscopic charge transport in the PM6:Y12 blends is limited. If hole transport through the acceptor phase was efficient, a much weaker dependence of $\sigmaoc$ and $\mueff$ on PM6 content would be expected. Instead, low-content PM6 devices exhibit a pronounced breakdown of both parameters, demonstrating that long-range hole transport remains primarily donor-mediated. As percolation if often connected to the percolation threshold that, in our case, is certainly below our lowest concentration of 1\%, we propose to refer to the low effective conductivity and low effective mobility as strongly topology limited.

We conclude that for comparable energetic disorder, the dependence of the effective conductivity -- determined by the harmonic mean of electron and hole conductivities -- on the donor fraction follows a simple three-dimensional percolation model. The strong decrease of the conductivity for low donor fractions make a sizable contribution of ambipolar transport -- reported for NFAs such as L8-BO,\cite{ortner2025donor} Y6,\cite{yao2021efficient,wang2022quasi} and IEICO-4F\cite{sharma2024semitransparent} -- very unlikely for Y12.

\subsection{Recombination}

To investigate nongeminate recombination in PM6:Y12 blends with varying donor content, we combine PL and TDCF measurements (see Supporting Information, sections~\ref{SI_sec:CTPL} and \ref{SI_sec:TDCF_rec}). For OSCs, the PL is dominated by photogenerated local excitons which usually recombine on the picosecond time scale -- their decay and impact was discussed already above. For nongeminate recombination, instead, we are interested in electrons and holes that recombine, forming CT excitons that can re-populate the acceptor singlet and emit from there.\cite{pranav2024}

\subsubsection{Charge carrier lifetime and recombination parameters}

\begin{figure}[t]
    \centering
    \includegraphics[width=0.9\linewidth]{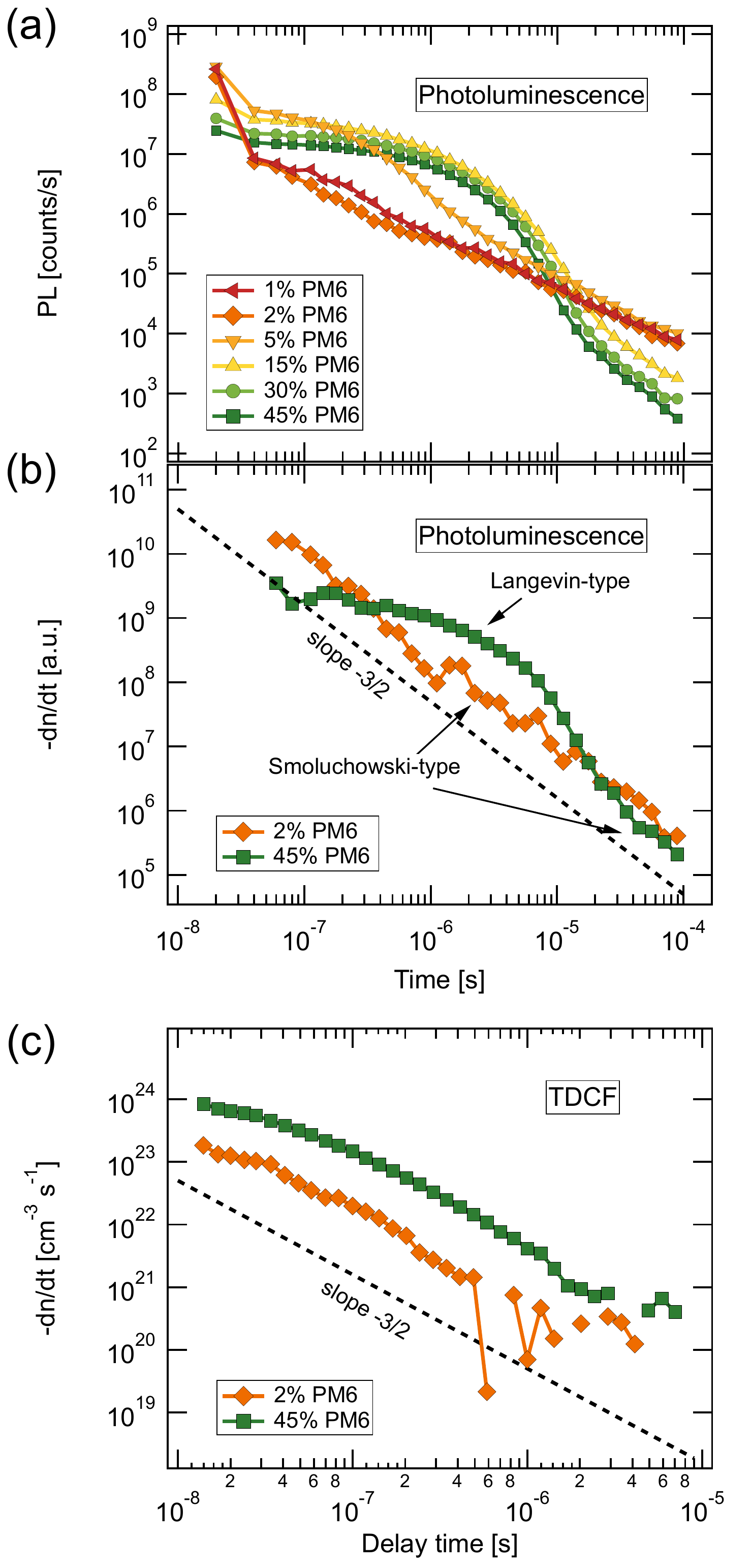}
    \caption{Nongeminate recombination dynamics in PM6:Y12 blends with varying PM6 content. (a) PL transients measured under 1~sun–equivalent excitation. (b) Recombination rates derived from PL assuming $\mathrm{PL}(t)\propto n^2(t)$, shown in arbitrary units. (c) Recombination rates extracted from TDCF. Slope $t^{-3/2}$ is indicative of Smoluchowski-type recombination.}
    \label{fig:nongeminate}
\end{figure}

The PL of mobile charge carriers was established earlier to investigate nongeminate recombination,\cite{list2023determination} and is distinguished from the sub-nanosecond local exciton decay by its much lower rates, which are laser fluence dependent. In our experiment, instead of a short laser pulse, the excitation is done by a laser operating with continuous wave before being switched off rapidly on the order of one nanosecond. Figure~\ref{fig:nongeminate}(a) shows these PL transients recorded under 1~sun-equivalent excitation. For blends with high PM6 fraction, the PL decay exhibits an initial exponential regime at below 10~ns, indicating the fast local exciton dominated emission. For the blends with donor fractions of 15\% and above, nongeminate recombination is clearly visible from \textmu{}s to ms. We will discuss the details further below.

The PL intensity is a measure of the radiative recombination rate. Previously, we reported that we could reconstruct the open-circuit voltage from the PL intensity.\cite{faisst2025implied} Consequently, $\text{PL}(t) = k_\mathrm{rad} n^2(t)$, where $k_\mathrm{rad}$ is the radiative recombination coefficient. $\Voc$, in turn, is given by the overall recombination rate -- the connection between the PL intensity and $\Voc$ is described in detail in the Supporting Information (Section~\ref{SI_sec:CTPL}). 

To discuss the nature of the PL decays, representative of the overall charge carrier decay in the solar cells, we start by discussing the recombination order $\delta$, which characterises how the recombination rate scales with charge carrier density. While power-law decays are commonly observed in organic semiconductors,\cite{nelson2003,gorenflot2014} they are most often attributed to time-dispersive transport arising from recombination of mobile with trapped charge carriers, i.e., a consequence of energetic disorder and a broad distribution of hopping times.\cite{scher1975anomalous,bassler1993charge,kurpiers2016dispersive} For trap-assisted recombination in an exponential DOS, the carrier concentration is expected to follow $n(t) \propto t^{-1/(\delta-1)}$, in which case the recombination order $\delta = \Eu/kT + 1$ is given by the characteristic energy $\Eu$. For a $PL(t) \propto n(t)^2$, we would expect a power-law exponent of $-2/(\delta-1)$. Some simple examples: If $\delta = 2$, that means $\Eu \sim kT$ and traps do not play a practical role, and the PL exponent would be --2. This example is the simplest, because it allows to understand this mode of electron--hole recombination as Langevin-type recombination: this encounter-limited mechanism implies that Coulomb-attraction of charge carriers, if larger than the thermal energy, finally leads to a recombination rate that is determined by the sum of electron and hole mobilities. In disordered systems in which electrons and holes have to travel to an interface, recombination orders higher than two are possible. For a recombination order of $\delta = 3$ at room temperature, this could mean $\Eu =$~50~meV. The PL exponent would be --1. Similarly, we can estimate the recombination order under the assumption of the trap-assisted recombination for a given PL exponent. If it was 3, the recombination order would have to be 1.67. However, if $\Eu$ is smaller than $kT$, the charge carriers are not trapped, in which case the recombination order remains at 2. We will come back to this point shortly.

We evaluate the PL transients shown Figure~\ref{fig:nongeminate}(a) with respect to the recombination order. For the blends with donor fractions of 15\% and above, we observe a recombination order of slightly below 2 at the transition from \textmu{}s to ms. For the sample with 5\%, this regime is already suppressed and shifted to shorter time scales, and for the lower fractions only a slight bump around 200~ns hints at this recombination mode. We will later discuss these decays in the context of the Langevin model. For all fractions, on longer time scales -- at 20~\textmu{}s for the higher donor fractions, but starting already below a \textmu{}s for the 1\% and 2\% samples, this is followed by a crossover to a power-law decay at longer times. The apparent recombination order (PL exponent) of this decay is at around 3. As discussed in the previous paragraph, this implies that $\Eu < kT$, in which case the recombination order should be 2. Therefore, we will propose an alternative model after the next section.

We take the integrated lifetime determined from the charge carrier-PL measurements as effective lifetime. It reflects the total recombination rate of all charge carriers, including both the ones recombining radiatively and nonradiatively. This relation allows the time-dependent carrier density $n(t)$ to be inferred from the PL decay, and the total recombination rate to be obtained as $R(t) = -\der n/\der t = n/\tau(n) = k_2 n^2$. Similarly, we determine the recombination parameters from the TDCF measurements of a subset of devices. We note that both recombination parameters are effective, and are useful to compare the recombination rates with predictions or literature. The 2nd order recombination coefficients $k_2$, are shown in Figure~\ref{fig:redcoef}(a). We will discuss them with respect to Langevin recombination in the next section. 

\subsubsection{Langevin-type recombination}

\begin{figure}[b]
    \centering
    \includegraphics[width=0.9\linewidth]{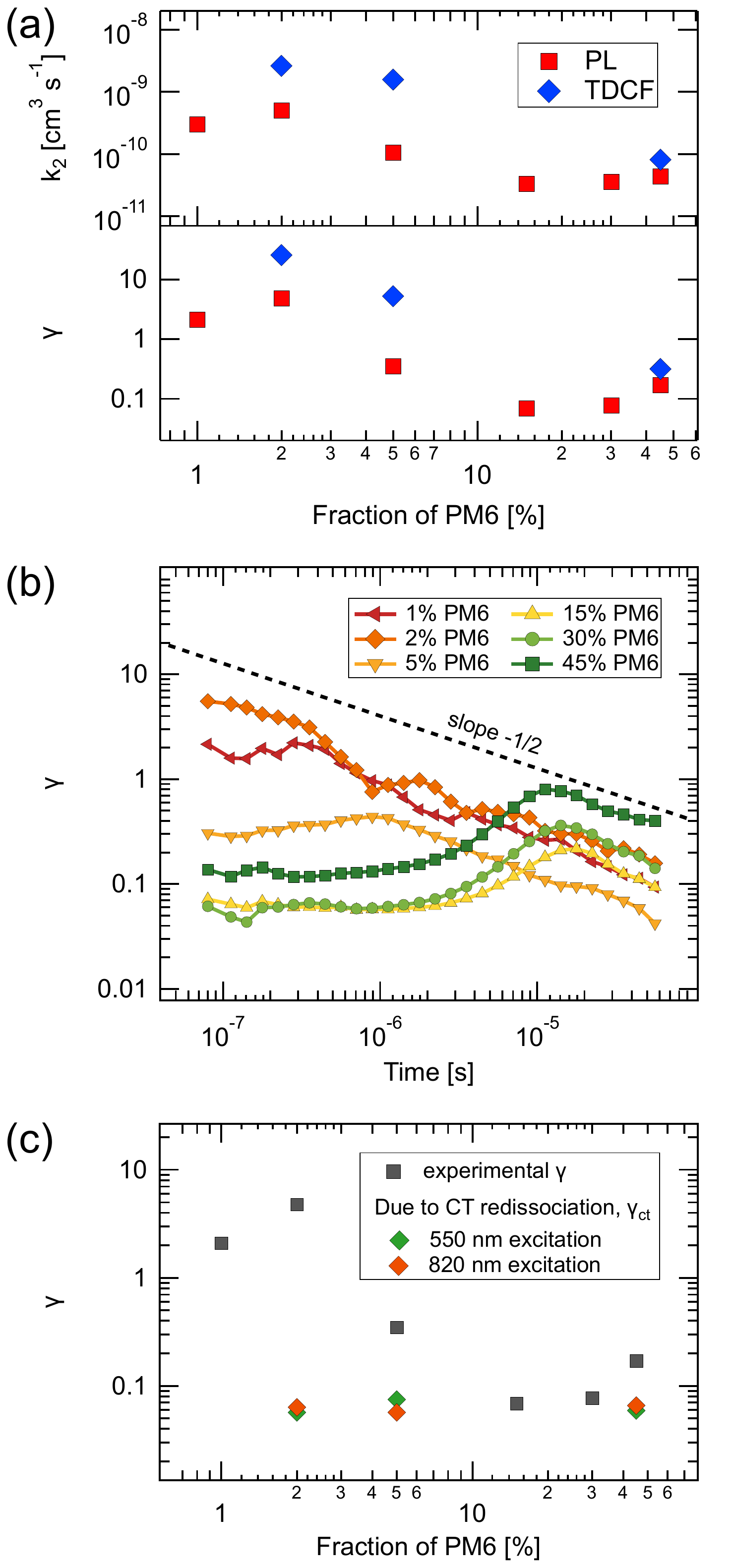}
    \caption{Recombination parameters of PM6:Y12 blends with varying PM6 content. (a) Effective $k_2$ and corresponding Langevin reduction factors $\gamma$ extracted from PL and TDCF. (b) Time-dependent $\gamma(t)$ obtained from PL. (c) Calculated $\gamma_\mathrm{ct}$, the Langevin reduction factor due to CT redissociation, estimated from $\etadiss$ at $\Voc$.}
    \label{fig:redcoef}
\end{figure}

We analyze the data within the Langevin model for encounter-limited recombination upon Coulomb attraction of the charge carriers. We determine the effective recombination coefficient $k_2$ from both PL and TDCF-derived recombination rates (see Supporting Information, sections~\ref{SI_sec:CTPL} and \ref{SI_sec:TDCF_rec} for details). They are shown in Figure~\ref{fig:redcoef}(a), together with the the Langevin reduction factor $\gamma$ that we calculated using the relation\cite{langevin1903recombinaison,koster2006bimolecular}
\begin{equation}
    k_2 = \gamma \cdot \frac{e\bl \mu_n + \mu_p \br}{\varepsilon\varepsilon_0}.
\end{equation}
Here, $e$ is the elementary charge, $\mu_n$ and $\mu_p$ are the electron and hole mobilities, respectively, $\varepsilon$ is the relative permittivity of the active layer, and $\varepsilon_0$ is the vacuum permittivity. The charge carrier mobilities were taken from SCLC measurements.

The extracted $k_2$ values and corresponding Langevin reduction factors are shown in Figure~\ref{fig:redcoef}(a) for PL (under 1~sun–equivalent excitation) and for TDCF (under comparable charge carrier density conditions). For blends with high PM6 content, the reduction factor $\gamma$ is significantly smaller than unity. This finding is consistent with reduced recombination relative to the Langevin prediction, and also consistent with reports in literature on PM6:Y12 blends with donor fractions of 15\% or higher.\cite{haffner2025high} In contrast, for low PM6 content, $\gamma$ approaches or even exceeds unity. Values of $\gamma>1$ indicate recombination rates that are faster than expected from a Langevin picture and point to the breakdown of the assumptions of Langevin theory.  We are not aware of reports of this behavior in literature. Usually, the effects discussed in literature mention \emph{reduced} recombination as compared to Langevin theory -- e.g., spatial inhomogeneities,\cite{deibel2009prb} phase separation,\cite{koster2006bimolecular,heiber2015prl} or redissociation\cite{koster2005light} -- but not higher rates than the Langevin ones. We speculate that the spatially trapped charge carriers, due to the low donor connectivity, generate local electric fields that lead to higher local carrier mobilities and could increase the recombination rate beyond Langevin. We discuss a Smoluchowski-type recombination framework in more detail later.

First, we will focus on the solar cells which fit the expectations of reduced Langevin recombination. A commonly assumed origin of reduced Langevin recombination is the redissociation of CT states into free electrons and holes, which leads to an effective reduction factor $\gamma_\mathrm{ct} = 1 - \etadiss$, as proposed by the Blom group.\cite{koster2005light} To assess whether this mechanism can account for the experimentally observed reduction factors, we determine the CT dissociation efficiency $\etadiss$ at $\Voc$ from TDCF measurements for devices with 2\%, 5\%, and 45\% PM6 content and calculate the corresponding $\gamma_\mathrm{ct}$. As shown in Figure~\ref{fig:redcoef}(c), the values of $\gamma_\mathrm{ct}$ are of the same order of magnitude as the experimentally extracted $\gamma$ for PM6 contents of 15\% and higher. This demonstrates that efficient charge photogeneration can mediate redissociation before recombination occurs, effectively reducing the loss rate. The deviation at 5\% PM6 may reflect the incomplete extraction of charge carriers in the TDCF method, a result of the low connectivity that is also represented by the low hole mobility, so that the carrier concentration that is extracted will underestimate the concentration present during recombination. For a given effective lifetime $\tau = 1/k_2n$, underestimation of $n$ leads to overestimation of $k_2$. For the most diluted blends (1\% and 2\% PM6), the experimentally derived $\gamma$ values exceeding one are not physically meaningful. In the next section, we will propose a rationalization that involved a different mode of recombination that is governed by Smoluchowski-type kinetics rather than Langevin recombination. 
The observed decrease of the experimentally derived apparent reduction factor with increasing PM6 content does not reflect a progressive suppression of recombination, but rather the gradual recovery of a regime in which a Langevin-type description becomes meaningful. At low PM6 fractions, recombination is governed by spatially dispersive kinetics, rendering the extracted reduction factors unphysical. As donor connectivity improves, charge transport becomes less constrained, spatial correlations weaken, and recombination increasingly approaches the scenario to which Langevin-type recombination fits. Consequently, the apparent reduction factor decreases and converges towards values consistent with CT-exciton redissociation, rather than indicating enhanced recombination suppression at high donor content.

As mentioned above, there is another explanation for power-law decays of nongeminately recombining charge carriers. A power-law decay can arise purely from diffusion statistics, even in a spatially homogeneous medium. In the long-time limit, recombination can be described by Smoluchowski-type kinetics, where charge carriers must diffuse through space before recombining. We will present this process in the next section, and discuss it with respect to our experimental results.

\subsubsection{Smoluchowski-type recombination}

In contrast to the nature of the Langevin-type, which makes the recombination rate proportional to the sum of well-defined electron and hole mobilities in the original theory, we interpret the Smoluchowski-type recombination as dispersive in nature.\cite{smoluchowski1918versuch} As diffusion increases the average separation between carriers, the encounter probability decreases with time. As a result, the classical Langevin recombination gives rise to characteristic timescales, while Smoluchowski recombination leads to scale-free, power-law kinetics. Solving the diffusion equation for recombination rate yields the characteristic asymptotic behavior\cite{mozumder1968theory,abell1972application,hong1978solution}
\begin{equation}
    t\to\infty,\quad R\bl t \br \to t^{-3/2} . 
\end{equation}
This $t^{-3/2}$ scaling reflects the universal statistics of such spatially driven diffusion-limited encounters\cite{hong1978solution} and arises independently of energetic disorder, trap distributions, or specific hopping parameters. It is therefore important to distinguish between these two modes of recombination.

Direct experimental reports explicitly identifying a $\tpl$ power-law in recombination kinetics of organic semiconductors are scarce. In studies of Me-LPPP:PCBM blends, Nikitenko et al.\ and later Kahle et al.\ reported PL decays of CT excitons with exponents close to $-3/2$, which were attributed to \emph{geminate} recombination of bound electron--hole pairs.\cite{nikitenko2001dispersive,kahle2018interpret} We have also observed the same decay exponents in PL of small molecule--fullerene blend systems at 4~K (unpublished data). All of these studies have in common that they were done upon pulsed laser excitation.
Similar decay exponents have also been observed in PL studies of amorphous inorganic semiconductors.\cite{noolandi1980geminate,murayama1985photoluminescence}

In contrast to these earlier studies, which focus on CT-state recombination in fullerene-based systems, the situation is fundamentally different in the NFA blends that we studied. Here, the energetic separation between CT excitons and local singlet excitons in the acceptor are small,\cite{morteani2005,classen2020} such that radiative recombination is no longer uniquely associated with a distinct CT manifold. Consequently, the PL signal -- on the time scale of 10~ns and longer -- predominantly reflects emission from spatially separated charge carriers that nongeminately meet at the donor--acceptor heterojunction and recombine. The previous reports of $PL \propto t^{-3/2}$ in organic semiconductor blends\cite{nikitenko2001dispersive,kahle2018interpret} represent geminate recombination. Here, we propose that in particular the blends with low donor fraction, but -- on long time-scales -- all devices that we studied show Smoluchowski-type \emph{nongeminate} recombination.

In a Smoluchowski regime the long-time kinetics obey $R(t) = -\der n/\der t \propto t^{-3/2}$, while the charge carrier density decays as $n(t) \propto t^{-1/2}$. Defining a 2nd order, nongeminate recombination coefficient as 
$$k_2(t) = \frac{R(t)}{n^2(t)}$$
therefore yields
$$k_2(t) \propto t^{-1/2} . $$
Consequently, the apparent Langevin reduction factor becomes time dependent, 
$$\gamma_\mathrm{app}(t) = \frac{k_2(t)}{k_L} \propto t^{-1/2} , $$
even if the underlying physical recombination mechanism is unchanged. Thus, in a Smoluchowski recombination regime, forcing a Langevin analysis necessarily produces a time-dependent effective recombination coefficient and reduction factor, which no longer reflect intrinsic material parameters but instead encode the dispersive encounter statistics. This behavior is directly observed in Figure~\ref{fig:redcoef}(b) for $\gamma(t)$, and in Figure~\ref{SI_fig:CTPL_k2} for $k_2(t)$. Both follow a $t^{-1/2}$ dependence at long times for higher PM6 concentrations, and over the whole time frame for 1\% and 2\% PM6 blends. Under these conditions, the characteristic $\tpl$ behavior emerges for nongeminate recombination, consistent with our observation that the \emph{total, nongeminate} recombination rate of charge carriers derived from PL obeys the same power-law decay. We propose that this dependence demonstrates that the $\tpl$ power law is not specific to CT exciton recombination, but represents a more general signature of dispersive charge carrier encounters in organic donor--acceptor blends. To our knowledge, this constitutes the first report of such behavior for nongeminate recombination in organic donor–acceptor systems.

Notably, all recombination rates in Figure~\ref{fig:nongeminate}(b) converge to a common slope of $s = -3/2$ at long times ($\gtrsim 10$~\textmu s), demonstrating that Smoluchowski-type recombination dominates the late-time kinetics even in blends with high PM6 content. This observation raises the question of how broadly applicable the Langevin framework is under typical experimental conditions. While early-time dynamics at high carrier densities may be approximated by Langevin-type kinetics, some experimental techniques probe long times, low charge carrier densities, or regimes where spatial correlations become important. Under such conditions, nongeminate recombination can be governed by diffusion-limited encounters, and a Langevin interpretation may no longer appropriate. 

Langevin-type nongeminate recombination in principle involves recombination coefficients that are time dependent: in case of higher order recombination $\delta > 2$ due to trap-assisted recombination, $k_2$ and $\gamma$  depend on carrier concentration $n(t)$ and, therefore, indirectly on time. Still, the resulting slopes would always be different than the ones for the Smoluchowski-type recombination. True Langevin-type 2nd order recombination is observed in the regime where $k_2$ and $\gamma$ are time independent, corresponding to the early-time, high-density region in Figures~\ref{SI_fig:CTPL_k2} and \ref{fig:redcoef}(b). The blend with 5\% PM6 represents an intermediate case. Here, a short time window of approximately time-independent $k_2$ and $\gamma$ can still be identified, however, this regime is rapidly followed by dispersive behavior, demonstrating that Smoluchowski-type recombination quickly dominates once charge carrier separation increases. Notably, RSoXS reveal the emergence of larger ordered structures at 5\% PM6, suggesting the presence of locally connected donor domains. This highlights the extreme sensitivity of the recombination mechanism to donor connectivity in the diluted regime. We expect Smoluchowski-type kinetics also to apply to other systems constrained by domain connectivity, e.g., in more amorphous materials where connectivity likely breaks down more rapidly with dilution, Smoluchowski-type recombination is probably more dominant. Further validation in different material systems remains to be explored.

Finally, the apparent difference between the recombination kinetics derived from PL and TDCF in Figures~\ref{fig:nongeminate}(b) and (c) originates from the fundamentally different charge carrier populations probed by the two techniques. PL monitors radiative recombination events and therefore is sensitive to locally recombining electron--hole pairs, which dominate the early-time dynamics and typically exhibit Langevin–type behavior. In contrast, TDCF does not probe recombination continuously during the delay but selectively measures the charge carriers that survive the delay and remain extractable. As a result, for connectivity-limited systems towards lower donor fractions, the recombination coefficients can easily be overestimated with TDCF, whereas for the well-performing solar cells, TDCF gives rather accurate results. The advantage of the PL spectroscopy is that, while only having indirect access to the carrier concentrations, it allows to observe recombination where and when it occurs.

\subsection{Collection efficiency: bringing transport and recombination together}

In this section, we combine charge transport and recombination to understand the influence of PM6 content on the device performance, especially the fill factor. The IQE loss analysis in Figure~\ref{fig:IQE}(b) indicates that charge collection is a limiting process for all devices, but becomes more limiting towards lower donor fractions. In OSCs, the intrinsically low carrier mobilities make the competition between charge transport and nongeminate recombination a decisive factor particularly for the $\FF$.\cite{neher2016new,wang2025transport} This corresponds to what was long discussed as the $\mu\tau$-product.\cite{hecht1932,crandall1983modeling} However, as the recombination is not a first order process in OSCs,\cite{deibel2010comment} and $\mu$ -- in the case of charge collection -- is an effective mobility that depends on the harmonic mean of electron and hole mobilities, we discuss the interplay of recombination and transport in more appropriate terms.


The figure‑of‑merit $\alpha$ quantifies the transport‑resistance loss that appears as a flattening of the illuminated JV curve near the open‑circuit voltage.\cite{neher2016new} The slope of an illuminated JV curve around $\Voc$ is reduced as compared to a hypothetical JV curve with infinite conductivity, as shown in Figure~\ref{fig:sigma}(a). In the analytical description, the slope of the JV characteristics at $\Voc$ is proportional to the apparent ideality factor $\nid+\alpha$. Generally, $\alpha$ is defined as\cite{saladina2024transport} 
\begin{align}\label{eq:alpha}
    \alpha = \frac{eL}{\kT} \cdot \frac{\jgen}{\sigmaoc}.
\end{align}
It captures the extra voltage drop caused by the finite conductivity of the active layer. As $\alpha$ is defined only at $\Voc$, it does not correctly describe the curvature of the JV curve at the maximum power point (mpp), where the voltage loss is larger. Accordingly, despite its merit, $\alpha$ is not a good and general predictor of the $\FF$.

To overcome this limitation we introduced the more general parameter $\beta$\cite{saladina2024transport} to replace $\alpha$ in the apparent ideality factor, which becomes $\nid+\beta$. $\beta$ can be determined for any voltage, not just $\Voc$. It is most relevant if evaluated at the voltage at which the solar cell is operating at mpp under illumination. Earlier, we have shown that this figure-of-merit predicts the $\FF$ universally for JV curves measured under a wide range of temperatures and light intensities, and is independent of specific material system.\cite{saladina2024transport} $\beta$ can be obtained iteratively: it requires the values of $\nid$, the corresponding ideality factor for the conductivity $\nsig$, $\alpha$, and $\Voc$. This procedure yields the figure-of-merit $\beta$ that accurately captures the transport‑induced series resistance throughout the whole JV curve. We refer the reader to Saladina et al.\cite{saladina2024transport} -- in particular section~S9B of its Supporting Information -- for the step-by-step explanation and demonstration of how well this method predicts the $\FF$.

\begin{figure}[t]
    \centering
    \includegraphics[width=0.9\linewidth]{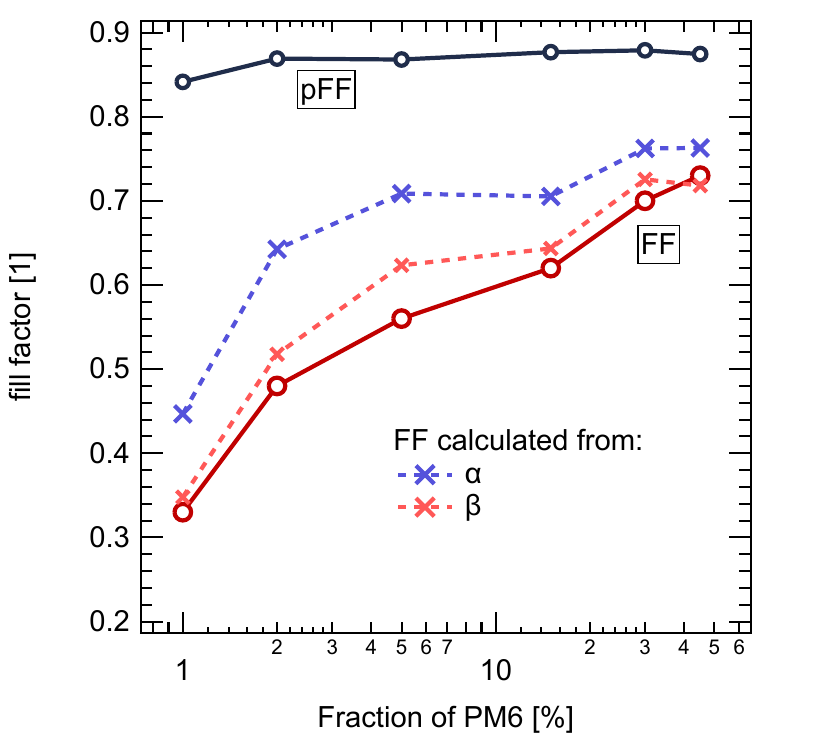}
    \caption{The measured and the predicted fill factors using the Green equation\cite{green_accuracy_1982} based on the figures-of-merit $\alpha$ and $\beta$ the reflect the impact of the effective charge carrier mobility at open circuit and mpp, respectively.}
    \label{fig:ff_alpha_beta}
\end{figure}

Figure~\ref{fig:ff_alpha_beta} shows the measured $\FF$ of devices as a function of PM6 fraction (circle with red solid line). Predicted $\FF$ values from the apparent ideality factor, including $\alpha$ (cross with blue dashed line) or $\beta$ (cross with red dashed line), are calculated using Green equation,\cite{green_accuracy_1982} 
\begin{equation}
\begin{aligned}
FF &= \frac{v_{oc} - \ln\!\left(v_{oc} + 0.72\right)}{v_{oc} + 1}, \\
v_{oc} &= \frac{e V_{oc}}{n_{\mathrm{app}} k_B T}.
\end{aligned}
\label{eq:FF_green}
\end{equation}
with the apparent ideality factor, $\napp$, contained in the normalised open-circuit voltage. We calculated $\alpha$ using $\sigmaoc$, which as determined by Equation~\ref{eq:sigmaoc} from the JV curves at 1~sun. 

We obtained a predictive $\FF$ from $\alpha$, as shown in Figure~\ref{fig:ff_alpha_beta}. The trend fits rather well, but generally overestimates the $\FF$. This discrepancy stems from the different conditions of solar cells when defining $\alpha$ (open-circuit) and the $\FF$ (mpp) as discussed above. To consider the figure-of-merit $\beta$ that can be determined at mpp, we applied the iterative approach,\cite{saladina2024transport} 
\begin{equation}
\begin{aligned}
&\beta = \alpha \,
\frac{v_{oc}\left(v_{oc}+1\right)^{\frac{\nid}{\nsig}-1}}
{\ln\!\left(v_{oc}+1\right)}, \\
&v_{oc}
= \frac{e V_{oc}}{\left(n_{\mathrm{id}}+\beta\right) \kT}.
\end{aligned}
\end{equation}
It can be seen that the calculated $\FF$ from $\beta$ predicts the device $\FF$ over the entire PM6 composition range rather well, and better than $\alpha$. This not only validates the feasibility of the iterative extraction of $\beta$ but also strongly supports that $\beta$ is an inherent figure-of-merit that directly reflects the transport/collection losses and dominates the device $\FF$.

In the diluted donor-content devices, electron and hole transport exhibit pronounced asymmetry, as shown in Figure~\ref{fig:percolation}(b). We recently experimentally confirmed that a harmonic mean rules the effective conductivity and mobility.\cite{wang2025contribution} This means that the slower charge carriers will determine the transport resistance and charge extraction yield, and therefore the $\FF$. The imbalanced conductivity caused by donor dilution leads to a rapid decrease in effective conductivity and an increase in $\beta$. Instead, when the lower mobility increases towards $\mu_n \approx \mu_p$, as we increase the PM6 content, the harmonic mean approaches the individual mobilities, $\beta$ becomes small, and the $\FF$ increases.

Accordingly, $\beta$ directly reflects the effective conductivity at mpp, the $\FF$ depends on the harmonic mean of electron and hole mobilities, $\mu_n$ and $\mu_p$, respectively.\cite{wang2025contribution} The transport loss is, therefore, governed by the slower carrier. A large imbalance raises $\beta$, enlarges the transport voltage loss, and reduces the $\FF$. When the value of the conductivities increases overall, the $\FF$ approaches the limit given by the $\pFF$, where recombination is determining the losses and charge collection is perfect with infinite effective conductivity.

Taken together, these results demonstrate that the $\FF$ in donor-diluted organic solar cells is primarily governed by transport and collection losses arising from an imbalance between electron and hole transport. Balancing—and simultaneously enhancing (both) mobilities minimise $\beta$ and enable the solar cells close to the $\pFF$ limit.

\section{Conclusion}

In this work, we have presented a thorough investigation of PM6:Y12 organic bulk heterojunction solar cells in which the donor (PM6) weight fraction was systematically varied from 1\% to 45\%. We combined complementary experimental techniques such as time-resolved photoluminescence not just for local excitons, but to monitor charge carrier recombination, X-ray scattering, photoelectron spectroscopy depth profiling, ellipsometry, and extensive optoelectronic characterization from external quantum efficiency spectroscopy over time-delayed collection field method and, last not least, temperature-dependent current--voltage measurements on solar cells.

The organic solar cells at 45\% efficiency have a power conversion efficiency corresponding to 15\% under AM1.5G conditions, whereas the devices with low donor fraction suffer from low fill factors and short circuit current densities. Nevertheless, even at the lowest donor fractions of 5\% or less, the PM6 phase begins to form a percolating network of nanofibrillar domains on the order around 200~nm lateral size. GIWAXS shows the emergence of a lamellar PM6 stacking peak already at 5\% donor content, while RSoXS confirms a heterogeneous but interconnected nanomorphology. We observe a vertical concentration gradient with UPS depth profiling, but fortunately, the thin PM6-rich surface layer does not impede charge extraction. 

Investigating the charge transport properties, we find that a network topology rather than a classical percolation threshold govern the dependence on the donor fraction. The effective conductivities of the active layer -- extracted from the slope of the JV curve at open-circuit voltage -- follow a simple three-dimensional percolation law $\sigma = \min(\sigma_{\min}(f - f_c)^p, \sigma_{\max})$, with a critical exponent $p \approx 2$ and an effectively vanishing percolation threshold $f_c \approx 0$. This demonstrates that the donor network retains continuous charge-transport pathways even at 1\% fraction. Nevertheless, we do not see evidence for reports that ambipolar charge transport in NFAs supports the hole transport at low donor concentrations by pathways through Y12 domains.

We observe a transition of the nongeminate recombination mechanism with donor content. At high donor fractions $\ge$~15\%, recombination is well described by the Langevin model, with reduction factors $\gamma < 1$ dominated by the high redissociation probability after encounter. When the donor fraction falls below 10\%, the recombination kinetics become dispersive: both the photoluminescence -- at times $> 10$~ns reflecting nongeminate charge carrier recombination rather than local exciton decay -- and TDCF measurements reveal a power-law decay $R(t) \propto t^{-3/2}$ for the \emph{total, nongeminate} recombination rate. Such decays are characteristic of Smoluchowski-type recombination. In this regime, the apparent Langevin reduction factor can exceed unity, strongly suggesting that the standard Langevin framework no longer captures the dominant loss pathway. While this Smoluchowski-type recombination has been observed previously for \emph{geminate} CT exciton recombination, we are not aware of other reports showing this behavior for nongeminate losses.

The fill factor is limited by transport resistance, which is not surprising as this is the case even for the best organic solar cells reported to date. Interesting is, rather, how this limitation changes with composition. The pseudo-fill factor ($\pFF$), calculated from the ideality factor, remains high, whereas the measured fill factor ($\FF$) drops markedly for low donor fractions. By applying the transport-resistance figure-of-merit $\beta$, evaluated at the maximum-power point, we show that the conductivity of the slower charge carrier, the hole on the topology-challenged donor, dominates the effective conductivity $\sigmaeff$ and the transport-resistance loss. The effective mobility $\mueff$ follows the same percolation scaling, supporting the idea that mobility imbalance and limited connectivity jointly lead to the $\FF$ reduction at lower donor fractions. 

Combining our findings on morphology, transport, and recombination of donor-diluted solar cells, we find that low donor fractions mainly preserve the interfacial donor--acceptor area while progressively reducing the connectivity of the hole transport pathways. This topology-controlled transport, together with the crossover from Langevin to Smoluchowski-type recombination, fully explains the observed reduction of the device performance towards 1\% dilution. 

Overall, our results demonstrate that organic bulk-heterojunction solar cells can tolerate very strong donor dilution without losing charge-generation efficiency, provided that a continuous donor network is maintained. The major limiting factors are the efficient charge extraction, reducing both the fill factor and the short circuit current density of the low donor fraction devices. Our findings allow to study the complex interplay of generation, recombination and transport within the same material system, thus providing new insights into the physics of organic photovoltaics.

\section*{Acknowledgements}

The authors thank the Deutsche Forschungsgemeinschaft (DFG) for funding this work (Research Unit FOR 5387 POPULAR, project no.~461909888). M.K.\ and E.M.H.\ also acknowledge use of DFG INST91/443-1 (438562776).

\section*{Competing interests}
We declare that we have no competing interests.

\section*{Data availability}
We will either upload the data to Zenodo and provide the link here, or provide the data upon request.

\bibliographystyle{apsrev4-2}
\bibliography{references}

\begin{thebibliography}{88}%
\makeatletter
\providecommand \@ifxundefined [1]{%
 \@ifx{#1\undefined}
}%
\providecommand \@ifnum [1]{%
 \ifnum #1\expandafter \@firstoftwo
 \else \expandafter \@secondoftwo
 \fi
}%
\providecommand \@ifx [1]{%
 \ifx #1\expandafter \@firstoftwo
 \else \expandafter \@secondoftwo
 \fi
}%
\providecommand \natexlab [1]{#1}%
\providecommand \enquote  [1]{``#1''}%
\providecommand \bibnamefont  [1]{#1}%
\providecommand \bibfnamefont [1]{#1}%
\providecommand \citenamefont [1]{#1}%
\providecommand \href@noop [0]{\@secondoftwo}%
\providecommand \href [0]{\begingroup \@sanitize@url \@href}%
\providecommand \@href[1]{\@@startlink{#1}\@@href}%
\providecommand \@@href[1]{\endgroup#1\@@endlink}%
\providecommand \@sanitize@url [0]{\catcode `\\12\catcode `\$12\catcode `\&12\catcode `\#12\catcode `\^12\catcode `\_12\catcode `\%12\relax}%
\providecommand \@@startlink[1]{}%
\providecommand \@@endlink[0]{}%
\providecommand \url  [0]{\begingroup\@sanitize@url \@url }%
\providecommand \@url [1]{\endgroup\@href {#1}{\urlprefix }}%
\providecommand \urlprefix  [0]{URL }%
\providecommand \Eprint [0]{\href }%
\providecommand \doibase [0]{https://doi.org/}%
\providecommand \selectlanguage [0]{\@gobble}%
\providecommand \bibinfo  [0]{\@secondoftwo}%
\providecommand \bibfield  [0]{\@secondoftwo}%
\providecommand \translation [1]{[#1]}%
\providecommand \BibitemOpen [0]{}%
\providecommand \bibitemStop [0]{}%
\providecommand \bibitemNoStop [0]{.\EOS\space}%
\providecommand \EOS [0]{\spacefactor3000\relax}%
\providecommand \BibitemShut  [1]{\csname bibitem#1\endcsname}%
\let\auto@bib@innerbib\@empty
\bibitem [{\citenamefont {Fu}\ \emph {et~al.}(2025)\citenamefont {Fu}, \citenamefont {Li}, \citenamefont {Liu}, \citenamefont {Huang}, \citenamefont {Chen}, \citenamefont {Fong}, \citenamefont {Dela~Pena}, \citenamefont {Li}, \citenamefont {Lu}, \citenamefont {Cheng}, \citenamefont {Xiao}, \citenamefont {Lu},\ and\ \citenamefont {Li}}]{fu2025two}%
  \BibitemOpen
  \bibfield  {author} {\bibinfo {author} {\bibfnamefont {J.}~\bibnamefont {Fu}}, \bibinfo {author} {\bibfnamefont {H.}~\bibnamefont {Li}}, \bibinfo {author} {\bibfnamefont {H.}~\bibnamefont {Liu}}, \bibinfo {author} {\bibfnamefont {P.}~\bibnamefont {Huang}}, \bibinfo {author} {\bibfnamefont {H.}~\bibnamefont {Chen}}, \bibinfo {author} {\bibfnamefont {P.~W.}\ \bibnamefont {Fong}}, \bibinfo {author} {\bibfnamefont {T.~A.}\ \bibnamefont {Dela~Pena}}, \bibinfo {author} {\bibfnamefont {M.}~\bibnamefont {Li}}, \bibinfo {author} {\bibfnamefont {X.}~\bibnamefont {Lu}}, \bibinfo {author} {\bibfnamefont {P.}~\bibnamefont {Cheng}}, \bibinfo {author} {\bibfnamefont {Z.}~\bibnamefont {Xiao}}, \bibinfo {author} {\bibfnamefont {S.}~\bibnamefont {Lu}},\ and\ \bibinfo {author} {\bibfnamefont {G.}~\bibnamefont {Li}},\ }\bibfield  {title} {\bibinfo {title} {Two-step crystallization modulated through acenaphthene enabling 21\% binary organic solar cells and 83.2\% fill factor},\ }\href@noop {} {\bibfield  {journal} {\bibinfo
  {journal} {Nature Energy}\ }\textbf {\bibinfo {volume} {10}},\ \bibinfo {pages} {1251} (\bibinfo {year} {2025})}\BibitemShut {NoStop}%
\bibitem [{\citenamefont {Jiang}\ \emph {et~al.}(2024)\citenamefont {Jiang}, \citenamefont {Sun}, \citenamefont {Xu}, \citenamefont {Liu}, \citenamefont {Miao}, \citenamefont {Ran}, \citenamefont {Liu}, \citenamefont {Yi}, \citenamefont {Zhang},\ and\ \citenamefont {Zhu}}]{jiang2024non}%
  \BibitemOpen
  \bibfield  {author} {\bibinfo {author} {\bibfnamefont {Y.}~\bibnamefont {Jiang}}, \bibinfo {author} {\bibfnamefont {S.}~\bibnamefont {Sun}}, \bibinfo {author} {\bibfnamefont {R.}~\bibnamefont {Xu}}, \bibinfo {author} {\bibfnamefont {F.}~\bibnamefont {Liu}}, \bibinfo {author} {\bibfnamefont {X.}~\bibnamefont {Miao}}, \bibinfo {author} {\bibfnamefont {G.}~\bibnamefont {Ran}}, \bibinfo {author} {\bibfnamefont {K.}~\bibnamefont {Liu}}, \bibinfo {author} {\bibfnamefont {Y.}~\bibnamefont {Yi}}, \bibinfo {author} {\bibfnamefont {W.}~\bibnamefont {Zhang}},\ and\ \bibinfo {author} {\bibfnamefont {X.}~\bibnamefont {Zhu}},\ }\bibfield  {title} {\bibinfo {title} {Non-fullerene acceptor with asymmetric structure and phenyl-substituted alkyl side chain for 20.2\% efficiency organic solar cells},\ }\href@noop {} {\bibfield  {journal} {\bibinfo  {journal} {Nature Energy}\ }\textbf {\bibinfo {volume} {9}},\ \bibinfo {pages} {975} (\bibinfo {year} {2024})}\BibitemShut {NoStop}%
\bibitem [{\citenamefont {Li}\ \emph {et~al.}(2025)\citenamefont {Li}, \citenamefont {Song}, \citenamefont {Lai}, \citenamefont {Zhang}, \citenamefont {Zhou}, \citenamefont {Xu}, \citenamefont {Huang}, \citenamefont {Liu}, \citenamefont {Gao}, \citenamefont {Li}, \citenamefont {Jee}, \citenamefont {Zheng}, \citenamefont {Liu}, \citenamefont {Yan}, \citenamefont {Chen}, \citenamefont {Tang}, \citenamefont {Zhang}, \citenamefont {Woo}, \citenamefont {He}, \citenamefont {Gao}, \citenamefont {Yan},\ and\ \citenamefont {Sun}}]{li2025non}%
  \BibitemOpen
  \bibfield  {author} {\bibinfo {author} {\bibfnamefont {C.}~\bibnamefont {Li}}, \bibinfo {author} {\bibfnamefont {J.}~\bibnamefont {Song}}, \bibinfo {author} {\bibfnamefont {H.}~\bibnamefont {Lai}}, \bibinfo {author} {\bibfnamefont {H.}~\bibnamefont {Zhang}}, \bibinfo {author} {\bibfnamefont {R.}~\bibnamefont {Zhou}}, \bibinfo {author} {\bibfnamefont {J.}~\bibnamefont {Xu}}, \bibinfo {author} {\bibfnamefont {H.}~\bibnamefont {Huang}}, \bibinfo {author} {\bibfnamefont {L.}~\bibnamefont {Liu}}, \bibinfo {author} {\bibfnamefont {J.}~\bibnamefont {Gao}}, \bibinfo {author} {\bibfnamefont {Y.}~\bibnamefont {Li}}, \bibinfo {author} {\bibfnamefont {M.~H.}\ \bibnamefont {Jee}}, \bibinfo {author} {\bibfnamefont {Z.}~\bibnamefont {Zheng}}, \bibinfo {author} {\bibfnamefont {S.}~\bibnamefont {Liu}}, \bibinfo {author} {\bibfnamefont {J.}~\bibnamefont {Yan}}, \bibinfo {author} {\bibfnamefont {X.-K.}\ \bibnamefont {Chen}}, \bibinfo {author} {\bibfnamefont {Z.}~\bibnamefont {Tang}}, \bibinfo {author} {\bibfnamefont
  {C.}~\bibnamefont {Zhang}}, \bibinfo {author} {\bibfnamefont {H.~Y.}\ \bibnamefont {Woo}}, \bibinfo {author} {\bibfnamefont {F.}~\bibnamefont {He}}, \bibinfo {author} {\bibfnamefont {F.}~\bibnamefont {Gao}}, \bibinfo {author} {\bibfnamefont {H.}~\bibnamefont {Yan}},\ and\ \bibinfo {author} {\bibfnamefont {Y.}~\bibnamefont {Sun}},\ }\bibfield  {title} {\bibinfo {title} {Non-fullerene acceptors with high crystallinity and photoluminescence quantum yield enable $>$ 20\% efficiency organic solar cells},\ }\href@noop {} {\bibfield  {journal} {\bibinfo  {journal} {Nature Materials}\ }\textbf {\bibinfo {volume} {24}},\ \bibinfo {pages} {433} (\bibinfo {year} {2025})}\BibitemShut {NoStop}%
\bibitem [{\citenamefont {Dolan}\ \emph {et~al.}(2024)\citenamefont {Dolan}, \citenamefont {Pan}, \citenamefont {Griffith}, \citenamefont {Sharma}, \citenamefont {de~la Perrelle}, \citenamefont {Baran}, \citenamefont {Metha}, \citenamefont {Huang}, \citenamefont {Kee},\ and\ \citenamefont {Andersson}}]{dolan2024enhanced}%
  \BibitemOpen
  \bibfield  {author} {\bibinfo {author} {\bibfnamefont {A.}~\bibnamefont {Dolan}}, \bibinfo {author} {\bibfnamefont {X.}~\bibnamefont {Pan}}, \bibinfo {author} {\bibfnamefont {M.~J.}\ \bibnamefont {Griffith}}, \bibinfo {author} {\bibfnamefont {A.}~\bibnamefont {Sharma}}, \bibinfo {author} {\bibfnamefont {J.~M.}\ \bibnamefont {de~la Perrelle}}, \bibinfo {author} {\bibfnamefont {D.}~\bibnamefont {Baran}}, \bibinfo {author} {\bibfnamefont {G.~F.}\ \bibnamefont {Metha}}, \bibinfo {author} {\bibfnamefont {D.~M.}\ \bibnamefont {Huang}}, \bibinfo {author} {\bibfnamefont {T.~W.}\ \bibnamefont {Kee}},\ and\ \bibinfo {author} {\bibfnamefont {M.~R.}\ \bibnamefont {Andersson}},\ }\bibfield  {title} {\bibinfo {title} {Enhanced photocatalytic and photovoltaic performance arising from unconventionally low donor--{Y6} ratios},\ }\href@noop {} {\bibfield  {journal} {\bibinfo  {journal} {Advanced Materials}\ }\textbf {\bibinfo {volume} {36}},\ \bibinfo {pages} {2309672} (\bibinfo {year} {2024})}\BibitemShut {NoStop}%
\bibitem [{\citenamefont {Sharma}\ \emph {et~al.}(2025)\citenamefont {Sharma}, \citenamefont {Gorenflot}, \citenamefont {Xu}, \citenamefont {Jurado}, \citenamefont {Alam}, \citenamefont {Rosas~Villalva}, \citenamefont {Pan}, \citenamefont {Bertrandie}, \citenamefont {Nayak}, \citenamefont {He}, \citenamefont {Alqurashi}, \citenamefont {Luo}, \citenamefont {Andersson}, \citenamefont {Sandberg}, \citenamefont {Laquai},\ and\ \citenamefont {Baran}}]{sharma2025elucidating}%
  \BibitemOpen
  \bibfield  {author} {\bibinfo {author} {\bibfnamefont {A.}~\bibnamefont {Sharma}}, \bibinfo {author} {\bibfnamefont {J.}~\bibnamefont {Gorenflot}}, \bibinfo {author} {\bibfnamefont {H.}~\bibnamefont {Xu}}, \bibinfo {author} {\bibfnamefont {J.~P.}\ \bibnamefont {Jurado}}, \bibinfo {author} {\bibfnamefont {S.}~\bibnamefont {Alam}}, \bibinfo {author} {\bibfnamefont {D.}~\bibnamefont {Rosas~Villalva}}, \bibinfo {author} {\bibfnamefont {X.}~\bibnamefont {Pan}}, \bibinfo {author} {\bibfnamefont {J.}~\bibnamefont {Bertrandie}}, \bibinfo {author} {\bibfnamefont {P.~D.}\ \bibnamefont {Nayak}}, \bibinfo {author} {\bibfnamefont {Y.}~\bibnamefont {He}}, \bibinfo {author} {\bibfnamefont {M.}~\bibnamefont {Alqurashi}}, \bibinfo {author} {\bibfnamefont {Y.}~\bibnamefont {Luo}}, \bibinfo {author} {\bibfnamefont {M.~R.}\ \bibnamefont {Andersson}}, \bibinfo {author} {\bibfnamefont {O.~J.}\ \bibnamefont {Sandberg}}, \bibinfo {author} {\bibfnamefont {F.}~\bibnamefont {Laquai}},\ and\ \bibinfo {author} {\bibfnamefont
  {D.}~\bibnamefont {Baran}},\ }\bibfield  {title} {\bibinfo {title} {Elucidating the role of heterojunction in pristine non-fullerene acceptor organic solar cells},\ }\href {https://doi.org/10.1039/d5ee02324f} {\bibfield  {journal} {\bibinfo  {journal} {Energy \& Environmental Science}\ }\textbf {\bibinfo {volume} {18}},\ \bibinfo {pages} {7610–7623} (\bibinfo {year} {2025})}\BibitemShut {NoStop}%
\bibitem [{\citenamefont {Wu}\ \emph {et~al.}(2025)\citenamefont {Wu}, \citenamefont {Xiao}, \citenamefont {Zhao}, \citenamefont {Wan}, \citenamefont {Gao}, \citenamefont {Sun},\ and\ \citenamefont {Min}}]{wu2025carrier}%
  \BibitemOpen
  \bibfield  {author} {\bibinfo {author} {\bibfnamefont {W.}~\bibnamefont {Wu}}, \bibinfo {author} {\bibfnamefont {B.}~\bibnamefont {Xiao}}, \bibinfo {author} {\bibfnamefont {J.}~\bibnamefont {Zhao}}, \bibinfo {author} {\bibfnamefont {J.}~\bibnamefont {Wan}}, \bibinfo {author} {\bibfnamefont {Y.}~\bibnamefont {Gao}}, \bibinfo {author} {\bibfnamefont {R.}~\bibnamefont {Sun}},\ and\ \bibinfo {author} {\bibfnamefont {J.}~\bibnamefont {Min}},\ }\bibfield  {title} {\bibinfo {title} {From carrier generation to collection: Unraveling the origin of composition tolerance differences in polymer/{NFA} and all-polymer solar cells},\ }\href@noop {} {\bibfield  {journal} {\bibinfo  {journal} {Advanced Functional Materials}\ ,\ \bibinfo {pages} {e13820}} (\bibinfo {year} {2025})}\BibitemShut {NoStop}%
\bibitem [{\citenamefont {Sharma}\ \emph {et~al.}(2024)\citenamefont {Sharma}, \citenamefont {Gasparini}, \citenamefont {Markina}, \citenamefont {Karuthedath}, \citenamefont {Gorenflot}, \citenamefont {Xu}, \citenamefont {Han}, \citenamefont {Balawi}, \citenamefont {Liu}, \citenamefont {Bryant}, \citenamefont {Bertrandie}, \citenamefont {Troughton}, \citenamefont {Paleti}, \citenamefont {Bristow}, \citenamefont {Laquai}, \citenamefont {Andrienko},\ and\ \citenamefont {Baran}}]{sharma2024semitransparent}%
  \BibitemOpen
  \bibfield  {author} {\bibinfo {author} {\bibfnamefont {A.}~\bibnamefont {Sharma}}, \bibinfo {author} {\bibfnamefont {N.}~\bibnamefont {Gasparini}}, \bibinfo {author} {\bibfnamefont {A.}~\bibnamefont {Markina}}, \bibinfo {author} {\bibfnamefont {S.}~\bibnamefont {Karuthedath}}, \bibinfo {author} {\bibfnamefont {J.}~\bibnamefont {Gorenflot}}, \bibinfo {author} {\bibfnamefont {H.}~\bibnamefont {Xu}}, \bibinfo {author} {\bibfnamefont {J.}~\bibnamefont {Han}}, \bibinfo {author} {\bibfnamefont {A.}~\bibnamefont {Balawi}}, \bibinfo {author} {\bibfnamefont {W.}~\bibnamefont {Liu}}, \bibinfo {author} {\bibfnamefont {D.}~\bibnamefont {Bryant}}, \bibinfo {author} {\bibfnamefont {J.}~\bibnamefont {Bertrandie}}, \bibinfo {author} {\bibfnamefont {J.}~\bibnamefont {Troughton}}, \bibinfo {author} {\bibfnamefont {S.~H.~K.}\ \bibnamefont {Paleti}}, \bibinfo {author} {\bibfnamefont {H.}~\bibnamefont {Bristow}}, \bibinfo {author} {\bibfnamefont {F.}~\bibnamefont {Laquai}}, \bibinfo {author} {\bibfnamefont {D.}~\bibnamefont
  {Andrienko}},\ and\ \bibinfo {author} {\bibfnamefont {D.}~\bibnamefont {Baran}},\ }\bibfield  {title} {\bibinfo {title} {Semitransparent organic photovoltaics utilizing intrinsic charge generation in non-fullerene acceptors},\ }\href@noop {} {\bibfield  {journal} {\bibinfo  {journal} {Advanced Materials}\ }\textbf {\bibinfo {volume} {36}},\ \bibinfo {pages} {2305367} (\bibinfo {year} {2024})}\BibitemShut {NoStop}%
\bibitem [{\citenamefont {Schopp}\ \emph {et~al.}(2022)\citenamefont {Schopp}, \citenamefont {Akhtanova}, \citenamefont {Panoy}, \citenamefont {Arbuz}, \citenamefont {Chae}, \citenamefont {Yi}, \citenamefont {Kim}, \citenamefont {Promarak}, \citenamefont {Nguyen},\ and\ \citenamefont {Brus}}]{schopp2022unraveling}%
  \BibitemOpen
  \bibfield  {author} {\bibinfo {author} {\bibfnamefont {N.}~\bibnamefont {Schopp}}, \bibinfo {author} {\bibfnamefont {G.}~\bibnamefont {Akhtanova}}, \bibinfo {author} {\bibfnamefont {P.}~\bibnamefont {Panoy}}, \bibinfo {author} {\bibfnamefont {A.}~\bibnamefont {Arbuz}}, \bibinfo {author} {\bibfnamefont {S.}~\bibnamefont {Chae}}, \bibinfo {author} {\bibfnamefont {A.}~\bibnamefont {Yi}}, \bibinfo {author} {\bibfnamefont {H.~J.}\ \bibnamefont {Kim}}, \bibinfo {author} {\bibfnamefont {V.}~\bibnamefont {Promarak}}, \bibinfo {author} {\bibfnamefont {T.-Q.}\ \bibnamefont {Nguyen}},\ and\ \bibinfo {author} {\bibfnamefont {V.~V.}\ \bibnamefont {Brus}},\ }\bibfield  {title} {\bibinfo {title} {Unraveling device physics of dilute-donor narrow-bandgap organic solar cells with highly transparent active layers},\ }\href@noop {} {\bibfield  {journal} {\bibinfo  {journal} {Advanced Materials}\ }\textbf {\bibinfo {volume} {34}},\ \bibinfo {pages} {2203796} (\bibinfo {year} {2022})}\BibitemShut {NoStop}%
\bibitem [{\citenamefont {Tang}\ \emph {et~al.}(2022)\citenamefont {Tang}, \citenamefont {Zheng}, \citenamefont {Zhou}, \citenamefont {Tang}, \citenamefont {Ma},\ and\ \citenamefont {Yan}}]{tang2022molecular}%
  \BibitemOpen
  \bibfield  {author} {\bibinfo {author} {\bibfnamefont {Y.}~\bibnamefont {Tang}}, \bibinfo {author} {\bibfnamefont {H.}~\bibnamefont {Zheng}}, \bibinfo {author} {\bibfnamefont {X.}~\bibnamefont {Zhou}}, \bibinfo {author} {\bibfnamefont {Z.}~\bibnamefont {Tang}}, \bibinfo {author} {\bibfnamefont {W.}~\bibnamefont {Ma}},\ and\ \bibinfo {author} {\bibfnamefont {H.}~\bibnamefont {Yan}},\ }\bibfield  {title} {\bibinfo {title} {Molecular doping increases the semitransparent photovoltaic performance of dilute bulk heterojunction film with discontinuous polymer donor networks},\ }\href@noop {} {\bibfield  {journal} {\bibinfo  {journal} {Small Methods}\ }\textbf {\bibinfo {volume} {6}},\ \bibinfo {pages} {2101570} (\bibinfo {year} {2022})}\BibitemShut {NoStop}%
\bibitem [{\citenamefont {Ortner}\ \emph {et~al.}(2025)\citenamefont {Ortner}, \citenamefont {Binter}, \citenamefont {H{\"o}nigsberger}, \citenamefont {Costa}, \citenamefont {Haberfehlner}, \citenamefont {Kothleitner}, \citenamefont {Amenitsch}, \citenamefont {Rath}, \citenamefont {Scharber},\ and\ \citenamefont {Trimmel}}]{ortner2025donor}%
  \BibitemOpen
  \bibfield  {author} {\bibinfo {author} {\bibfnamefont {B.~C.}\ \bibnamefont {Ortner}}, \bibinfo {author} {\bibfnamefont {K.}~\bibnamefont {Binter}}, \bibinfo {author} {\bibfnamefont {J.}~\bibnamefont {H{\"o}nigsberger}}, \bibinfo {author} {\bibfnamefont {S.~F.}\ \bibnamefont {Costa}}, \bibinfo {author} {\bibfnamefont {G.}~\bibnamefont {Haberfehlner}}, \bibinfo {author} {\bibfnamefont {G.}~\bibnamefont {Kothleitner}}, \bibinfo {author} {\bibfnamefont {H.}~\bibnamefont {Amenitsch}}, \bibinfo {author} {\bibfnamefont {T.}~\bibnamefont {Rath}}, \bibinfo {author} {\bibfnamefont {M.~C.}\ \bibnamefont {Scharber}},\ and\ \bibinfo {author} {\bibfnamefont {G.}~\bibnamefont {Trimmel}},\ }\bibfield  {title} {\bibinfo {title} {Donor dilution in {D18:L8-BO} organic solar cells: visualization of morphology and effects on device characteristics},\ }\href@noop {} {\bibfield  {journal} {\bibinfo  {journal} {Journal of Materials Chemistry C}\ }\textbf {\bibinfo {volume} {13}},\ \bibinfo {pages} {18981} (\bibinfo {year}
  {2025})}\BibitemShut {NoStop}%
\bibitem [{\citenamefont {Wang}\ \emph {et~al.}(2022)\citenamefont {Wang}, \citenamefont {Price}, \citenamefont {Bobba}, \citenamefont {Lu}, \citenamefont {Xue}, \citenamefont {Wang}, \citenamefont {Li}, \citenamefont {Ilina}, \citenamefont {Hume}, \citenamefont {Jia}, \citenamefont {Li}, \citenamefont {Zhang}, \citenamefont {Davis}, \citenamefont {Tang}, \citenamefont {Ma}, \citenamefont {Qiao}, \citenamefont {Hodgkiss},\ and\ \citenamefont {Zhan}}]{wang2022quasi}%
  \BibitemOpen
  \bibfield  {author} {\bibinfo {author} {\bibfnamefont {Y.}~\bibnamefont {Wang}}, \bibinfo {author} {\bibfnamefont {M.~B.}\ \bibnamefont {Price}}, \bibinfo {author} {\bibfnamefont {R.~S.}\ \bibnamefont {Bobba}}, \bibinfo {author} {\bibfnamefont {H.}~\bibnamefont {Lu}}, \bibinfo {author} {\bibfnamefont {J.}~\bibnamefont {Xue}}, \bibinfo {author} {\bibfnamefont {Y.}~\bibnamefont {Wang}}, \bibinfo {author} {\bibfnamefont {M.}~\bibnamefont {Li}}, \bibinfo {author} {\bibfnamefont {A.}~\bibnamefont {Ilina}}, \bibinfo {author} {\bibfnamefont {P.~A.}\ \bibnamefont {Hume}}, \bibinfo {author} {\bibfnamefont {B.}~\bibnamefont {Jia}}, \bibinfo {author} {\bibfnamefont {T.}~\bibnamefont {Li}}, \bibinfo {author} {\bibfnamefont {Y.}~\bibnamefont {Zhang}}, \bibinfo {author} {\bibfnamefont {N.~J. L.~K.}\ \bibnamefont {Davis}}, \bibinfo {author} {\bibfnamefont {Z.}~\bibnamefont {Tang}}, \bibinfo {author} {\bibfnamefont {W.}~\bibnamefont {Ma}}, \bibinfo {author} {\bibfnamefont {Q.}~\bibnamefont {Qiao}}, \bibinfo {author}
  {\bibfnamefont {J.~M.}\ \bibnamefont {Hodgkiss}},\ and\ \bibinfo {author} {\bibfnamefont {X.}~\bibnamefont {Zhan}},\ }\bibfield  {title} {\bibinfo {title} {Quasi-homojunction organic nonfullerene photovoltaics featuring fundamentals distinct from bulk heterojunctions},\ }\href@noop {} {\bibfield  {journal} {\bibinfo  {journal} {Advanced Materials}\ }\textbf {\bibinfo {volume} {34}},\ \bibinfo {pages} {2206717} (\bibinfo {year} {2022})}\BibitemShut {NoStop}%
\bibitem [{\citenamefont {Yao}\ \emph {et~al.}(2021)\citenamefont {Yao}, \citenamefont {Wang}, \citenamefont {Chen}, \citenamefont {Bian}, \citenamefont {Xia}, \citenamefont {Zhang}, \citenamefont {Zhang}, \citenamefont {Qin}, \citenamefont {Zhu}, \citenamefont {Zhang},\ and\ \citenamefont {Zhang}}]{yao2021efficient}%
  \BibitemOpen
  \bibfield  {author} {\bibinfo {author} {\bibfnamefont {N.}~\bibnamefont {Yao}}, \bibinfo {author} {\bibfnamefont {J.}~\bibnamefont {Wang}}, \bibinfo {author} {\bibfnamefont {Z.}~\bibnamefont {Chen}}, \bibinfo {author} {\bibfnamefont {Q.}~\bibnamefont {Bian}}, \bibinfo {author} {\bibfnamefont {Y.}~\bibnamefont {Xia}}, \bibinfo {author} {\bibfnamefont {R.}~\bibnamefont {Zhang}}, \bibinfo {author} {\bibfnamefont {J.}~\bibnamefont {Zhang}}, \bibinfo {author} {\bibfnamefont {L.}~\bibnamefont {Qin}}, \bibinfo {author} {\bibfnamefont {H.}~\bibnamefont {Zhu}}, \bibinfo {author} {\bibfnamefont {Y.}~\bibnamefont {Zhang}},\ and\ \bibinfo {author} {\bibfnamefont {F.}~\bibnamefont {Zhang}},\ }\bibfield  {title} {\bibinfo {title} {Efficient charge transport enables high efficiency in dilute donor organic solar cells},\ }\href@noop {} {\bibfield  {journal} {\bibinfo  {journal} {The Journal of Physical Chemistry Letters}\ }\textbf {\bibinfo {volume} {12}},\ \bibinfo {pages} {5039} (\bibinfo {year} {2021})}\BibitemShut
  {NoStop}%
\bibitem [{\citenamefont {Haffner‐Schirmer}\ \emph {et~al.}(2025)\citenamefont {Haffner‐Schirmer}, \citenamefont {Le~Corre}, \citenamefont {Forberich}, \citenamefont {Egelhaaf}, \citenamefont {Osterrieder}, \citenamefont {Wortmann}, \citenamefont {Liu}, \citenamefont {Weitz}, \citenamefont {Heum\"{u}ller}, \citenamefont {Bornschlegl}, \citenamefont {Wachsmuth}, \citenamefont {Distler}, \citenamefont {Wagner}, \citenamefont {Peng}, \citenamefont {L\"{u}er},\ and\ \citenamefont {Brabec}}]{haffner2025high}%
  \BibitemOpen
  \bibfield  {author} {\bibinfo {author} {\bibfnamefont {J.~M.}\ \bibnamefont {Haffner‐Schirmer}}, \bibinfo {author} {\bibfnamefont {V.~M.}\ \bibnamefont {Le~Corre}}, \bibinfo {author} {\bibfnamefont {K.}~\bibnamefont {Forberich}}, \bibinfo {author} {\bibfnamefont {H.~J.}\ \bibnamefont {Egelhaaf}}, \bibinfo {author} {\bibfnamefont {T.}~\bibnamefont {Osterrieder}}, \bibinfo {author} {\bibfnamefont {J.}~\bibnamefont {Wortmann}}, \bibinfo {author} {\bibfnamefont {C.}~\bibnamefont {Liu}}, \bibinfo {author} {\bibfnamefont {P.}~\bibnamefont {Weitz}}, \bibinfo {author} {\bibfnamefont {T.}~\bibnamefont {Heum\"{u}ller}}, \bibinfo {author} {\bibfnamefont {A.~J.}\ \bibnamefont {Bornschlegl}}, \bibinfo {author} {\bibfnamefont {J.}~\bibnamefont {Wachsmuth}}, \bibinfo {author} {\bibfnamefont {A.}~\bibnamefont {Distler}}, \bibinfo {author} {\bibfnamefont {M.}~\bibnamefont {Wagner}}, \bibinfo {author} {\bibfnamefont {Z.}~\bibnamefont {Peng}}, \bibinfo {author} {\bibfnamefont {L.}~\bibnamefont {L\"{u}er}},\ and\ \bibinfo
  {author} {\bibfnamefont {C.~J.}\ \bibnamefont {Brabec}},\ }\bibfield  {title} {\bibinfo {title} {A high throughput platform to minimize voltage and fill factor losses},\ }\href@noop {} {\bibfield  {journal} {\bibinfo  {journal} {Advanced Energy Materials}\ }\textbf {\bibinfo {volume} {15}},\ \bibinfo {pages} {2403479} (\bibinfo {year} {2025})}\BibitemShut {NoStop}%
\bibitem [{\citenamefont {Ma}\ \emph {et~al.}(2020)\citenamefont {Ma}, \citenamefont {Wang}, \citenamefont {Gao}, \citenamefont {Hu}, \citenamefont {Xu}, \citenamefont {Zhang},\ and\ \citenamefont {Zhang}}]{ma2020achieving}%
  \BibitemOpen
  \bibfield  {author} {\bibinfo {author} {\bibfnamefont {X.}~\bibnamefont {Ma}}, \bibinfo {author} {\bibfnamefont {J.}~\bibnamefont {Wang}}, \bibinfo {author} {\bibfnamefont {J.}~\bibnamefont {Gao}}, \bibinfo {author} {\bibfnamefont {Z.}~\bibnamefont {Hu}}, \bibinfo {author} {\bibfnamefont {C.}~\bibnamefont {Xu}}, \bibinfo {author} {\bibfnamefont {X.}~\bibnamefont {Zhang}},\ and\ \bibinfo {author} {\bibfnamefont {F.}~\bibnamefont {Zhang}},\ }\bibfield  {title} {\bibinfo {title} {Achieving 17.4\% efficiency of ternary organic photovoltaics with two well-compatible nonfullerene acceptors for minimizing energy loss},\ }\href@noop {} {\bibfield  {journal} {\bibinfo  {journal} {Advanced Energy Materials}\ }\textbf {\bibinfo {volume} {10}},\ \bibinfo {pages} {2001404} (\bibinfo {year} {2020})}\BibitemShut {NoStop}%
\bibitem [{\citenamefont {Kramdi}\ \emph {et~al.}(2025)\citenamefont {Kramdi}, \citenamefont {Karahan}, \citenamefont {Abbassi}, \citenamefont {Watanabe}, \citenamefont {Sekimoto}, \citenamefont {Margeat}, \citenamefont {Ackermann}, \citenamefont {Ruiz~Herrero},\ and\ \citenamefont {Videlot-Ackermann}}]{kramdi2025blade}%
  \BibitemOpen
  \bibfield  {author} {\bibinfo {author} {\bibfnamefont {M.~E.~A.}\ \bibnamefont {Kramdi}}, \bibinfo {author} {\bibfnamefont {A.}~\bibnamefont {Karahan}}, \bibinfo {author} {\bibfnamefont {L.}~\bibnamefont {Abbassi}}, \bibinfo {author} {\bibfnamefont {T.}~\bibnamefont {Watanabe}}, \bibinfo {author} {\bibfnamefont {H.}~\bibnamefont {Sekimoto}}, \bibinfo {author} {\bibfnamefont {O.}~\bibnamefont {Margeat}}, \bibinfo {author} {\bibfnamefont {J.}~\bibnamefont {Ackermann}}, \bibinfo {author} {\bibfnamefont {C.~M.}\ \bibnamefont {Ruiz~Herrero}},\ and\ \bibinfo {author} {\bibfnamefont {C.}~\bibnamefont {Videlot-Ackermann}},\ }\bibfield  {title} {\bibinfo {title} {Blade-coated all-polymer organic solar cells with 15\% efficiency using eco-friendly solvent systems},\ }\href@noop {} {\bibfield  {journal} {\bibinfo  {journal} {ACS Applied Materials \& Interfaces}\ }\textbf {\bibinfo {volume} {17}},\ \bibinfo {pages} {58479} (\bibinfo {year} {2025})}\BibitemShut {NoStop}%
\bibitem [{\citenamefont {Kirchartz}\ \emph {et~al.}(2013)\citenamefont {Kirchartz}, \citenamefont {Deledalle}, \citenamefont {Tuladhar}, \citenamefont {Durrant},\ and\ \citenamefont {Nelson}}]{kirchartz2013differences}%
  \BibitemOpen
  \bibfield  {author} {\bibinfo {author} {\bibfnamefont {T.}~\bibnamefont {Kirchartz}}, \bibinfo {author} {\bibfnamefont {F.}~\bibnamefont {Deledalle}}, \bibinfo {author} {\bibfnamefont {P.~S.}\ \bibnamefont {Tuladhar}}, \bibinfo {author} {\bibfnamefont {J.~R.}\ \bibnamefont {Durrant}},\ and\ \bibinfo {author} {\bibfnamefont {J.}~\bibnamefont {Nelson}},\ }\bibfield  {title} {\bibinfo {title} {On the differences between dark and light ideality factor in polymer: fullerene solar cells},\ }\href@noop {} {\bibfield  {journal} {\bibinfo  {journal} {The Journal of Physical Chemistry Letters}\ }\textbf {\bibinfo {volume} {4}},\ \bibinfo {pages} {2371} (\bibinfo {year} {2013})}\BibitemShut {NoStop}%
\bibitem [{\citenamefont {Neher}\ \emph {et~al.}(2016)\citenamefont {Neher}, \citenamefont {Kniepert}, \citenamefont {Elimelech},\ and\ \citenamefont {Koster}}]{neher2016new}%
  \BibitemOpen
  \bibfield  {author} {\bibinfo {author} {\bibfnamefont {D.}~\bibnamefont {Neher}}, \bibinfo {author} {\bibfnamefont {J.}~\bibnamefont {Kniepert}}, \bibinfo {author} {\bibfnamefont {A.}~\bibnamefont {Elimelech}},\ and\ \bibinfo {author} {\bibfnamefont {L.~J.~A.}\ \bibnamefont {Koster}},\ }\bibfield  {title} {\bibinfo {title} {A new figure of merit for organic solar cells with transport-limited photocurrents},\ }\href@noop {} {\bibfield  {journal} {\bibinfo  {journal} {Scientific Reports}\ }\textbf {\bibinfo {volume} {6}},\ \bibinfo {pages} {24861} (\bibinfo {year} {2016})}\BibitemShut {NoStop}%
\bibitem [{\citenamefont {Saladina}\ and\ \citenamefont {Deibel}(2025)}]{saladina2024transport}%
  \BibitemOpen
  \bibfield  {author} {\bibinfo {author} {\bibfnamefont {M.}~\bibnamefont {Saladina}}\ and\ \bibinfo {author} {\bibfnamefont {C.}~\bibnamefont {Deibel}},\ }\bibfield  {title} {\bibinfo {title} {Transport resistance strikes back: unveiling its impact on fill factor losses in organic solar cells},\ }\href@noop {} {\bibfield  {journal} {\bibinfo  {journal} {Reports on Progress in Physics}\ }\textbf {\bibinfo {volume} {88}},\ \bibinfo {pages} {038001} (\bibinfo {year} {2025})}\BibitemShut {NoStop}%
\bibitem [{\citenamefont {Green}(1982)}]{green_accuracy_1982}%
  \BibitemOpen
  \bibfield  {author} {\bibinfo {author} {\bibfnamefont {M.~A.}\ \bibnamefont {Green}},\ }\bibfield  {title} {\bibinfo {title} {Accuracy of analytical expressions for solar cell fill factors},\ }\href@noop {} {\bibfield  {journal} {\bibinfo  {journal} {Solar Cells}\ }\textbf {\bibinfo {volume} {7}},\ \bibinfo {pages} {337} (\bibinfo {year} {1982})}\BibitemShut {NoStop}%
\bibitem [{\citenamefont {Vandewal}\ \emph {et~al.}(2014)\citenamefont {Vandewal}, \citenamefont {Widmer}, \citenamefont {Heumueller}, \citenamefont {Brabec}, \citenamefont {McGehee}, \citenamefont {Leo}, \citenamefont {Riede},\ and\ \citenamefont {Salleo}}]{vandewal2014increased}%
  \BibitemOpen
  \bibfield  {author} {\bibinfo {author} {\bibfnamefont {K.}~\bibnamefont {Vandewal}}, \bibinfo {author} {\bibfnamefont {J.}~\bibnamefont {Widmer}}, \bibinfo {author} {\bibfnamefont {T.}~\bibnamefont {Heumueller}}, \bibinfo {author} {\bibfnamefont {C.~J.}\ \bibnamefont {Brabec}}, \bibinfo {author} {\bibfnamefont {M.~D.}\ \bibnamefont {McGehee}}, \bibinfo {author} {\bibfnamefont {K.}~\bibnamefont {Leo}}, \bibinfo {author} {\bibfnamefont {M.}~\bibnamefont {Riede}},\ and\ \bibinfo {author} {\bibfnamefont {A.}~\bibnamefont {Salleo}},\ }\bibfield  {title} {\bibinfo {title} {Increased open-circuit voltage of organic solar cells by reduced donor-acceptor interface area},\ }\href@noop {} {\bibfield  {journal} {\bibinfo  {journal} {Advanced Materials}\ }\textbf {\bibinfo {volume} {26}},\ \bibinfo {pages} {3839} (\bibinfo {year} {2014})}\BibitemShut {NoStop}%
\bibitem [{\citenamefont {Rau}(2007)}]{rau2007reciprocity}%
  \BibitemOpen
  \bibfield  {author} {\bibinfo {author} {\bibfnamefont {U.}~\bibnamefont {Rau}},\ }\bibfield  {title} {\bibinfo {title} {Reciprocity relation between photovoltaic quantum efficiency and electroluminescent emission of solar cells},\ }\href@noop {} {\bibfield  {journal} {\bibinfo  {journal} {Physical Review B—Condensed Matter and Materials Physics}\ }\textbf {\bibinfo {volume} {76}},\ \bibinfo {pages} {085303} (\bibinfo {year} {2007})}\BibitemShut {NoStop}%
\bibitem [{\citenamefont {Liu}\ and\ \citenamefont {Vandewal}(2023)}]{liu2023understanding}%
  \BibitemOpen
  \bibfield  {author} {\bibinfo {author} {\bibfnamefont {Q.}~\bibnamefont {Liu}}\ and\ \bibinfo {author} {\bibfnamefont {K.}~\bibnamefont {Vandewal}},\ }\bibfield  {title} {\bibinfo {title} {Understanding and suppressing non-radiative recombination losses in non-fullerene organic solar cells},\ }\href@noop {} {\bibfield  {journal} {\bibinfo  {journal} {Advanced Materials}\ }\textbf {\bibinfo {volume} {35}},\ \bibinfo {pages} {2302452} (\bibinfo {year} {2023})}\BibitemShut {NoStop}%
\bibitem [{\citenamefont {Vaynzof}\ \emph {et~al.}(2011)\citenamefont {Vaynzof}, \citenamefont {Kabra}, \citenamefont {Zhao}, \citenamefont {Chua}, \citenamefont {Steiner},\ and\ \citenamefont {Friend}}]{Vaynzof2011}%
  \BibitemOpen
  \bibfield  {author} {\bibinfo {author} {\bibfnamefont {Y.}~\bibnamefont {Vaynzof}}, \bibinfo {author} {\bibfnamefont {D.}~\bibnamefont {Kabra}}, \bibinfo {author} {\bibfnamefont {L.}~\bibnamefont {Zhao}}, \bibinfo {author} {\bibfnamefont {L.~L.}\ \bibnamefont {Chua}}, \bibinfo {author} {\bibfnamefont {U.}~\bibnamefont {Steiner}},\ and\ \bibinfo {author} {\bibfnamefont {R.~H.}\ \bibnamefont {Friend}},\ }\bibfield  {title} {\bibinfo {title} {Surface-directed spinodal decomposition in poly[3-hexylthiophene] and {C61}-butyric acid methyl ester blends},\ }\href {https://doi.org/10.1021/nn102899g} {\bibfield  {journal} {\bibinfo  {journal} {ACS Nano}\ }\textbf {\bibinfo {volume} {5}},\ \bibinfo {pages} {329} (\bibinfo {year} {2011})}\BibitemShut {NoStop}%
\bibitem [{\citenamefont {Lami}\ \emph {et~al.}(2020)\citenamefont {Lami}, \citenamefont {Hofstetter}, \citenamefont {Butscher},\ and\ \citenamefont {Vaynzof}}]{lami2020}%
  \BibitemOpen
  \bibfield  {author} {\bibinfo {author} {\bibfnamefont {V.}~\bibnamefont {Lami}}, \bibinfo {author} {\bibfnamefont {Y.~J.}\ \bibnamefont {Hofstetter}}, \bibinfo {author} {\bibfnamefont {J.~F.}\ \bibnamefont {Butscher}},\ and\ \bibinfo {author} {\bibfnamefont {Y.}~\bibnamefont {Vaynzof}},\ }\bibfield  {title} {\bibinfo {title} {Energy level alignment in ternary organic solar cells},\ }\href {https://doi.org/10.1002/aelm.202000213} {\bibfield  {journal} {\bibinfo  {journal} {Advanced Electronic Materials}\ }\textbf {\bibinfo {volume} {6}},\ \bibinfo {pages} {2000213} (\bibinfo {year} {2020})}\BibitemShut {NoStop}%
\bibitem [{\citenamefont {Cai}\ \emph {et~al.}(2022)\citenamefont {Cai}, \citenamefont {Li}, \citenamefont {Lu}, \citenamefont {Ryu}, \citenamefont {Li}, \citenamefont {Jin}, \citenamefont {Chen}, \citenamefont {Tang}, \citenamefont {Lu}, \citenamefont {Hao}, \citenamefont {Woo}, \citenamefont {Zhang},\ and\ \citenamefont {Sun}}]{cai2022vertically}%
  \BibitemOpen
  \bibfield  {author} {\bibinfo {author} {\bibfnamefont {Y.}~\bibnamefont {Cai}}, \bibinfo {author} {\bibfnamefont {Q.}~\bibnamefont {Li}}, \bibinfo {author} {\bibfnamefont {G.}~\bibnamefont {Lu}}, \bibinfo {author} {\bibfnamefont {H.~S.}\ \bibnamefont {Ryu}}, \bibinfo {author} {\bibfnamefont {Y.}~\bibnamefont {Li}}, \bibinfo {author} {\bibfnamefont {H.}~\bibnamefont {Jin}}, \bibinfo {author} {\bibfnamefont {Z.}~\bibnamefont {Chen}}, \bibinfo {author} {\bibfnamefont {Z.}~\bibnamefont {Tang}}, \bibinfo {author} {\bibfnamefont {G.}~\bibnamefont {Lu}}, \bibinfo {author} {\bibfnamefont {X.}~\bibnamefont {Hao}}, \bibinfo {author} {\bibfnamefont {H.~Y.}\ \bibnamefont {Woo}}, \bibinfo {author} {\bibfnamefont {C.}~\bibnamefont {Zhang}},\ and\ \bibinfo {author} {\bibfnamefont {Y.}~\bibnamefont {Sun}},\ }\bibfield  {title} {\bibinfo {title} {Vertically optimized phase separation with improved exciton diffusion enables efficient organic solar cells with thick active layers},\ }\href@noop {} {\bibfield  {journal}
  {\bibinfo  {journal} {Nature Communications}\ }\textbf {\bibinfo {volume} {13}},\ \bibinfo {pages} {2369} (\bibinfo {year} {2022})}\BibitemShut {NoStop}%
\bibitem [{\citenamefont {Karuthedath}\ \emph {et~al.}(2021)\citenamefont {Karuthedath}, \citenamefont {Gorenflot}, \citenamefont {Firdaus}, \citenamefont {Chaturvedi}, \citenamefont {De~Castro}, \citenamefont {Harrison}, \citenamefont {Khan}, \citenamefont {Markina}, \citenamefont {Balawi}, \citenamefont {Peña}, \citenamefont {Liu}, \citenamefont {Liang}, \citenamefont {Sharma}, \citenamefont {Paleti}, \citenamefont {Zhang}, \citenamefont {Lin}, \citenamefont {Alarousu}, \citenamefont {Lopatin}, \citenamefont {Anjum}, \citenamefont {Beaujuge}, \citenamefont {De~Wolf}, \citenamefont {McCulloch}, \citenamefont {Anthopoulos}, \citenamefont {Baran}, \citenamefont {Andrienko},\ and\ \citenamefont {Laquai}}]{karuthedath2021intrinsic}%
  \BibitemOpen
  \bibfield  {author} {\bibinfo {author} {\bibfnamefont {S.}~\bibnamefont {Karuthedath}}, \bibinfo {author} {\bibfnamefont {J.}~\bibnamefont {Gorenflot}}, \bibinfo {author} {\bibfnamefont {Y.}~\bibnamefont {Firdaus}}, \bibinfo {author} {\bibfnamefont {N.}~\bibnamefont {Chaturvedi}}, \bibinfo {author} {\bibfnamefont {C.~S.~P.}\ \bibnamefont {De~Castro}}, \bibinfo {author} {\bibfnamefont {G.~T.}\ \bibnamefont {Harrison}}, \bibinfo {author} {\bibfnamefont {J.~I.}\ \bibnamefont {Khan}}, \bibinfo {author} {\bibfnamefont {A.}~\bibnamefont {Markina}}, \bibinfo {author} {\bibfnamefont {A.~H.}\ \bibnamefont {Balawi}}, \bibinfo {author} {\bibfnamefont {T.~A.~D.}\ \bibnamefont {Peña}}, \bibinfo {author} {\bibfnamefont {W.}~\bibnamefont {Liu}}, \bibinfo {author} {\bibfnamefont {R.-Z.}\ \bibnamefont {Liang}}, \bibinfo {author} {\bibfnamefont {A.}~\bibnamefont {Sharma}}, \bibinfo {author} {\bibfnamefont {S.~H.~K.}\ \bibnamefont {Paleti}}, \bibinfo {author} {\bibfnamefont {W.}~\bibnamefont {Zhang}}, \bibinfo {author}
  {\bibfnamefont {Y.}~\bibnamefont {Lin}}, \bibinfo {author} {\bibfnamefont {E.}~\bibnamefont {Alarousu}}, \bibinfo {author} {\bibfnamefont {S.}~\bibnamefont {Lopatin}}, \bibinfo {author} {\bibfnamefont {D.~H.}\ \bibnamefont {Anjum}}, \bibinfo {author} {\bibfnamefont {P.~M.}\ \bibnamefont {Beaujuge}}, \bibinfo {author} {\bibfnamefont {S.}~\bibnamefont {De~Wolf}}, \bibinfo {author} {\bibfnamefont {I.}~\bibnamefont {McCulloch}}, \bibinfo {author} {\bibfnamefont {T.~D.}\ \bibnamefont {Anthopoulos}}, \bibinfo {author} {\bibfnamefont {D.}~\bibnamefont {Baran}}, \bibinfo {author} {\bibfnamefont {D.}~\bibnamefont {Andrienko}},\ and\ \bibinfo {author} {\bibfnamefont {F.}~\bibnamefont {Laquai}},\ }\bibfield  {title} {\bibinfo {title} {Intrinsic efficiency limits in low-bandgap non-fullerene acceptor organic solar cells},\ }\href@noop {} {\bibfield  {journal} {\bibinfo  {journal} {Nature Materials}\ }\textbf {\bibinfo {volume} {20}},\ \bibinfo {pages} {378} (\bibinfo {year} {2021})}\BibitemShut {NoStop}%
\bibitem [{\citenamefont {MacKenzie}\ \emph {et~al.}(2012)\citenamefont {MacKenzie}, \citenamefont {Shuttle}, \citenamefont {Chabinyc},\ and\ \citenamefont {Nelson}}]{mackenzie2012extracting}%
  \BibitemOpen
  \bibfield  {author} {\bibinfo {author} {\bibfnamefont {R.~C.}\ \bibnamefont {MacKenzie}}, \bibinfo {author} {\bibfnamefont {C.~G.}\ \bibnamefont {Shuttle}}, \bibinfo {author} {\bibfnamefont {M.~L.}\ \bibnamefont {Chabinyc}},\ and\ \bibinfo {author} {\bibfnamefont {J.}~\bibnamefont {Nelson}},\ }\bibfield  {title} {\bibinfo {title} {Extracting microscopic device parameters from transient photocurrent measurements of {P3HT: PCBM} solar cells},\ }\href@noop {} {\bibfield  {journal} {\bibinfo  {journal} {Advanced Energy Materials}\ }\textbf {\bibinfo {volume} {2}},\ \bibinfo {pages} {662} (\bibinfo {year} {2012})}\BibitemShut {NoStop}%
\bibitem [{\citenamefont {Dattani}\ \emph {et~al.}(2014)\citenamefont {Dattani}, \citenamefont {Bannock}, \citenamefont {Fei}, \citenamefont {MacKenzie}, \citenamefont {Guilbert}, \citenamefont {Vezie}, \citenamefont {Nelson}, \citenamefont {de~Mello}, \citenamefont {Heeney}, \citenamefont {Cabral},\ and\ \citenamefont {Nedoma}}]{dattani2014general}%
  \BibitemOpen
  \bibfield  {author} {\bibinfo {author} {\bibfnamefont {R.}~\bibnamefont {Dattani}}, \bibinfo {author} {\bibfnamefont {J.~H.}\ \bibnamefont {Bannock}}, \bibinfo {author} {\bibfnamefont {Z.}~\bibnamefont {Fei}}, \bibinfo {author} {\bibfnamefont {R.~C.~I.}\ \bibnamefont {MacKenzie}}, \bibinfo {author} {\bibfnamefont {A.~A.~Y.}\ \bibnamefont {Guilbert}}, \bibinfo {author} {\bibfnamefont {M.~S.}\ \bibnamefont {Vezie}}, \bibinfo {author} {\bibfnamefont {J.}~\bibnamefont {Nelson}}, \bibinfo {author} {\bibfnamefont {J.~C.}\ \bibnamefont {de~Mello}}, \bibinfo {author} {\bibfnamefont {M.}~\bibnamefont {Heeney}}, \bibinfo {author} {\bibfnamefont {J.~T.}\ \bibnamefont {Cabral}},\ and\ \bibinfo {author} {\bibfnamefont {A.~J.}\ \bibnamefont {Nedoma}},\ }\bibfield  {title} {\bibinfo {title} {A general mechanism for controlling thin film structures in all-conjugated block copolymer: fullerene blends},\ }\href@noop {} {\bibfield  {journal} {\bibinfo  {journal} {Journal of Materials Chemistry A}\ }\textbf {\bibinfo {volume}
  {2}},\ \bibinfo {pages} {14711} (\bibinfo {year} {2014})}\BibitemShut {NoStop}%
\bibitem [{\citenamefont {Liu}\ \emph {et~al.}(2015)\citenamefont {Liu}, \citenamefont {MacKenzie}, \citenamefont {Xu}, \citenamefont {Gao}, \citenamefont {Gimeno-Fabra}, \citenamefont {Grant}, \citenamefont {van Loosdrecht},\ and\ \citenamefont {Tian}}]{liu2015organic}%
  \BibitemOpen
  \bibfield  {author} {\bibinfo {author} {\bibfnamefont {Y.}~\bibnamefont {Liu}}, \bibinfo {author} {\bibfnamefont {R.~C.}\ \bibnamefont {MacKenzie}}, \bibinfo {author} {\bibfnamefont {B.}~\bibnamefont {Xu}}, \bibinfo {author} {\bibfnamefont {Y.}~\bibnamefont {Gao}}, \bibinfo {author} {\bibfnamefont {M.}~\bibnamefont {Gimeno-Fabra}}, \bibinfo {author} {\bibfnamefont {D.}~\bibnamefont {Grant}}, \bibinfo {author} {\bibfnamefont {P.~H.}\ \bibnamefont {van Loosdrecht}},\ and\ \bibinfo {author} {\bibfnamefont {W.}~\bibnamefont {Tian}},\ }\bibfield  {title} {\bibinfo {title} {Organic semiconductors with a charge carrier life time of over 2 hours at room temperature},\ }\href@noop {} {\bibfield  {journal} {\bibinfo  {journal} {Journal of Materials Chemistry C}\ }\textbf {\bibinfo {volume} {3}},\ \bibinfo {pages} {12260} (\bibinfo {year} {2015})}\BibitemShut {NoStop}%
\bibitem [{\citenamefont {Bickerdike}\ \emph {et~al.}(2025)\citenamefont {Bickerdike}, \citenamefont {Mackenzie},\ and\ \citenamefont {Chaudhry}}]{bickerdike2025unravelling}%
  \BibitemOpen
  \bibfield  {author} {\bibinfo {author} {\bibfnamefont {A.}~\bibnamefont {Bickerdike}}, \bibinfo {author} {\bibfnamefont {R.~C.}\ \bibnamefont {Mackenzie}},\ and\ \bibinfo {author} {\bibfnamefont {M.~U.}\ \bibnamefont {Chaudhry}},\ }\bibfield  {title} {\bibinfo {title} {Unravelling the spatiotemporal exciton dynamics in electrically pumped organic laser diodes},\ }\href@noop {} {\bibfield  {journal} {\bibinfo  {journal} {Laser \& Photonics Reviews}\ }\textbf {\bibinfo {volume} {20}},\ \bibinfo {pages} {e00189} (\bibinfo {year} {2025})}\BibitemShut {NoStop}%
\bibitem [{\citenamefont {Nelson}(2003{\natexlab{a}})}]{nelson2003physics}%
  \BibitemOpen
  \bibfield  {author} {\bibinfo {author} {\bibfnamefont {J.~A.}\ \bibnamefont {Nelson}},\ }\href@noop {} {\emph {\bibinfo {title} {The physics of solar cells}}}\ (\bibinfo  {publisher} {World Scientific Publishing Company},\ \bibinfo {year} {2003})\BibitemShut {NoStop}%
\bibitem [{\citenamefont {Heeger}(2010)}]{heeger2010semiconducting}%
  \BibitemOpen
  \bibfield  {author} {\bibinfo {author} {\bibfnamefont {A.~J.}\ \bibnamefont {Heeger}},\ }\bibfield  {title} {\bibinfo {title} {Semiconducting polymers: the third generation},\ }\href@noop {} {\bibfield  {journal} {\bibinfo  {journal} {Chemical Society Reviews}\ }\textbf {\bibinfo {volume} {39}},\ \bibinfo {pages} {2354} (\bibinfo {year} {2010})}\BibitemShut {NoStop}%
\bibitem [{\citenamefont {Bartelt}\ \emph {et~al.}(2013)\citenamefont {Bartelt}, \citenamefont {Beiley}, \citenamefont {Hoke}, \citenamefont {Mateker}, \citenamefont {Douglas}, \citenamefont {Collins}, \citenamefont {Tumbleston}, \citenamefont {Graham}, \citenamefont {Amassian}, \citenamefont {Ade}, \citenamefont {Fréchet}, \citenamefont {Toney},\ and\ \citenamefont {McGehee}}]{bartelt2013importance}%
  \BibitemOpen
  \bibfield  {author} {\bibinfo {author} {\bibfnamefont {J.~A.}\ \bibnamefont {Bartelt}}, \bibinfo {author} {\bibfnamefont {Z.~M.}\ \bibnamefont {Beiley}}, \bibinfo {author} {\bibfnamefont {E.~T.}\ \bibnamefont {Hoke}}, \bibinfo {author} {\bibfnamefont {W.~R.}\ \bibnamefont {Mateker}}, \bibinfo {author} {\bibfnamefont {J.~D.}\ \bibnamefont {Douglas}}, \bibinfo {author} {\bibfnamefont {B.~A.}\ \bibnamefont {Collins}}, \bibinfo {author} {\bibfnamefont {J.~R.}\ \bibnamefont {Tumbleston}}, \bibinfo {author} {\bibfnamefont {K.~R.}\ \bibnamefont {Graham}}, \bibinfo {author} {\bibfnamefont {A.}~\bibnamefont {Amassian}}, \bibinfo {author} {\bibfnamefont {H.}~\bibnamefont {Ade}}, \bibinfo {author} {\bibfnamefont {J.~M.~J.}\ \bibnamefont {Fréchet}}, \bibinfo {author} {\bibfnamefont {M.~F.}\ \bibnamefont {Toney}},\ and\ \bibinfo {author} {\bibfnamefont {M.~D.}\ \bibnamefont {McGehee}},\ }\bibfield  {title} {\bibinfo {title} {The importance of fullerene percolation in the mixed regions of polymer--fullerene bulk
  heterojunction solar cells},\ }\href@noop {} {\bibfield  {journal} {\bibinfo  {journal} {Advanced Energy Materials}\ }\textbf {\bibinfo {volume} {3}},\ \bibinfo {pages} {364} (\bibinfo {year} {2013})}\BibitemShut {NoStop}%
\bibitem [{\citenamefont {B{\"a}ssler}(1993)}]{bassler1993charge}%
  \BibitemOpen
  \bibfield  {author} {\bibinfo {author} {\bibfnamefont {H.}~\bibnamefont {B{\"a}ssler}},\ }\bibfield  {title} {\bibinfo {title} {Charge transport in disordered organic photoconductors. a monte carlo simulation study},\ }\href@noop {} {\bibfield  {journal} {\bibinfo  {journal} {Physica Status Solidi (b)}\ }\textbf {\bibinfo {volume} {175}} (\bibinfo {year} {1993})}\BibitemShut {NoStop}%
\bibitem [{\citenamefont {Baranovskii}(2014)}]{baranovskii2014theoretical}%
  \BibitemOpen
  \bibfield  {author} {\bibinfo {author} {\bibfnamefont {S.}~\bibnamefont {Baranovskii}},\ }\bibfield  {title} {\bibinfo {title} {Theoretical description of charge transport in disordered organic semiconductors},\ }\href@noop {} {\bibfield  {journal} {\bibinfo  {journal} {Physica Status Solidi (b)}\ }\textbf {\bibinfo {volume} {251}},\ \bibinfo {pages} {487} (\bibinfo {year} {2014})}\BibitemShut {NoStop}%
\bibitem [{\citenamefont {Deibel}\ \emph {et~al.}(2009{\natexlab{a}})\citenamefont {Deibel}, \citenamefont {Strobel},\ and\ \citenamefont {Dyakonov}}]{deibel2009prl}%
  \BibitemOpen
  \bibfield  {author} {\bibinfo {author} {\bibfnamefont {C.}~\bibnamefont {Deibel}}, \bibinfo {author} {\bibfnamefont {T.}~\bibnamefont {Strobel}},\ and\ \bibinfo {author} {\bibfnamefont {V.}~\bibnamefont {Dyakonov}},\ }\bibfield  {title} {\bibinfo {title} {Origin of the efficient polaron-pair dissociation in polymer-fullerene blends},\ }\href {https://doi.org/10.1103/physrevlett.103.036402} {\bibfield  {journal} {\bibinfo  {journal} {Physical Review Letters}\ }\textbf {\bibinfo {volume} {103}},\ \bibinfo {pages} {036402} (\bibinfo {year} {2009}{\natexlab{a}})}\BibitemShut {NoStop}%
\bibitem [{\citenamefont {Wang}\ \emph {et~al.}(2025{\natexlab{a}})\citenamefont {Wang}, \citenamefont {MacKenzie}, \citenamefont {W{\"u}rfel}, \citenamefont {Neher}, \citenamefont {Kirchartz}, \citenamefont {Deibel},\ and\ \citenamefont {Saladina}}]{wang2025transport}%
  \BibitemOpen
  \bibfield  {author} {\bibinfo {author} {\bibfnamefont {C.}~\bibnamefont {Wang}}, \bibinfo {author} {\bibfnamefont {R.~C.}\ \bibnamefont {MacKenzie}}, \bibinfo {author} {\bibfnamefont {U.}~\bibnamefont {W{\"u}rfel}}, \bibinfo {author} {\bibfnamefont {D.}~\bibnamefont {Neher}}, \bibinfo {author} {\bibfnamefont {T.}~\bibnamefont {Kirchartz}}, \bibinfo {author} {\bibfnamefont {C.}~\bibnamefont {Deibel}},\ and\ \bibinfo {author} {\bibfnamefont {M.}~\bibnamefont {Saladina}},\ }\bibfield  {title} {\bibinfo {title} {Transport resistance dominates the fill factor losses in record organic solar cells},\ }\href@noop {} {\bibfield  {journal} {\bibinfo  {journal} {Advanced Energy Materials}\ }\textbf {\bibinfo {volume} {16}},\ \bibinfo {pages} {2405889} (\bibinfo {year} {2025}{\natexlab{a}})}\BibitemShut {NoStop}%
\bibitem [{\citenamefont {Schiefer}\ \emph {et~al.}(2014)\citenamefont {Schiefer}, \citenamefont {Zimmermann}, \citenamefont {Glunz},\ and\ \citenamefont {Wurfel}}]{schiefer2014}%
  \BibitemOpen
  \bibfield  {author} {\bibinfo {author} {\bibfnamefont {S.}~\bibnamefont {Schiefer}}, \bibinfo {author} {\bibfnamefont {B.}~\bibnamefont {Zimmermann}}, \bibinfo {author} {\bibfnamefont {S.~W.}\ \bibnamefont {Glunz}},\ and\ \bibinfo {author} {\bibfnamefont {U.}~\bibnamefont {Wurfel}},\ }\bibfield  {title} {\bibinfo {title} {Applicability of the {Suns-V$_{\rm OC}$} method on organic solar cells},\ }\href {https://doi.org/10.1109/jphotov.2013.2288527} {\bibfield  {journal} {\bibinfo  {journal} {IEEE Journal of Photovoltaics}\ }\textbf {\bibinfo {volume} {4}},\ \bibinfo {pages} {271} (\bibinfo {year} {2014})}\BibitemShut {NoStop}%
\bibitem [{\citenamefont {Hofacker}\ and\ \citenamefont {Neher}(2017)}]{hofacker2017}%
  \BibitemOpen
  \bibfield  {author} {\bibinfo {author} {\bibfnamefont {A.}~\bibnamefont {Hofacker}}\ and\ \bibinfo {author} {\bibfnamefont {D.}~\bibnamefont {Neher}},\ }\bibfield  {title} {\bibinfo {title} {Dispersive and steady-state recombination in organic disordered semiconductors},\ }\href {https://doi.org/10.1103/physrevb.96.245204} {\bibfield  {journal} {\bibinfo  {journal} {Physical Review B}\ }\textbf {\bibinfo {volume} {96}},\ \bibinfo {pages} {245204} (\bibinfo {year} {2017})}\BibitemShut {NoStop}%
\bibitem [{\citenamefont {Saladina}\ \emph {et~al.}(2023)\citenamefont {Saladina}, \citenamefont {W\"{o}pke}, \citenamefont {G\"{o}hler}, \citenamefont {Ramirez}, \citenamefont {Gerdes}, \citenamefont {Liu}, \citenamefont {Li}, \citenamefont {Heum\"{u}ller}, \citenamefont {Brabec}, \citenamefont {Walzer}, \citenamefont {Pfeiffer},\ and\ \citenamefont {Deibel}}]{saladina2023power}%
  \BibitemOpen
  \bibfield  {author} {\bibinfo {author} {\bibfnamefont {M.}~\bibnamefont {Saladina}}, \bibinfo {author} {\bibfnamefont {C.}~\bibnamefont {W\"{o}pke}}, \bibinfo {author} {\bibfnamefont {C.}~\bibnamefont {G\"{o}hler}}, \bibinfo {author} {\bibfnamefont {I.}~\bibnamefont {Ramirez}}, \bibinfo {author} {\bibfnamefont {O.}~\bibnamefont {Gerdes}}, \bibinfo {author} {\bibfnamefont {C.}~\bibnamefont {Liu}}, \bibinfo {author} {\bibfnamefont {N.}~\bibnamefont {Li}}, \bibinfo {author} {\bibfnamefont {T.}~\bibnamefont {Heum\"{u}ller}}, \bibinfo {author} {\bibfnamefont {C.~J.}\ \bibnamefont {Brabec}}, \bibinfo {author} {\bibfnamefont {K.}~\bibnamefont {Walzer}}, \bibinfo {author} {\bibfnamefont {M.}~\bibnamefont {Pfeiffer}},\ and\ \bibinfo {author} {\bibfnamefont {C.}~\bibnamefont {Deibel}},\ }\bibfield  {title} {\bibinfo {title} {Power-law density of states in organic solar cells revealed by the open-circuit voltage dependence of the ideality factor},\ }\href@noop {} {\bibfield  {journal} {\bibinfo  {journal} {Physical
  Review Letters}\ }\textbf {\bibinfo {volume} {130}},\ \bibinfo {pages} {236403} (\bibinfo {year} {2023})}\BibitemShut {NoStop}%
\bibitem [{\citenamefont {Miller}\ and\ \citenamefont {Abrahams}(1960)}]{miller1960impurity}%
  \BibitemOpen
  \bibfield  {author} {\bibinfo {author} {\bibfnamefont {A.}~\bibnamefont {Miller}}\ and\ \bibinfo {author} {\bibfnamefont {E.}~\bibnamefont {Abrahams}},\ }\bibfield  {title} {\bibinfo {title} {Impurity conduction at low concentrations},\ }\href@noop {} {\bibfield  {journal} {\bibinfo  {journal} {Physical Review}\ }\textbf {\bibinfo {volume} {120}},\ \bibinfo {pages} {745} (\bibinfo {year} {1960})}\BibitemShut {NoStop}%
\bibitem [{\citenamefont {Ambegaokar}\ \emph {et~al.}(1971)\citenamefont {Ambegaokar}, \citenamefont {Halperin},\ and\ \citenamefont {Langer}}]{ambegaokar1971hopping}%
  \BibitemOpen
  \bibfield  {author} {\bibinfo {author} {\bibfnamefont {V.}~\bibnamefont {Ambegaokar}}, \bibinfo {author} {\bibfnamefont {B.}~\bibnamefont {Halperin}},\ and\ \bibinfo {author} {\bibfnamefont {J.}~\bibnamefont {Langer}},\ }\bibfield  {title} {\bibinfo {title} {Hopping conductivity in disordered systems},\ }\href@noop {} {\bibfield  {journal} {\bibinfo  {journal} {Physical Review B}\ }\textbf {\bibinfo {volume} {4}},\ \bibinfo {pages} {2612} (\bibinfo {year} {1971})}\BibitemShut {NoStop}%
\bibitem [{\citenamefont {Ambrosetti}\ \emph {et~al.}(2010)\citenamefont {Ambrosetti}, \citenamefont {Grimaldi}, \citenamefont {Balberg}, \citenamefont {Maeder}, \citenamefont {Danani},\ and\ \citenamefont {Ryser}}]{ambrosetti2010solution}%
  \BibitemOpen
  \bibfield  {author} {\bibinfo {author} {\bibfnamefont {G.}~\bibnamefont {Ambrosetti}}, \bibinfo {author} {\bibfnamefont {C.}~\bibnamefont {Grimaldi}}, \bibinfo {author} {\bibfnamefont {I.}~\bibnamefont {Balberg}}, \bibinfo {author} {\bibfnamefont {T.}~\bibnamefont {Maeder}}, \bibinfo {author} {\bibfnamefont {A.}~\bibnamefont {Danani}},\ and\ \bibinfo {author} {\bibfnamefont {P.}~\bibnamefont {Ryser}},\ }\bibfield  {title} {\bibinfo {title} {Solution of the tunneling-percolation problem in the nanocomposite regime},\ }\href@noop {} {\bibfield  {journal} {\bibinfo  {journal} {Physical Review B—Condensed Matter and Materials Physics}\ }\textbf {\bibinfo {volume} {81}},\ \bibinfo {pages} {155434} (\bibinfo {year} {2010})}\BibitemShut {NoStop}%
\bibitem [{\citenamefont {Hofacker}(2020)}]{hofacker2020critical}%
  \BibitemOpen
  \bibfield  {author} {\bibinfo {author} {\bibfnamefont {A.}~\bibnamefont {Hofacker}},\ }\bibfield  {title} {\bibinfo {title} {Critical charge transport networks in doped organic semiconductors},\ }\href@noop {} {\bibfield  {journal} {\bibinfo  {journal} {Communications Materials}\ }\textbf {\bibinfo {volume} {1}},\ \bibinfo {pages} {88} (\bibinfo {year} {2020})}\BibitemShut {NoStop}%
\bibitem [{\citenamefont {Hofacker}(2021)}]{hofacker2021modelling}%
  \BibitemOpen
  \bibfield  {author} {\bibinfo {author} {\bibfnamefont {A.}~\bibnamefont {Hofacker}},\ }\emph {\bibinfo {title} {Modelling charge carrier dynamics in organic semiconductors}},\ \href@noop {} {Ph.D. thesis},\ \bibinfo  {school} {Technische Universit{\"a}t Dresden} (\bibinfo {year} {2021})\BibitemShut {NoStop}%
\bibitem [{\citenamefont {Wang}\ \emph {et~al.}(2025{\natexlab{b}})\citenamefont {Wang}, \citenamefont {Seiler}, \citenamefont {Sun}, \citenamefont {Shoaee}, \citenamefont {Saladina},\ and\ \citenamefont {Deibel}}]{wang2025contribution}%
  \BibitemOpen
  \bibfield  {author} {\bibinfo {author} {\bibfnamefont {C.}~\bibnamefont {Wang}}, \bibinfo {author} {\bibfnamefont {T.}~\bibnamefont {Seiler}}, \bibinfo {author} {\bibfnamefont {D.}~\bibnamefont {Sun}}, \bibinfo {author} {\bibfnamefont {S.}~\bibnamefont {Shoaee}}, \bibinfo {author} {\bibfnamefont {M.}~\bibnamefont {Saladina}},\ and\ \bibinfo {author} {\bibfnamefont {C.}~\bibnamefont {Deibel}},\ }\bibfield  {title} {\bibinfo {title} {The contribution of electron and hole conductivity to the transport loss in organic solar cells},\ }\href@noop {} {\bibfield  {journal} {\bibinfo  {journal} {arXiv preprint arXiv:2508.11399}\ } (\bibinfo {year} {2025}{\natexlab{b}})}\BibitemShut {NoStop}%
\bibitem [{\citenamefont {Stauffer}\ and\ \citenamefont {Aharony}(2018)}]{stauffer2018introduction}%
  \BibitemOpen
  \bibfield  {author} {\bibinfo {author} {\bibfnamefont {D.}~\bibnamefont {Stauffer}}\ and\ \bibinfo {author} {\bibfnamefont {A.}~\bibnamefont {Aharony}},\ }\href@noop {} {\emph {\bibinfo {title} {Introduction to percolation theory}}}\ (\bibinfo  {publisher} {Taylor \& Francis},\ \bibinfo {year} {2018})\BibitemShut {NoStop}%
\bibitem [{\citenamefont {Kim}\ and\ \citenamefont {Jenekhe}(2012)}]{kim2012charge}%
  \BibitemOpen
  \bibfield  {author} {\bibinfo {author} {\bibfnamefont {F.~S.}\ \bibnamefont {Kim}}\ and\ \bibinfo {author} {\bibfnamefont {S.~A.}\ \bibnamefont {Jenekhe}},\ }\bibfield  {title} {\bibinfo {title} {Charge transport in poly (3-butylthiophene) nanowires and their nanocomposites with an insulating polymer},\ }\href@noop {} {\bibfield  {journal} {\bibinfo  {journal} {Macromolecules}\ }\textbf {\bibinfo {volume} {45}},\ \bibinfo {pages} {7514} (\bibinfo {year} {2012})}\BibitemShut {NoStop}%
\bibitem [{\citenamefont {Goffri}\ \emph {et~al.}(2006)\citenamefont {Goffri}, \citenamefont {M\"{u}ller}, \citenamefont {Stingelin-Stutzmann}, \citenamefont {Breiby}, \citenamefont {Radano}, \citenamefont {Andreasen}, \citenamefont {Thompson}, \citenamefont {Janssen}, \citenamefont {Nielsen}, \citenamefont {Smith},\ and\ \citenamefont {Sirringhaus}}]{goffri2006multicomponent}%
  \BibitemOpen
  \bibfield  {author} {\bibinfo {author} {\bibfnamefont {S.}~\bibnamefont {Goffri}}, \bibinfo {author} {\bibfnamefont {C.}~\bibnamefont {M\"{u}ller}}, \bibinfo {author} {\bibfnamefont {N.}~\bibnamefont {Stingelin-Stutzmann}}, \bibinfo {author} {\bibfnamefont {D.~W.}\ \bibnamefont {Breiby}}, \bibinfo {author} {\bibfnamefont {C.~P.}\ \bibnamefont {Radano}}, \bibinfo {author} {\bibfnamefont {J.~W.}\ \bibnamefont {Andreasen}}, \bibinfo {author} {\bibfnamefont {R.}~\bibnamefont {Thompson}}, \bibinfo {author} {\bibfnamefont {R.~A.~J.}\ \bibnamefont {Janssen}}, \bibinfo {author} {\bibfnamefont {M.~M.}\ \bibnamefont {Nielsen}}, \bibinfo {author} {\bibfnamefont {P.}~\bibnamefont {Smith}},\ and\ \bibinfo {author} {\bibfnamefont {H.}~\bibnamefont {Sirringhaus}},\ }\bibfield  {title} {\bibinfo {title} {Multicomponent semiconducting polymer systems with low crystallization-induced percolation threshold},\ }\href@noop {} {\bibfield  {journal} {\bibinfo  {journal} {Nature Materials}\ }\textbf {\bibinfo {volume} {5}},\
  \bibinfo {pages} {950} (\bibinfo {year} {2006})}\BibitemShut {NoStop}%
\bibitem [{\citenamefont {Mott}\ and\ \citenamefont {Gurney}(1948)}]{mott1948electronic}%
  \BibitemOpen
  \bibfield  {author} {\bibinfo {author} {\bibfnamefont {N.~F.}\ \bibnamefont {Mott}}\ and\ \bibinfo {author} {\bibfnamefont {R.~W.}\ \bibnamefont {Gurney}},\ }\bibfield  {title} {\bibinfo {title} {Electronic processes in ionic crystals},\ }\href@noop {} {\bibfield  {journal} {\bibinfo  {journal} {Clarendon Press, Oxford}\ } (\bibinfo {year} {1948})}\BibitemShut {NoStop}%
\bibitem [{\citenamefont {Mark}\ and\ \citenamefont {Helfrich}(1962)}]{mark1962space}%
  \BibitemOpen
  \bibfield  {author} {\bibinfo {author} {\bibfnamefont {P.}~\bibnamefont {Mark}}\ and\ \bibinfo {author} {\bibfnamefont {W.}~\bibnamefont {Helfrich}},\ }\bibfield  {title} {\bibinfo {title} {Space-charge-limited currents in organic crystals},\ }\href@noop {} {\bibfield  {journal} {\bibinfo  {journal} {Journal of Applied Physics}\ }\textbf {\bibinfo {volume} {33}},\ \bibinfo {pages} {205} (\bibinfo {year} {1962})}\BibitemShut {NoStop}%
\bibitem [{\citenamefont {Nicolai}\ \emph {et~al.}(2011)\citenamefont {Nicolai}, \citenamefont {Mandoc},\ and\ \citenamefont {Blom}}]{nicolai2011electron}%
  \BibitemOpen
  \bibfield  {author} {\bibinfo {author} {\bibfnamefont {H.}~\bibnamefont {Nicolai}}, \bibinfo {author} {\bibfnamefont {M.}~\bibnamefont {Mandoc}},\ and\ \bibinfo {author} {\bibfnamefont {P.}~\bibnamefont {Blom}},\ }\bibfield  {title} {\bibinfo {title} {Electron traps in semiconducting polymers: Exponential versus gaussian trap distribution},\ }\href@noop {} {\bibfield  {journal} {\bibinfo  {journal} {Physical Review B—Condensed Matter and Materials Physics}\ }\textbf {\bibinfo {volume} {83}},\ \bibinfo {pages} {195204} (\bibinfo {year} {2011})}\BibitemShut {NoStop}%
\bibitem [{\citenamefont {Zhang}\ \emph {et~al.}(2020)\citenamefont {Zhang}, \citenamefont {Chen}, \citenamefont {Xiao}, \citenamefont {Chow}, \citenamefont {Ren}, \citenamefont {Kupgan}, \citenamefont {Jiao}, \citenamefont {Chan}, \citenamefont {Du}, \citenamefont {Xia}, \citenamefont {Chen}, \citenamefont {Yuan}, \citenamefont {Zhang}, \citenamefont {Zhang}, \citenamefont {Liu}, \citenamefont {Zou}, \citenamefont {Yan}, \citenamefont {Wong}, \citenamefont {Coropceanu}, \citenamefont {Li}, \citenamefont {Brabec}, \citenamefont {Bredas}, \citenamefont {Yip},\ and\ \citenamefont {Cao}}]{zhang2020delocalization}%
  \BibitemOpen
  \bibfield  {author} {\bibinfo {author} {\bibfnamefont {G.}~\bibnamefont {Zhang}}, \bibinfo {author} {\bibfnamefont {X.-K.}\ \bibnamefont {Chen}}, \bibinfo {author} {\bibfnamefont {J.}~\bibnamefont {Xiao}}, \bibinfo {author} {\bibfnamefont {P.~C.~Y.}\ \bibnamefont {Chow}}, \bibinfo {author} {\bibfnamefont {M.}~\bibnamefont {Ren}}, \bibinfo {author} {\bibfnamefont {G.}~\bibnamefont {Kupgan}}, \bibinfo {author} {\bibfnamefont {X.}~\bibnamefont {Jiao}}, \bibinfo {author} {\bibfnamefont {C.~C.~S.}\ \bibnamefont {Chan}}, \bibinfo {author} {\bibfnamefont {X.}~\bibnamefont {Du}}, \bibinfo {author} {\bibfnamefont {R.}~\bibnamefont {Xia}}, \bibinfo {author} {\bibfnamefont {Z.}~\bibnamefont {Chen}}, \bibinfo {author} {\bibfnamefont {J.}~\bibnamefont {Yuan}}, \bibinfo {author} {\bibfnamefont {Y.}~\bibnamefont {Zhang}}, \bibinfo {author} {\bibfnamefont {S.}~\bibnamefont {Zhang}}, \bibinfo {author} {\bibfnamefont {Y.}~\bibnamefont {Liu}}, \bibinfo {author} {\bibfnamefont {Y.}~\bibnamefont {Zou}}, \bibinfo {author}
  {\bibfnamefont {H.}~\bibnamefont {Yan}}, \bibinfo {author} {\bibfnamefont {K.~S.}\ \bibnamefont {Wong}}, \bibinfo {author} {\bibfnamefont {V.}~\bibnamefont {Coropceanu}}, \bibinfo {author} {\bibfnamefont {N.}~\bibnamefont {Li}}, \bibinfo {author} {\bibfnamefont {C.~J.}\ \bibnamefont {Brabec}}, \bibinfo {author} {\bibfnamefont {J.-L.}\ \bibnamefont {Bredas}}, \bibinfo {author} {\bibfnamefont {H.-L.}\ \bibnamefont {Yip}},\ and\ \bibinfo {author} {\bibfnamefont {Y.}~\bibnamefont {Cao}},\ }\bibfield  {title} {\bibinfo {title} {Delocalization of exciton and electron wavefunction in non-fullerene acceptor molecules enables efficient organic solar cells},\ }\href@noop {} {\bibfield  {journal} {\bibinfo  {journal} {Nature Communications}\ }\textbf {\bibinfo {volume} {11}},\ \bibinfo {pages} {3943} (\bibinfo {year} {2020})}\BibitemShut {NoStop}%
\bibitem [{\citenamefont {Pranav}\ \emph {et~al.}(2024)\citenamefont {Pranav}, \citenamefont {Shukla}, \citenamefont {Moser}, \citenamefont {Rumeney}, \citenamefont {Liu}, \citenamefont {Wang}, \citenamefont {Sun}, \citenamefont {Smeets}, \citenamefont {Tokmoldin}, \citenamefont {Cao}, \citenamefont {He}, \citenamefont {Beitz}, \citenamefont {Jaiser}, \citenamefont {Hultzsch}, \citenamefont {Shoaee}, \citenamefont {Maes}, \citenamefont {L\"{u}er}, \citenamefont {Brabec}, \citenamefont {Vandewal}, \citenamefont {Andrienko}, \citenamefont {Ludwigs},\ and\ \citenamefont {Neher}}]{pranav2024}%
  \BibitemOpen
  \bibfield  {author} {\bibinfo {author} {\bibfnamefont {M.}~\bibnamefont {Pranav}}, \bibinfo {author} {\bibfnamefont {A.}~\bibnamefont {Shukla}}, \bibinfo {author} {\bibfnamefont {D.}~\bibnamefont {Moser}}, \bibinfo {author} {\bibfnamefont {J.}~\bibnamefont {Rumeney}}, \bibinfo {author} {\bibfnamefont {W.}~\bibnamefont {Liu}}, \bibinfo {author} {\bibfnamefont {R.}~\bibnamefont {Wang}}, \bibinfo {author} {\bibfnamefont {B.}~\bibnamefont {Sun}}, \bibinfo {author} {\bibfnamefont {S.}~\bibnamefont {Smeets}}, \bibinfo {author} {\bibfnamefont {N.}~\bibnamefont {Tokmoldin}}, \bibinfo {author} {\bibfnamefont {Y.}~\bibnamefont {Cao}}, \bibinfo {author} {\bibfnamefont {G.}~\bibnamefont {He}}, \bibinfo {author} {\bibfnamefont {T.}~\bibnamefont {Beitz}}, \bibinfo {author} {\bibfnamefont {F.}~\bibnamefont {Jaiser}}, \bibinfo {author} {\bibfnamefont {T.}~\bibnamefont {Hultzsch}}, \bibinfo {author} {\bibfnamefont {S.}~\bibnamefont {Shoaee}}, \bibinfo {author} {\bibfnamefont {W.}~\bibnamefont {Maes}}, \bibinfo {author}
  {\bibfnamefont {L.}~\bibnamefont {L\"{u}er}}, \bibinfo {author} {\bibfnamefont {C.}~\bibnamefont {Brabec}}, \bibinfo {author} {\bibfnamefont {K.}~\bibnamefont {Vandewal}}, \bibinfo {author} {\bibfnamefont {D.}~\bibnamefont {Andrienko}}, \bibinfo {author} {\bibfnamefont {S.}~\bibnamefont {Ludwigs}},\ and\ \bibinfo {author} {\bibfnamefont {D.}~\bibnamefont {Neher}},\ }\bibfield  {title} {\bibinfo {title} {On the critical competition between singlet exciton decay and free charge generation in non-fullerene based organic solar cells with low energetic offsets},\ }\href {https://doi.org/10.1039/d4ee01409j} {\bibfield  {journal} {\bibinfo  {journal} {Energy \& Environmental Science}\ }\textbf {\bibinfo {volume} {17}},\ \bibinfo {pages} {6676} (\bibinfo {year} {2024})}\BibitemShut {NoStop}%
\bibitem [{\citenamefont {List}\ \emph {et~al.}(2023)\citenamefont {List}, \citenamefont {Faisst}, \citenamefont {Heinz},\ and\ \citenamefont {W{\"u}rfel}}]{list2023determination}%
  \BibitemOpen
  \bibfield  {author} {\bibinfo {author} {\bibfnamefont {M.}~\bibnamefont {List}}, \bibinfo {author} {\bibfnamefont {J.}~\bibnamefont {Faisst}}, \bibinfo {author} {\bibfnamefont {F.}~\bibnamefont {Heinz}},\ and\ \bibinfo {author} {\bibfnamefont {U.}~\bibnamefont {W{\"u}rfel}},\ }\bibfield  {title} {\bibinfo {title} {Determination of free charge carrier luminescence and quasi-fermi level separation in organic solar cells via transient photoluminescence measurements},\ }\href@noop {} {\bibfield  {journal} {\bibinfo  {journal} {Advanced Optical Materials}\ }\textbf {\bibinfo {volume} {11}},\ \bibinfo {pages} {2300895} (\bibinfo {year} {2023})}\BibitemShut {NoStop}%
\bibitem [{\citenamefont {Faisst}\ \emph {et~al.}(2025)\citenamefont {Faisst}, \citenamefont {List}, \citenamefont {Baretzky}, \citenamefont {Bett},\ and\ \citenamefont {W{\"u}rfel}}]{faisst2025implied}%
  \BibitemOpen
  \bibfield  {author} {\bibinfo {author} {\bibfnamefont {J.}~\bibnamefont {Faisst}}, \bibinfo {author} {\bibfnamefont {M.}~\bibnamefont {List}}, \bibinfo {author} {\bibfnamefont {C.}~\bibnamefont {Baretzky}}, \bibinfo {author} {\bibfnamefont {A.~W.}\ \bibnamefont {Bett}},\ and\ \bibinfo {author} {\bibfnamefont {U.}~\bibnamefont {W{\"u}rfel}},\ }\bibfield  {title} {\bibinfo {title} {Implied voltage and current characterization in organic solar cells using transient photoluminescence},\ }\href@noop {} {\bibfield  {journal} {\bibinfo  {journal} {Advanced Energy Materials}\ }\textbf {\bibinfo {volume} {15}},\ \bibinfo {pages} {2501348} (\bibinfo {year} {2025})}\BibitemShut {NoStop}%
\bibitem [{\citenamefont {Nelson}(2003{\natexlab{b}})}]{nelson2003}%
  \BibitemOpen
  \bibfield  {author} {\bibinfo {author} {\bibfnamefont {J.}~\bibnamefont {Nelson}},\ }\bibfield  {title} {\bibinfo {title} {Diffusion-limited recombination in polymer-fullerene blends and its influence on photocurrent collection},\ }\href {https://doi.org/10.1103/physrevb.67.155209} {\bibfield  {journal} {\bibinfo  {journal} {Physical Review B}\ }\textbf {\bibinfo {volume} {67}},\ \bibinfo {pages} {155209} (\bibinfo {year} {2003}{\natexlab{b}})}\BibitemShut {NoStop}%
\bibitem [{\citenamefont {Gorenflot}\ \emph {et~al.}(2014)\citenamefont {Gorenflot}, \citenamefont {Heiber}, \citenamefont {Baumann}, \citenamefont {Lorrmann}, \citenamefont {Gunz}, \citenamefont {K\"{a}mpgen}, \citenamefont {Dyakonov},\ and\ \citenamefont {Deibel}}]{gorenflot2014}%
  \BibitemOpen
  \bibfield  {author} {\bibinfo {author} {\bibfnamefont {J.}~\bibnamefont {Gorenflot}}, \bibinfo {author} {\bibfnamefont {M.~C.}\ \bibnamefont {Heiber}}, \bibinfo {author} {\bibfnamefont {A.}~\bibnamefont {Baumann}}, \bibinfo {author} {\bibfnamefont {J.}~\bibnamefont {Lorrmann}}, \bibinfo {author} {\bibfnamefont {M.}~\bibnamefont {Gunz}}, \bibinfo {author} {\bibfnamefont {A.}~\bibnamefont {K\"{a}mpgen}}, \bibinfo {author} {\bibfnamefont {V.}~\bibnamefont {Dyakonov}},\ and\ \bibinfo {author} {\bibfnamefont {C.}~\bibnamefont {Deibel}},\ }\bibfield  {title} {\bibinfo {title} {Nongeminate recombination in neat {P3HT} and {P3HT:PCBM} blend films},\ }\href {https://doi.org/10.1063/1.4870805} {\bibfield  {journal} {\bibinfo  {journal} {Journal of Applied Physics}\ }\textbf {\bibinfo {volume} {115}},\ \bibinfo {pages} {144502} (\bibinfo {year} {2014})}\BibitemShut {NoStop}%
\bibitem [{\citenamefont {Scher}\ and\ \citenamefont {Montroll}(1975)}]{scher1975anomalous}%
  \BibitemOpen
  \bibfield  {author} {\bibinfo {author} {\bibfnamefont {H.}~\bibnamefont {Scher}}\ and\ \bibinfo {author} {\bibfnamefont {E.~W.}\ \bibnamefont {Montroll}},\ }\bibfield  {title} {\bibinfo {title} {Anomalous transit-time dispersion in amorphous solids},\ }\href@noop {} {\bibfield  {journal} {\bibinfo  {journal} {Physical Review B}\ }\textbf {\bibinfo {volume} {12}},\ \bibinfo {pages} {2455} (\bibinfo {year} {1975})}\BibitemShut {NoStop}%
\bibitem [{\citenamefont {Kurpiers}\ and\ \citenamefont {Neher}(2016)}]{kurpiers2016dispersive}%
  \BibitemOpen
  \bibfield  {author} {\bibinfo {author} {\bibfnamefont {J.}~\bibnamefont {Kurpiers}}\ and\ \bibinfo {author} {\bibfnamefont {D.}~\bibnamefont {Neher}},\ }\bibfield  {title} {\bibinfo {title} {Dispersive non-geminate recombination in an amorphous polymer: fullerene blend},\ }\href@noop {} {\bibfield  {journal} {\bibinfo  {journal} {Scientific Reports}\ }\textbf {\bibinfo {volume} {6}},\ \bibinfo {pages} {26832} (\bibinfo {year} {2016})}\BibitemShut {NoStop}%
\bibitem [{\citenamefont {Langevin}(1903)}]{langevin1903recombinaison}%
  \BibitemOpen
  \bibfield  {author} {\bibinfo {author} {\bibfnamefont {P.}~\bibnamefont {Langevin}},\ }\bibfield  {title} {\bibinfo {title} {Recombinaison et mobilit{\'e}s des ions dans les gaz},\ }\href@noop {} {\bibfield  {journal} {\bibinfo  {journal} {Annales de Chimie et de Physique}\ }\textbf {\bibinfo {volume} {28}},\ \bibinfo {pages} {433} (\bibinfo {year} {1903})}\BibitemShut {NoStop}%
\bibitem [{\citenamefont {Koster}\ \emph {et~al.}(2006)\citenamefont {Koster}, \citenamefont {Mihailetchi},\ and\ \citenamefont {Blom}}]{koster2006bimolecular}%
  \BibitemOpen
  \bibfield  {author} {\bibinfo {author} {\bibfnamefont {L.}~\bibnamefont {Koster}}, \bibinfo {author} {\bibfnamefont {V.}~\bibnamefont {Mihailetchi}},\ and\ \bibinfo {author} {\bibfnamefont {P.}~\bibnamefont {Blom}},\ }\bibfield  {title} {\bibinfo {title} {Bimolecular recombination in polymer/fullerene bulk heterojunction solar cells},\ }\href@noop {} {\bibfield  {journal} {\bibinfo  {journal} {Applied Physics Letters}\ }\textbf {\bibinfo {volume} {88}},\ \bibinfo {pages} {052104} (\bibinfo {year} {2006})}\BibitemShut {NoStop}%
\bibitem [{\citenamefont {Deibel}\ \emph {et~al.}(2009{\natexlab{b}})\citenamefont {Deibel}, \citenamefont {Wagenpfahl},\ and\ \citenamefont {Dyakonov}}]{deibel2009prb}%
  \BibitemOpen
  \bibfield  {author} {\bibinfo {author} {\bibfnamefont {C.}~\bibnamefont {Deibel}}, \bibinfo {author} {\bibfnamefont {A.}~\bibnamefont {Wagenpfahl}},\ and\ \bibinfo {author} {\bibfnamefont {V.}~\bibnamefont {Dyakonov}},\ }\bibfield  {title} {\bibinfo {title} {Origin of reduced polaron recombination in organic semiconductor devices},\ }\href {https://doi.org/10.1103/physrevb.80.075203} {\bibfield  {journal} {\bibinfo  {journal} {Physical Review B}\ }\textbf {\bibinfo {volume} {80}},\ \bibinfo {pages} {075203} (\bibinfo {year} {2009}{\natexlab{b}})}\BibitemShut {NoStop}%
\bibitem [{\citenamefont {Heiber}\ \emph {et~al.}(2015)\citenamefont {Heiber}, \citenamefont {Baumbach}, \citenamefont {Dyakonov},\ and\ \citenamefont {Deibel}}]{heiber2015prl}%
  \BibitemOpen
  \bibfield  {author} {\bibinfo {author} {\bibfnamefont {M.~C.}\ \bibnamefont {Heiber}}, \bibinfo {author} {\bibfnamefont {C.}~\bibnamefont {Baumbach}}, \bibinfo {author} {\bibfnamefont {V.}~\bibnamefont {Dyakonov}},\ and\ \bibinfo {author} {\bibfnamefont {C.}~\bibnamefont {Deibel}},\ }\bibfield  {title} {\bibinfo {title} {Encounter-limited charge-carrier recombination in phase-separated organic semiconductor blends},\ }\href {https://doi.org/10.1103/physrevlett.114.136602} {\bibfield  {journal} {\bibinfo  {journal} {Physical Review Letters}\ }\textbf {\bibinfo {volume} {114}},\ \bibinfo {pages} {136602} (\bibinfo {year} {2015})}\BibitemShut {NoStop}%
\bibitem [{\citenamefont {Koster}\ \emph {et~al.}(2005)\citenamefont {Koster}, \citenamefont {Mihailetchi}, \citenamefont {Ramaker},\ and\ \citenamefont {Blom}}]{koster2005light}%
  \BibitemOpen
  \bibfield  {author} {\bibinfo {author} {\bibfnamefont {L.~J.~A.}\ \bibnamefont {Koster}}, \bibinfo {author} {\bibfnamefont {V.~D.}\ \bibnamefont {Mihailetchi}}, \bibinfo {author} {\bibfnamefont {R.}~\bibnamefont {Ramaker}},\ and\ \bibinfo {author} {\bibfnamefont {P.~W.}\ \bibnamefont {Blom}},\ }\bibfield  {title} {\bibinfo {title} {Light intensity dependence of open-circuit voltage of polymer: fullerene solar cells},\ }\href@noop {} {\bibfield  {journal} {\bibinfo  {journal} {Applied Physics Letters}\ }\textbf {\bibinfo {volume} {86}} (\bibinfo {year} {2005})}\BibitemShut {NoStop}%
\bibitem [{\citenamefont {Smoluchowski}(1918)}]{smoluchowski1918versuch}%
  \BibitemOpen
  \bibfield  {author} {\bibinfo {author} {\bibfnamefont {M.~v.}\ \bibnamefont {Smoluchowski}},\ }\bibfield  {title} {\bibinfo {title} {Versuch einer mathematischen {T}heorie der {K}oagulationskinetik kolloider {L}{\"o}sungen},\ }\href@noop {} {\bibfield  {journal} {\bibinfo  {journal} {Zeitschrift f{\"u}r Physikalische Chemie}\ }\textbf {\bibinfo {volume} {92}},\ \bibinfo {pages} {129} (\bibinfo {year} {1918})}\BibitemShut {NoStop}%
\bibitem [{\citenamefont {Mozumder}(1968)}]{mozumder1968theory}%
  \BibitemOpen
  \bibfield  {author} {\bibinfo {author} {\bibfnamefont {A.}~\bibnamefont {Mozumder}},\ }\bibfield  {title} {\bibinfo {title} {Theory of neutralization of an isolated ion pair: application of the method of prescribed diffusion to random walk in a {C}oulomb field},\ }\href@noop {} {\bibfield  {journal} {\bibinfo  {journal} {The Journal of Chemical Physics}\ }\textbf {\bibinfo {volume} {48}},\ \bibinfo {pages} {1659} (\bibinfo {year} {1968})}\BibitemShut {NoStop}%
\bibitem [{\citenamefont {Abell}\ and\ \citenamefont {Mozumder}(1972)}]{abell1972application}%
  \BibitemOpen
  \bibfield  {author} {\bibinfo {author} {\bibfnamefont {G.}~\bibnamefont {Abell}}\ and\ \bibinfo {author} {\bibfnamefont {A.}~\bibnamefont {Mozumder}},\ }\bibfield  {title} {\bibinfo {title} {Application of diffusion model for recombination of isolated pairs in condensed media},\ }\href@noop {} {\bibfield  {journal} {\bibinfo  {journal} {The Journal of Chemical Physics}\ }\textbf {\bibinfo {volume} {56}},\ \bibinfo {pages} {4079} (\bibinfo {year} {1972})}\BibitemShut {NoStop}%
\bibitem [{\citenamefont {Hong}\ and\ \citenamefont {Noolandi}(1978)}]{hong1978solution}%
  \BibitemOpen
  \bibfield  {author} {\bibinfo {author} {\bibfnamefont {K.}~\bibnamefont {Hong}}\ and\ \bibinfo {author} {\bibfnamefont {J.}~\bibnamefont {Noolandi}},\ }\bibfield  {title} {\bibinfo {title} {Solution of the {S}moluchowski equation with a {C}oulomb potential. {I}. {G}eneral results},\ }\href@noop {} {\bibfield  {journal} {\bibinfo  {journal} {The Journal of Chemical Physics}\ }\textbf {\bibinfo {volume} {68}},\ \bibinfo {pages} {5163} (\bibinfo {year} {1978})}\BibitemShut {NoStop}%
\bibitem [{\citenamefont {Nikitenko}\ \emph {et~al.}(2001)\citenamefont {Nikitenko}, \citenamefont {Hertel},\ and\ \citenamefont {B{\"a}ssler}}]{nikitenko2001dispersive}%
  \BibitemOpen
  \bibfield  {author} {\bibinfo {author} {\bibfnamefont {V.}~\bibnamefont {Nikitenko}}, \bibinfo {author} {\bibfnamefont {D.}~\bibnamefont {Hertel}},\ and\ \bibinfo {author} {\bibfnamefont {H.}~\bibnamefont {B{\"a}ssler}},\ }\bibfield  {title} {\bibinfo {title} {Dispersive geminate recombination in a conjugated polymer},\ }\href@noop {} {\bibfield  {journal} {\bibinfo  {journal} {Chemical Physics Letters}\ }\textbf {\bibinfo {volume} {348}},\ \bibinfo {pages} {89} (\bibinfo {year} {2001})}\BibitemShut {NoStop}%
\bibitem [{\citenamefont {Kahle}\ \emph {et~al.}(2018)\citenamefont {Kahle}, \citenamefont {Rudnick}, \citenamefont {B{\"a}ssler},\ and\ \citenamefont {K{\"o}hler}}]{kahle2018interpret}%
  \BibitemOpen
  \bibfield  {author} {\bibinfo {author} {\bibfnamefont {F.-J.}\ \bibnamefont {Kahle}}, \bibinfo {author} {\bibfnamefont {A.}~\bibnamefont {Rudnick}}, \bibinfo {author} {\bibfnamefont {H.}~\bibnamefont {B{\"a}ssler}},\ and\ \bibinfo {author} {\bibfnamefont {A.}~\bibnamefont {K{\"o}hler}},\ }\bibfield  {title} {\bibinfo {title} {How to interpret absorption and fluorescence spectra of charge transfer states in an organic solar cell},\ }\href@noop {} {\bibfield  {journal} {\bibinfo  {journal} {Materials Horizons}\ }\textbf {\bibinfo {volume} {5}},\ \bibinfo {pages} {837} (\bibinfo {year} {2018})}\BibitemShut {NoStop}%
\bibitem [{\citenamefont {Noolandi}\ \emph {et~al.}(1980)\citenamefont {Noolandi}, \citenamefont {Hong},\ and\ \citenamefont {Street}}]{noolandi1980geminate}%
  \BibitemOpen
  \bibfield  {author} {\bibinfo {author} {\bibfnamefont {J.}~\bibnamefont {Noolandi}}, \bibinfo {author} {\bibfnamefont {K.}~\bibnamefont {Hong}},\ and\ \bibinfo {author} {\bibfnamefont {R.}~\bibnamefont {Street}},\ }\bibfield  {title} {\bibinfo {title} {A geminate recombination model for photoluminescence decay in plasma-deposited amorphous {Si: H}},\ }\href@noop {} {\bibfield  {journal} {\bibinfo  {journal} {Solid State Communications}\ }\textbf {\bibinfo {volume} {34}},\ \bibinfo {pages} {45} (\bibinfo {year} {1980})}\BibitemShut {NoStop}%
\bibitem [{\citenamefont {Murayama}\ and\ \citenamefont {Ninomiya}(1985)}]{murayama1985photoluminescence}%
  \BibitemOpen
  \bibfield  {author} {\bibinfo {author} {\bibfnamefont {K.}~\bibnamefont {Murayama}}\ and\ \bibinfo {author} {\bibfnamefont {T.}~\bibnamefont {Ninomiya}},\ }\bibfield  {title} {\bibinfo {title} {Photoluminescence decay affected by variable range hopping in band tail in amorphous {As2S3}},\ }\href@noop {} {\bibfield  {journal} {\bibinfo  {journal} {Solid State Communications}\ }\textbf {\bibinfo {volume} {53}},\ \bibinfo {pages} {125} (\bibinfo {year} {1985})}\BibitemShut {NoStop}%
\bibitem [{\citenamefont {Morteani}\ \emph {et~al.}(2005)\citenamefont {Morteani}, \citenamefont {Friend},\ and\ \citenamefont {Silva}}]{morteani2005}%
  \BibitemOpen
  \bibfield  {author} {\bibinfo {author} {\bibfnamefont {A.~C.}\ \bibnamefont {Morteani}}, \bibinfo {author} {\bibfnamefont {R.~H.}\ \bibnamefont {Friend}},\ and\ \bibinfo {author} {\bibfnamefont {C.}~\bibnamefont {Silva}},\ }\bibfield  {title} {\bibinfo {title} {Exciton trapping at heterojunctions in polymer blends},\ }\href {https://doi.org/10.1063/1.1924504} {\bibfield  {journal} {\bibinfo  {journal} {The Journal of Chemical Physics}\ }\textbf {\bibinfo {volume} {122}},\ \bibinfo {pages} {244906} (\bibinfo {year} {2005})}\BibitemShut {NoStop}%
\bibitem [{\citenamefont {Classen}\ \emph {et~al.}(2020)\citenamefont {Classen}, \citenamefont {Chochos}, \citenamefont {L\"{u}er}, \citenamefont {Gregoriou}, \citenamefont {Wortmann}, \citenamefont {Osvet}, \citenamefont {Forberich}, \citenamefont {McCulloch}, \citenamefont {Heum\"{u}ller},\ and\ \citenamefont {Brabec}}]{classen2020}%
  \BibitemOpen
  \bibfield  {author} {\bibinfo {author} {\bibfnamefont {A.}~\bibnamefont {Classen}}, \bibinfo {author} {\bibfnamefont {C.~L.}\ \bibnamefont {Chochos}}, \bibinfo {author} {\bibfnamefont {L.}~\bibnamefont {L\"{u}er}}, \bibinfo {author} {\bibfnamefont {V.~G.}\ \bibnamefont {Gregoriou}}, \bibinfo {author} {\bibfnamefont {J.}~\bibnamefont {Wortmann}}, \bibinfo {author} {\bibfnamefont {A.}~\bibnamefont {Osvet}}, \bibinfo {author} {\bibfnamefont {K.}~\bibnamefont {Forberich}}, \bibinfo {author} {\bibfnamefont {I.}~\bibnamefont {McCulloch}}, \bibinfo {author} {\bibfnamefont {T.}~\bibnamefont {Heum\"{u}ller}},\ and\ \bibinfo {author} {\bibfnamefont {C.~J.}\ \bibnamefont {Brabec}},\ }\bibfield  {title} {\bibinfo {title} {The role of exciton lifetime for charge generation in organic solar cells at negligible energy-level offsets},\ }\href {https://doi.org/10.1038/s41560-020-00684-7} {\bibfield  {journal} {\bibinfo  {journal} {Nature Energy}\ }\textbf {\bibinfo {volume} {5}},\ \bibinfo {pages} {711} (\bibinfo {year}
  {2020})}\BibitemShut {NoStop}%
\bibitem [{\citenamefont {Hecht}(1932)}]{hecht1932}%
  \BibitemOpen
  \bibfield  {author} {\bibinfo {author} {\bibfnamefont {K.}~\bibnamefont {Hecht}},\ }\bibfield  {title} {\bibinfo {title} {{Zum Mechanismus des lichtelektrischen Prim{\"a}rstromes in isolierenden Kristallen}},\ }\href {https://doi.org/10.1007/bf01338917} {\bibfield  {journal} {\bibinfo  {journal} {Zeitschrift f{\"u}r Physik}\ }\textbf {\bibinfo {volume} {77}},\ \bibinfo {pages} {235} (\bibinfo {year} {1932})}\BibitemShut {NoStop}%
\bibitem [{\citenamefont {Crandall}(1983)}]{crandall1983modeling}%
  \BibitemOpen
  \bibfield  {author} {\bibinfo {author} {\bibfnamefont {R.~S.}\ \bibnamefont {Crandall}},\ }\bibfield  {title} {\bibinfo {title} {Modeling of thin film solar cells: Uniform field approximation},\ }\href@noop {} {\bibfield  {journal} {\bibinfo  {journal} {Journal of Applied Physics}\ }\textbf {\bibinfo {volume} {54}},\ \bibinfo {pages} {7176} (\bibinfo {year} {1983})}\BibitemShut {NoStop}%
\bibitem [{\citenamefont {Deibel}\ and\ \citenamefont {Wagenpfahl}(2010)}]{deibel2010comment}%
  \BibitemOpen
  \bibfield  {author} {\bibinfo {author} {\bibfnamefont {C.}~\bibnamefont {Deibel}}\ and\ \bibinfo {author} {\bibfnamefont {A.}~\bibnamefont {Wagenpfahl}},\ }\bibfield  {title} {\bibinfo {title} {Comment on “{I}nterface state recombination in organic solar cells”},\ }\href {https://doi.org/10.1103/physrevb.82.207301} {\bibfield  {journal} {\bibinfo  {journal} {Physical Review B}\ }\textbf {\bibinfo {volume} {82}},\ \bibinfo {pages} {207301} (\bibinfo {year} {2010})}\BibitemShut {NoStop}%
\bibitem [{\citenamefont {Qin}\ \emph {et~al.}(2020)\citenamefont {Qin}, \citenamefont {Wang}, \citenamefont {Sun}, \citenamefont {Jiang}, \citenamefont {Hu}, \citenamefont {Xiong}, \citenamefont {Liu}, \citenamefont {Dong}, \citenamefont {Li}, \citenamefont {Jiang}, \citenamefont {Hou}, \citenamefont {Fukuda}, \citenamefont {Someya},\ and\ \citenamefont {Zhou}}]{qin2020robust}%
  \BibitemOpen
  \bibfield  {author} {\bibinfo {author} {\bibfnamefont {F.}~\bibnamefont {Qin}}, \bibinfo {author} {\bibfnamefont {W.}~\bibnamefont {Wang}}, \bibinfo {author} {\bibfnamefont {L.}~\bibnamefont {Sun}}, \bibinfo {author} {\bibfnamefont {X.}~\bibnamefont {Jiang}}, \bibinfo {author} {\bibfnamefont {L.}~\bibnamefont {Hu}}, \bibinfo {author} {\bibfnamefont {S.}~\bibnamefont {Xiong}}, \bibinfo {author} {\bibfnamefont {T.}~\bibnamefont {Liu}}, \bibinfo {author} {\bibfnamefont {X.}~\bibnamefont {Dong}}, \bibinfo {author} {\bibfnamefont {J.}~\bibnamefont {Li}}, \bibinfo {author} {\bibfnamefont {Y.}~\bibnamefont {Jiang}}, \bibinfo {author} {\bibfnamefont {J.}~\bibnamefont {Hou}}, \bibinfo {author} {\bibfnamefont {K.}~\bibnamefont {Fukuda}}, \bibinfo {author} {\bibfnamefont {T.}~\bibnamefont {Someya}},\ and\ \bibinfo {author} {\bibfnamefont {Y.}~\bibnamefont {Zhou}},\ }\bibfield  {title} {\bibinfo {title} {Robust metal ion-chelated polymer interfacial layer for ultraflexible non-fullerene organic solar cells},\
  }\href@noop {} {\bibfield  {journal} {\bibinfo  {journal} {Nature Communications}\ }\textbf {\bibinfo {volume} {11}},\ \bibinfo {pages} {4508} (\bibinfo {year} {2020})}\BibitemShut {NoStop}%
\bibitem [{\citenamefont {Kaiser}\ \emph {et~al.}(2021)\citenamefont {Kaiser}, \citenamefont {Sandberg}, \citenamefont {Zarrabi}, \citenamefont {Li}, \citenamefont {Meredith},\ and\ \citenamefont {Armin}}]{kaiser2021universal}%
  \BibitemOpen
  \bibfield  {author} {\bibinfo {author} {\bibfnamefont {C.}~\bibnamefont {Kaiser}}, \bibinfo {author} {\bibfnamefont {O.~J.}\ \bibnamefont {Sandberg}}, \bibinfo {author} {\bibfnamefont {N.}~\bibnamefont {Zarrabi}}, \bibinfo {author} {\bibfnamefont {W.}~\bibnamefont {Li}}, \bibinfo {author} {\bibfnamefont {P.}~\bibnamefont {Meredith}},\ and\ \bibinfo {author} {\bibfnamefont {A.}~\bibnamefont {Armin}},\ }\bibfield  {title} {\bibinfo {title} {A universal urbach rule for disordered organic semiconductors},\ }\href@noop {} {\bibfield  {journal} {\bibinfo  {journal} {Nature Communications}\ }\textbf {\bibinfo {volume} {12}},\ \bibinfo {pages} {3988} (\bibinfo {year} {2021})}\BibitemShut {NoStop}%
\bibitem [{\citenamefont {Rau}\ \emph {et~al.}(2017)\citenamefont {Rau}, \citenamefont {Blank}, \citenamefont {M{\"u}ller},\ and\ \citenamefont {Kirchartz}}]{rau2017efficiency}%
  \BibitemOpen
  \bibfield  {author} {\bibinfo {author} {\bibfnamefont {U.}~\bibnamefont {Rau}}, \bibinfo {author} {\bibfnamefont {B.}~\bibnamefont {Blank}}, \bibinfo {author} {\bibfnamefont {T.~C.}\ \bibnamefont {M{\"u}ller}},\ and\ \bibinfo {author} {\bibfnamefont {T.}~\bibnamefont {Kirchartz}},\ }\bibfield  {title} {\bibinfo {title} {Efficiency potential of photovoltaic materials and devices unveiled by detailed-balance analysis},\ }\href@noop {} {\bibfield  {journal} {\bibinfo  {journal} {Physical Review Applied}\ }\textbf {\bibinfo {volume} {7}},\ \bibinfo {pages} {044016} (\bibinfo {year} {2017})}\BibitemShut {NoStop}%
\bibitem [{\citenamefont {M{\"u}ller}\ \emph {et~al.}(2023)\citenamefont {M{\"u}ller}, \citenamefont {Comi}, \citenamefont {Eisner}, \citenamefont {Azzouzi}, \citenamefont {Herrera~Ruiz}, \citenamefont {Yan}, \citenamefont {Attar}, \citenamefont {Al-Hashimi},\ and\ \citenamefont {Nelson}}]{mueller2023charge}%
  \BibitemOpen
  \bibfield  {author} {\bibinfo {author} {\bibfnamefont {J.~S.}\ \bibnamefont {M{\"u}ller}}, \bibinfo {author} {\bibfnamefont {M.}~\bibnamefont {Comi}}, \bibinfo {author} {\bibfnamefont {F.}~\bibnamefont {Eisner}}, \bibinfo {author} {\bibfnamefont {M.}~\bibnamefont {Azzouzi}}, \bibinfo {author} {\bibfnamefont {D.}~\bibnamefont {Herrera~Ruiz}}, \bibinfo {author} {\bibfnamefont {J.}~\bibnamefont {Yan}}, \bibinfo {author} {\bibfnamefont {S.~S.}\ \bibnamefont {Attar}}, \bibinfo {author} {\bibfnamefont {M.}~\bibnamefont {Al-Hashimi}},\ and\ \bibinfo {author} {\bibfnamefont {J.}~\bibnamefont {Nelson}},\ }\bibfield  {title} {\bibinfo {title} {Charge-transfer state dissociation efficiency can limit free charge generation in low-offset organic solar cells},\ }\href@noop {} {\bibfield  {journal} {\bibinfo  {journal} {ACS Energy Letters}\ }\textbf {\bibinfo {volume} {8}},\ \bibinfo {pages} {3387} (\bibinfo {year} {2023})}\BibitemShut {NoStop}%
\bibitem [{\citenamefont {Lami}\ \emph {et~al.}(2019)\citenamefont {Lami}, \citenamefont {Weu}, \citenamefont {Zhang}, \citenamefont {Chen}, \citenamefont {Fei}, \citenamefont {Heeney}, \citenamefont {Friend},\ and\ \citenamefont {Vaynzof}}]{Lami2019}%
  \BibitemOpen
  \bibfield  {author} {\bibinfo {author} {\bibfnamefont {V.}~\bibnamefont {Lami}}, \bibinfo {author} {\bibfnamefont {A.}~\bibnamefont {Weu}}, \bibinfo {author} {\bibfnamefont {J.}~\bibnamefont {Zhang}}, \bibinfo {author} {\bibfnamefont {Y.}~\bibnamefont {Chen}}, \bibinfo {author} {\bibfnamefont {Z.}~\bibnamefont {Fei}}, \bibinfo {author} {\bibfnamefont {M.}~\bibnamefont {Heeney}}, \bibinfo {author} {\bibfnamefont {R.~H.}\ \bibnamefont {Friend}},\ and\ \bibinfo {author} {\bibfnamefont {Y.}~\bibnamefont {Vaynzof}},\ }\bibfield  {title} {\bibinfo {title} {Visualizing the vertical energetic landscape in organic photovoltaics},\ }\href {https://doi.org/10.1016/j.joule.2019.06.018} {\bibfield  {journal} {\bibinfo  {journal} {Joule}\ }\textbf {\bibinfo {volume} {3}},\ \bibinfo {pages} {2513} (\bibinfo {year} {2019})}\BibitemShut {NoStop}%
\bibitem [{\citenamefont {Roth}(1988)}]{roth1988introduction}%
  \BibitemOpen
  \bibfield  {author} {\bibinfo {author} {\bibfnamefont {A.~E.}\ \bibnamefont {Roth}},\ }\bibfield  {title} {\bibinfo {title} {Introduction to the shapley value},\ }\href@noop {} {\bibfield  {journal} {\bibinfo  {journal} {The Shapley value}\ }\textbf {\bibinfo {volume} {1}} (\bibinfo {year} {1988})}\BibitemShut {NoStop}%
\bibitem [{\citenamefont {Mazzolini}\ \emph {et~al.}(2025)\citenamefont {Mazzolini}, \citenamefont {Qiao}, \citenamefont {Muller}, \citenamefont {Furlan}, \citenamefont {Sanviti}, \citenamefont {Nodari}, \citenamefont {Rimmele}, \citenamefont {Collauto}, \citenamefont {Deibel}, \citenamefont {Heeney}, \citenamefont {Martin}, \citenamefont {Eisner}, \citenamefont {Nelson}, \citenamefont {Gasparini},\ and\ \citenamefont {Panidi}}]{mazzolini2025discerning}%
  \BibitemOpen
  \bibfield  {author} {\bibinfo {author} {\bibfnamefont {E.}~\bibnamefont {Mazzolini}}, \bibinfo {author} {\bibfnamefont {Z.}~\bibnamefont {Qiao}}, \bibinfo {author} {\bibfnamefont {J.}~\bibnamefont {Muller}}, \bibinfo {author} {\bibfnamefont {F.}~\bibnamefont {Furlan}}, \bibinfo {author} {\bibfnamefont {M.}~\bibnamefont {Sanviti}}, \bibinfo {author} {\bibfnamefont {D.}~\bibnamefont {Nodari}}, \bibinfo {author} {\bibfnamefont {M.}~\bibnamefont {Rimmele}}, \bibinfo {author} {\bibfnamefont {A.}~\bibnamefont {Collauto}}, \bibinfo {author} {\bibfnamefont {C.}~\bibnamefont {Deibel}}, \bibinfo {author} {\bibfnamefont {M.}~\bibnamefont {Heeney}}, \bibinfo {author} {\bibfnamefont {J.}~\bibnamefont {Martin}}, \bibinfo {author} {\bibfnamefont {F.}~\bibnamefont {Eisner}}, \bibinfo {author} {\bibfnamefont {J.}~\bibnamefont {Nelson}}, \bibinfo {author} {\bibfnamefont {N.}~\bibnamefont {Gasparini}},\ and\ \bibinfo {author} {\bibfnamefont {J.}~\bibnamefont {Panidi}},\ }\bibfield  {title} {\bibinfo {title} {Discerning blend
  microstructure and charge recombination for stable biorenewable-based organic photovoltaics},\ }\href@noop {} {\bibfield  {journal} {\bibinfo  {journal} {Advanced Energy Materials}\ }\textbf {\bibinfo {volume} {16}},\ \bibinfo {pages} {2405635} (\bibinfo {year} {2025})}\BibitemShut {NoStop}%
\bibitem [{\citenamefont {Thor}\ \emph {et~al.}(2025)\citenamefont {Thor}, \citenamefont {B{\"u}nzli}, \citenamefont {Wong},\ and\ \citenamefont {Tanner}}]{thor2025shedding}%
  \BibitemOpen
  \bibfield  {author} {\bibinfo {author} {\bibfnamefont {W.}~\bibnamefont {Thor}}, \bibinfo {author} {\bibfnamefont {J.-C.~G.}\ \bibnamefont {B{\"u}nzli}}, \bibinfo {author} {\bibfnamefont {K.-L.}\ \bibnamefont {Wong}},\ and\ \bibinfo {author} {\bibfnamefont {P.~A.}\ \bibnamefont {Tanner}},\ }\bibfield  {title} {\bibinfo {title} {Shedding light on luminescence lifetime measurement and associated data treatment},\ }\href@noop {} {\bibfield  {journal} {\bibinfo  {journal} {Advanced Photonics Research}\ }\textbf {\bibinfo {volume} {6}},\ \bibinfo {pages} {2400081} (\bibinfo {year} {2025})}\BibitemShut {NoStop}%
\bibitem [{\citenamefont {van Berkel}\ \emph {et~al.}(1993)\citenamefont {van Berkel}, \citenamefont {Powell}, \citenamefont {Franklin},\ and\ \citenamefont {French}}]{vanBerkel1993}%
  \BibitemOpen
  \bibfield  {author} {\bibinfo {author} {\bibfnamefont {C.}~\bibnamefont {van Berkel}}, \bibinfo {author} {\bibfnamefont {M.~J.}\ \bibnamefont {Powell}}, \bibinfo {author} {\bibfnamefont {A.~R.}\ \bibnamefont {Franklin}},\ and\ \bibinfo {author} {\bibfnamefont {I.~D.}\ \bibnamefont {French}},\ }\bibfield  {title} {\bibinfo {title} {Quality factor in {a-Si:H} nip and pin diodes},\ }\href {https://doi.org/10.1063/1.353755} {\bibfield  {journal} {\bibinfo  {journal} {Journal of Applied Physics}\ }\textbf {\bibinfo {volume} {73}},\ \bibinfo {pages} {5264–5268} (\bibinfo {year} {1993})}\BibitemShut {NoStop}%
\bibitem [{\citenamefont {Kniepert}\ \emph {et~al.}(2014)\citenamefont {Kniepert}, \citenamefont {Lange}, \citenamefont {Van Der~Kaap}, \citenamefont {Koster},\ and\ \citenamefont {Neher}}]{kniepert2014conclusive}%
  \BibitemOpen
  \bibfield  {author} {\bibinfo {author} {\bibfnamefont {J.}~\bibnamefont {Kniepert}}, \bibinfo {author} {\bibfnamefont {I.}~\bibnamefont {Lange}}, \bibinfo {author} {\bibfnamefont {N.~J.}\ \bibnamefont {Van Der~Kaap}}, \bibinfo {author} {\bibfnamefont {L.~J.~A.}\ \bibnamefont {Koster}},\ and\ \bibinfo {author} {\bibfnamefont {D.}~\bibnamefont {Neher}},\ }\bibfield  {title} {\bibinfo {title} {A conclusive view on charge generation, recombination, and extraction in as-prepared and annealed {P3HT: PCBM} blends: combined experimental and simulation work},\ }\href@noop {} {\bibfield  {journal} {\bibinfo  {journal} {Advanced Energy Materials}\ }\textbf {\bibinfo {volume} {4}},\ \bibinfo {pages} {1301401} (\bibinfo {year} {2014})}\BibitemShut {NoStop}%
\end{thebibliography}%

\clearpage

\onecolumngrid 
\graphicspath{{SI_figures/}}
\renewcommand{\thepage}{S\arabic{page}}  
\renewcommand{\thesection}{S\arabic{section}}   
\renewcommand{\thetable}{S\arabic{table}}   
\renewcommand{\thefigure}{S\arabic{figure}}
\renewcommand{\theequation}{S\arabic{equation}}
\renewcommand{\figurename}{Figure}
\renewcommand{\tablename}{Table}
\setlength{\parskip}{0.3cm}
\setlength{\parindent}{0pt}
\setcounter{page}{1}
\setcounter{figure}{0}
\setcounter{table}{0}
\setcounter{equation}{0}
\setcounter{section}{0}

\addtocontents{toc}{\protect\setcounter{tocdepth}{1}}
\setlength{\cftsecnumwidth}{3em}
\startcontents
\clearpage
\begin{center}
    \large
    Supporting Information\\
    \vspace{0.2cm}
    {\bfseries Rethinking charge transport and recombination in donor-diluted organic solar cells}\\
    \vspace{0.2cm}
    {\normalsize Chen Wang$^{1}$, Christopher Wöpke$^{1}$, Toni Seiler$^{1}$, Jared Faisst$^{2, 3}$, Mathias List$^{3,4}$, Meike Kuhn$^{5}$, Bekcy Joseph$^{6}$, Alexander Ehm$^{1}$, Dietrich R.\ T.\ Zahn$^{1}$, Yana Vaynzof$^{6}$, Eva M. Herzig$^{5}$, Roderick C.\ I.\ Mackenzie$^{7}$, Uli Würfel$^{3,4}$, Maria Saladina$^{1}$, Carsten Deibel$^{1}$\textsuperscript{*}\\
    \vspace{0.2cm}
    $^{1}$Institut für Physik, Technische Universität Chemnitz, 09126 Chemnitz, Germany\\
    $^{2}$Institute of Physics, University of Freiburg, 79104 Freiburg, Germany\\
    $^{3}$Fraunhofer Institute for Solar Energy Systems ISE, 79110 Freiburg, Germany\\
    $^{4}$Freiburg Materials Research Center FMF, University of Freiburg, 79104 Freiburg, Germany\\
    $^{5}$Dynamics and Structure Formation -- Herzig Group, University of Bayreuth, 95447 Bayreuth, Germany\\
    $^{6}$Leibniz-Institut für Festkörper- und Werkstoffforschung Dresden (IFW), 01069 Dresden, Germany\\
    $^{7}$Department of Engineering, Durham University, Lower Mount Joy, South Road, Durham DH1 3LE, United Kingdom
    }
\end{center}

\clearpage
\printcontents{}{1}{\setcounter{tocdepth}{2}}
\newpage

\section{Materials and device fabrication}\label{SI_sec:fabrication}

\textbf{Materials}: PM6, Y12, and PEDOT:F were obtained from Brilliant Matters Inc (Canada). ZnO nanoparticles (N10) were sourced from Avantama AG, and PDINO was supplied by ONE-material. Zinc acetate dihydrate, branched PEI (Mw 25\,000~$\mathrm{g\ \cdot mol^{-1}}$), 2-methoxyethanol, chloroform, and methanol were purchased from Merck. PEDOT:PSS (4083) was acquired from Heraeus Deutschland GmbH \& Co.\ KG. All materials and solvents were used as received without further purification. ZnO and PEDOT:F dispersions were ultrasonically bathed and filtered before use.

\textbf{Film preparation}: PM6 and Y12 were separately dissolved in chloroform (each at 12~$\mathrm{mg\ ml^{-1}}$) and blended at varying volume ratios to achieve PM6 volume fraction ranging from 1\% to 45\%. The total volume of the mixed solutions was kept constant at 200~$\text{\textmu}\mathrm{l}$ for all ratios. The uncertainty in the PM6 fraction comes from the volumetric uncertainty during blending solutions. The error bars of the PM6 fraction were estimated according to the typical volumetric error of the pipettes (Eppendorf, 10~$\text{\textmu}\mathrm{l}$ and 100~$\text{\textmu}\mathrm{l}$ range). The neat PM6 and Y12 films, and the blend films were spin-coated (spin speeds: 1000 to 3000~rpm, annealed at $100 \,^\circ\mathrm{C}$ for 10~min) on the various substrates for film characterizations. Glass substrate was used in ellipsometry, photoluminescence (PL), and short-time transient PL measurements. ITO(non-patterned)/ZnO was used in UPS/XPS profiling and GIWAXS measurements. UV fused silica substrate was used in long-time transient PL measurement. In RSoXs measurement, the films were coated on a rigid substrate, then peeled off and transferred to silicon nitride membranes.

\textbf{Device fabrication}: Inverted solar cells with the structure ITO/ZnO/PM6:Y12/PEDOT:F/Ag were fabricated. ITO substrates were sequentially cleaned in an ultrasonic bath using diluted Hellmanex, deionized water, acetone, and 2-propanol, followed by drying under a nitrogen stream. A dispersion of ZnO nanoparticles (2.5~wt\% in 2-propanol) was spin-coated at 2500~rpm for 60~s and then thermally annealed at $200 \,^\circ\mathrm{C}$ for 30~min. Subsequent layers were deposited in a nitrogen-filled glovebox. The PM6:Y12 solutions were spin-coated onto the ITO/ZnO substrates under different spin speeds (ranging from 1000 to 3000~rpm) to obtain active layers with a similar thickness of $\sim$60--70~nm. The films were thermally annealed at $100 \,^\circ\mathrm{C}$ for 10~min. A PEDOT:F layer was then spin-coated at 3000~rpm for 50~s and annealed at $100 \,^\circ\mathrm{C}$ for 5~min. Finally, 150 nm of Ag was thermally evaporated through a shadow mask under vacuum ($\approx 2\times10^{-6}$ Torr) to complete the devices, with an active area of 0.04~cm$^2$.

For TDCF measurements, devices with very low leakage currents were needed. To this end, the ZnO layer was replaced by PEI-Zn.\cite{qin2020robust} PEI-Zn 2-methoxyethanol solution was spin-coated onto clean ITO substrates at 4000~rpm for 45~s, followed by thermal annealing at $150\,^\circ\mathrm{C}$ for 10~min. The concentrations of both PM6 and Y12 solutions were increased to 15~$\mathrm{mg\ ml^{-1}}$ to achieve thicker active layers (thickness of $\sim$75--87~nm) and suppress shunt pathways. A smaller active area of 0.005~cm$^2$ was used in TDCF measurement to reduce the RC constant.

A configuration of ITO/ZnO/PM6:Y12/PDINO/Al and ITO/PEDOT:PSS/PM6:Y12/PEDOT:F/Ag were used for electron-only and hole-only devices, respectively. For electron-only devices, a PDINO layer was deposited by spin-coating a methanol solution (1~$\mathrm{mg\ ml^{-1}}$) at 4000~rpm for 15~s, followed by thermal treatment at $60 \,^\circ\mathrm{C}$ for 5~s. For hole-only devices, oxygen plasma-treated ITO substrates were coated with PEDOT:PSS (spin-coated at 3000~rpm for 50~s and annealed at $140 \,^\circ\mathrm{C}$ for 15~min). All other layers were prepared as described for the normal solar cell devices.

\textbf{Statistical analysis:} The statistical solar cell parameters were obtained from 20 independently fabricated cells. The statistics of active layer thickness were obtained by measuring three different positions on the film. Both were presented in mean $\pm$ standard deviation.

\clearpage
\section{Experimental methods}\label{SI_sec:exp_methods}

\subsection{Current density--voltage measurements}

The current density--voltage (JV) characteristics of the photovoltaic and single-carrier devices containing different PM6 fractions were measured inside a nitrogen-filled glovebox. JV curves were recorded using a Keithley 236 source-measure unit under both dark and 1~sun (100~mW\,cm$^{-2}$, AM1.5G) equivalent illumination provided by a Wavelabs LS-2 solar simulator. No aperture was used. To minimize device heating during the measurement, the illumination duration for each JV scan was limited to 1.2~s.

\subsection{External quantum efficiency}

External quantum efficiency (EQE) spectra were measured by illuminating the devices with a monochromatic, spectrally tunable light beam generated by a Bentham monochromator equipped with an optical chopper. The resulting photocurrent was detected using a lock-in amplifier (Stanford Research Systems SR830) synchronized to the chopper frequency. The signal from a calibrated ThorlabsFDS1010 silicon photodiode was measured in parallel under identical conditions to account for spectral variations of the light intensity and to obtain the absolute EQE response.

\subsection{Steady-state and transient photoluminescence}

Steady-state and time-resolved photoluminescence measurements were performed using an Edinburgh Instruments FLS1000 modular spectrometer equipped with excitation and emission double monochromators. All measurements were carried out at room temperature, with the sample mounted under ambient conditions.

For steady-state PL, the sample was excited by light from a 450~W xenon lamp passed through the excitation monochromator. Excitation wavelengths of 820~nm (acceptor excitation) and 510~nm (donor excitation) were used. The emission was dispersed by the double monochromator and detected with a NIR photomultiplier tube (nitrogen-flow-cooled housing, 500-1400~nm range, time resolution 800~ps).

Time-resolved photoluminescence (TRPL) was recorded using the TCSPC module of the FLS1000. Excitation was provided by an Agile white-light laser with pulses of $\leq$350~ps duration. Monochromatic excitation at 820~nm and 556~nm was obtained using the excitation monochromator. The repetition rate was set to 1~MHz. The instrument response function (IRF) was measured separately using a reflective substrate under identical optical conditions. TRPL decay curves were analysed by reconvolution with the measured IRF.

\subsection{Sensitive external quantum efficiency}

Sensitive EQE measurements were performed to obtain the photocurrent spectrum of the solar cell. Monochromatic illumination was produced using an MSHD-300 double monochromator (LOT Quantum Design) with a 20~nm spectral bandwidth, coupled to a 100~W quartz–tungsten–halogen lamp. The device was exposed to low-intensity monochromatic light, and the generated photocurrent was measured using a Zurich Instruments MFLI lock-in amplifier.

The light beam was modulated at 223~Hz with a mechanical chopper (Thorlabs MC2000B-EC), and a set of OD4 long-pass filters with progressively increasing cut-on wavelengths was employed to suppress stray-light contributions. After spectral selection, the light was transmitted through a liquid light guide (Newport 77638) and focused onto the sample with a lens. An additional filter (OD4 950nm) was inserted at $\lambda>1000$\,nm to reduce stray light artefacts.

The recorded photocurrent was normalized to the incident photon flux at each wavelength. The light intensity was measured using a Hamamatsu K1718-B dual-color photodiode comprising silicon and InGaAs detectors, with the signals amplified by Thorlabs  AMP120 transimpedance amplifiers.

All measurements were conducted with the devices placed in a nitrogen environment inside a Linkam Scientific LTS420 cryostat. The final measured EQE value was corrected using the responsivity of a calibrated Thorlabs FDS1010 silicon photodiode.

\subsection{Light-intensity- and temperature-dependent JV}

Light-intensity- and temperature-dependent JV measurements were performed using a home-built setup. The sample was mounted in a Linkam Scientific LTS420 cryostat, where the temperature was controlled by a Linkam Scientific T96-S controller via a combination of liquid nitrogen flow and a heating plate. Excitation was provided by an Omicron~LDM~A350 continuous wave 515~nm laser. The illumination intensity incident on the sample was adjusted using Thorlabs FW102C and FW112C neutral-density filters and monitored with a Newport 818-BB-40 biased silicon photodiode. A Keithley~2634b source measure unit was used to supply voltage bias to the solar cell as well as measure the current from both, the solar cell and photodiode.

\subsection{Long-time transient photoluminescence}

\begin{figure}[h]
    \centering
    \includegraphics[width=0.9\linewidth]{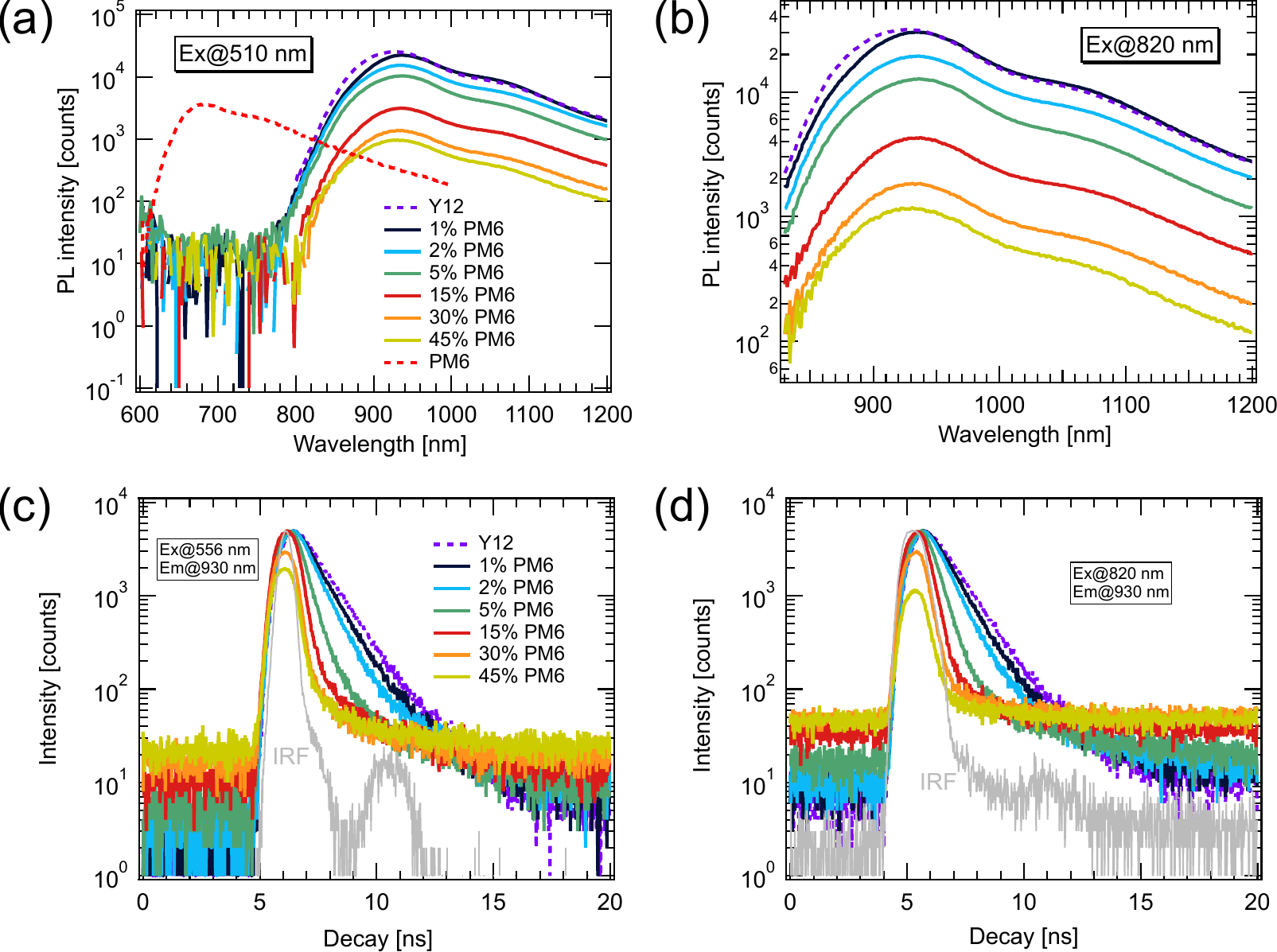}
    \caption{Steady-state PL and time-resolved TRPL results for different PM6 content. Steady-state PL spectrum excited at (a) 510~nm and (b) 820~nm. PL decay spectrum excited at (c) 556~nm and (d) 820~nm, 930~nm emission was detected in both cases.}
    \label{SI_fig:PL_TRPL}
\end{figure}

For transient PL measurements the layers deposited on fused silica substrates were illuminated using a \textit{PhoxX+515-150} Laser (manufactured by \textit{Omicron}) widened to a circular spot of $1\,\text{cm}^2$ using a TopHat converter. For PL data acquisition a single photon counting photomultiplier (manufactured by \textit{Hamamatsu}) in combination with a \textit{PicoQuant Time Harp 260} time-correlated single-photon counting card as employed. The detector is sensitive in the wavelength range 950-1200\,nm which is well suitable for the studied material systems (see Figure \ref{SI_fig:PL_TRPL}). The illumination intensity for "1 sun" equivalent was determined by adjusting the laser intensity to yield the same $\jsc$ as measured using the solar simulator (AM1.5G, corrected for spectral mismatch) for the solar cells. Time-resolved luminescence data was acquired with a time resolution of 20\,ns. The laser on-time was set to 50\,\textmu{}s and the cycle time was set to 150\,\textmu{}s. The integration time was set to 300\,s resulting in $2\times 10^{6}$ periods. The acquired data was log-smoothed to 10 points per temporal decade for $t>10^{-7}$s.

\subsection{Time-delayed collection field}

The sample was excited with a Light Conversion PHAROS femtosecond laser combined with a Light Conversion ORPHEUS optical parametric amplifier, providing tunable wavelengths from 315 to 2600 nm. The laser repetition rate was reduced from 50 kHz to 5 kHz using the built-in pulse picker. Unwanted residual laser light was removed using short- and long-pass filters so that only the selected wavelength reached the sample.

Before the optical parametric amplifier, the pulse energy was reduced to about 200 µJ using a 50/50 beam splitter. Laser fluence was roughly adjusted using a continuous neutral density filter wheel. During the experiment, fluence was further altered with two motorized filter wheels (Thorlabs FW102C and FW112C). Laser fluence and stability were monitored during the experiment with a Newport 818-BB-40 photodetector. Throughout the measurements, the sample was illuminated through an optical diffuser and kept in a custom-built holder incorporating an integrated amplifier circuit.

The voltage applied to the sample was supplied by a Keysight 81160A pattern generator, with the output passed through a voltage amplifier. The device response was measured by recording the current as a voltage drop across a 2~$\Upomega$ resistor.
This voltage signal was amplified using a differential amplifier and recorded with a GaGe~CS121G2 digitizer.

\subsection{X-ray and ultraviolet photoemission spectroscopy}

Samples were transferred to an ultrahigh vacuum chamber (ESCALAB 250Xi, Thermo Scientific) with a base pressure of $2\E{-10}$~mbar for X-ray photoemission spectroscopy (XPS) and ultraviolet photoemission spectroscopy (UPS) measurements. UPS was conducted utilizing a double-differentially pumped helium discharge lamp ($h\nu = 21.22$~eV), set at a pass energy of $2$~eV and a bias of $-5$~V. UPS and XPS depth profiling were performed using the MAGCIS dual mode ion source (ESCALAB 250Xi, Thermo Scientific), operated in argon gas cluster ion beam mode, a technique demonstrated to reduce material damage caused by etching. The cluster mode was applied at an energy of $4$~keV, producing an etch crater with dimensions of $2 \times 2$~mm$^2$. The UPS spectra were sequentially collected after each etch step.

\subsection{Grazing-incidence wide-angle x-ray scattering}

The grazing-incidence wide-angle x-ray scattering (GIWAXS) measurements were carried out using a laboratory system at the University of Bayreuth (Xeuss 3.0, Xenocs SAS, Grenoble, France) with a high flux Cu K $\alpha$ source ($\lambda$ = 1.54 \AA) and a Dectris EIGER 2R 1 m detector. A sample-to-detector distance of 73~mm was used. All measurements were conducted in vacuum using varying incident angles of 0.12°, 0.18° and 0.20°. The presented line cuts in q$_r$ were obtained by performing a horizontal line cut between q$_z$ = 0.02 and 0.1~\AA$^{-1}$. Data processing and analysis were performed using custom-written Python scripts.
The RSoXS measurements were performed at the carbon edge of the samples, which were measured at beamline 11.0.1.2 at the Advanced Light Source (ALS), Lawrence Berkeley National Lab (LBNL). The recorded data were normalized with respect to the incident photon flux, local sample thickness, the energy-dependent transmittance, and the exposure time. The final 1D scattering profiles were obtained by azimuthally averaging the 2D patterns and stitching the low-q and high-q detector images in their overlap region.

\subsection{Variable-angle spectroscopic ellipsometry}

The real and imaginary parts of the dielectric function ($\varepsilon_1$, $\varepsilon_2$) and the absorption coefficient ($\alpha$) of neat PM6, Y12, and PM6:Y12 blend films with 1–45 \% PM6 contents were determined by variable-angle spectroscopic ellipsometry (VASE). Films were spin-coated on glass substrates, with backside reflections suppressed by roughening the rear surface. Measurements were performed using a J.\ A.\ Woollam M2000 T-Solar ellipsometer at angles of incidence between 45$^{\circ}$ and 75$^{\circ}$ over a spectral range of 0.7-5~eV. 

Optical modelling and fitting were carried out using CompleteEASE. Film thickness and surface roughness were obtained from sub-bandgap Cauchy fits, with roughness modelled via a Bruggeman EMA layer. The dielectric functions were parameterised using Kramers-Kronig consistent oscillator models, employing a Cody-Lorentz oscillator for the first absorption peak and Gaussian oscillators for higher-energy features. Blend films were modelled as effective homogeneous layers. Conventional EMA approaches were tested but failed to adequately reproduce the measured optical response.

\clearpage
\section{Basic device characterization: EQE and JV}\label{SI_sec:JVEQE}

\begin{figure}[h]
    \centering
    \includegraphics[width=0.78\linewidth]{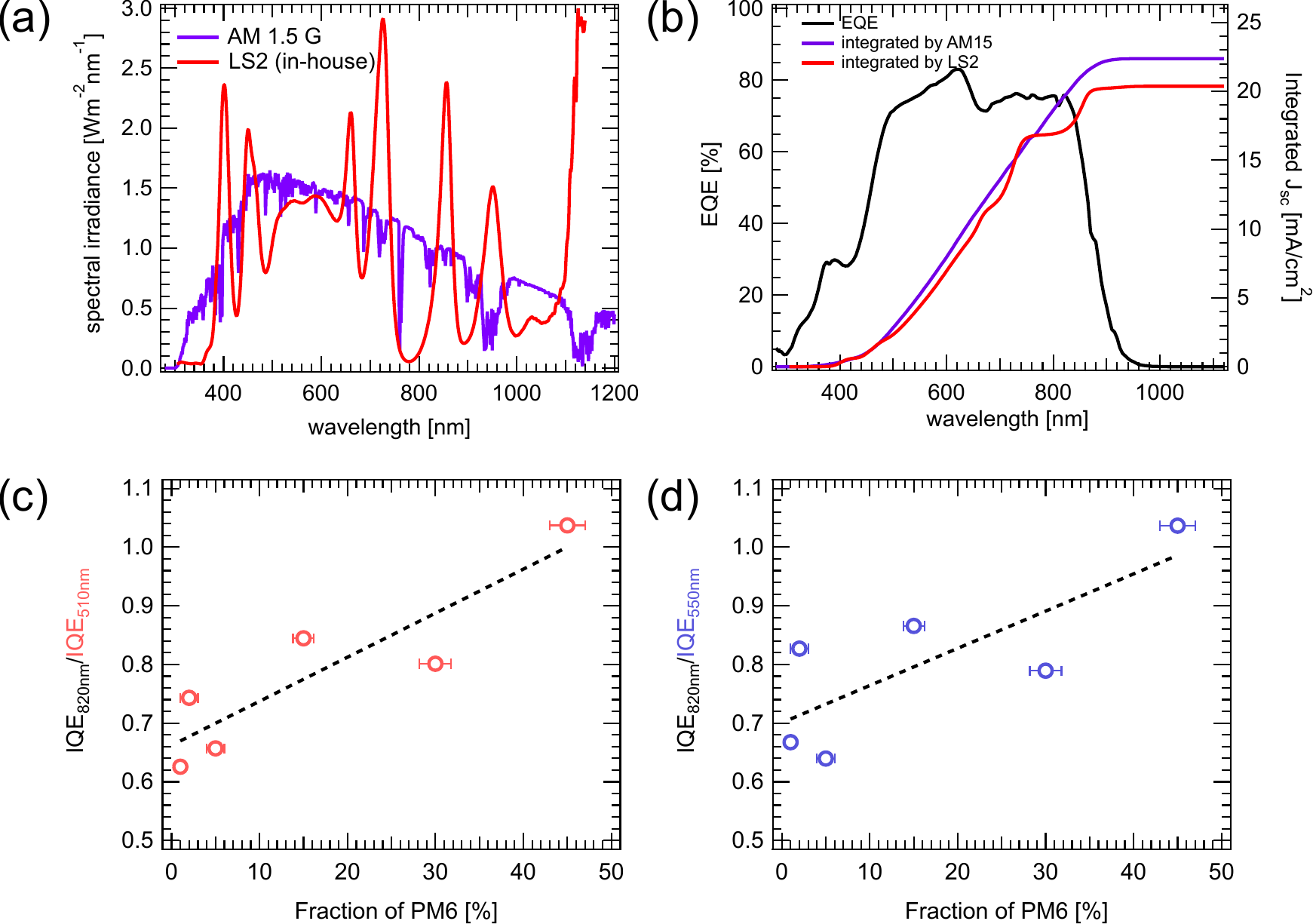}
    \caption{(a) Spectral irradiance of the Wavelabs LS-2 solar simulator compared to the AM1.5G reference spectrum. (b) Integrated $\jsc$ calculated from the LS-2 spectrum relative to AM1.5G, illustrating the average $\sim$10\% lower $\jsc$ due to spectral mismatch. The calculated IQE response ratio between 820~nm and (c) 510~nm or (d) 550~nm.}
    \label{SI_fig:EQEJV}
\end{figure}

The spectra of in-house LS-2 solar simulator and AM1.5G are shown in Figure~\ref{SI_fig:EQEJV}(a). The spectral mismatch causes a $\sim$10\% lower $\jsc$ on average, compared to the AM1.5G spectrum (see Figure~\ref{SI_fig:EQEJV}(b)). This leads to a discrepancy in the solar cell efficiencies reported in this study compared to the literature values.\cite{ma2020achieving}

\begin{figure}[b]
    \centering
    \includegraphics[width=0.45\linewidth]{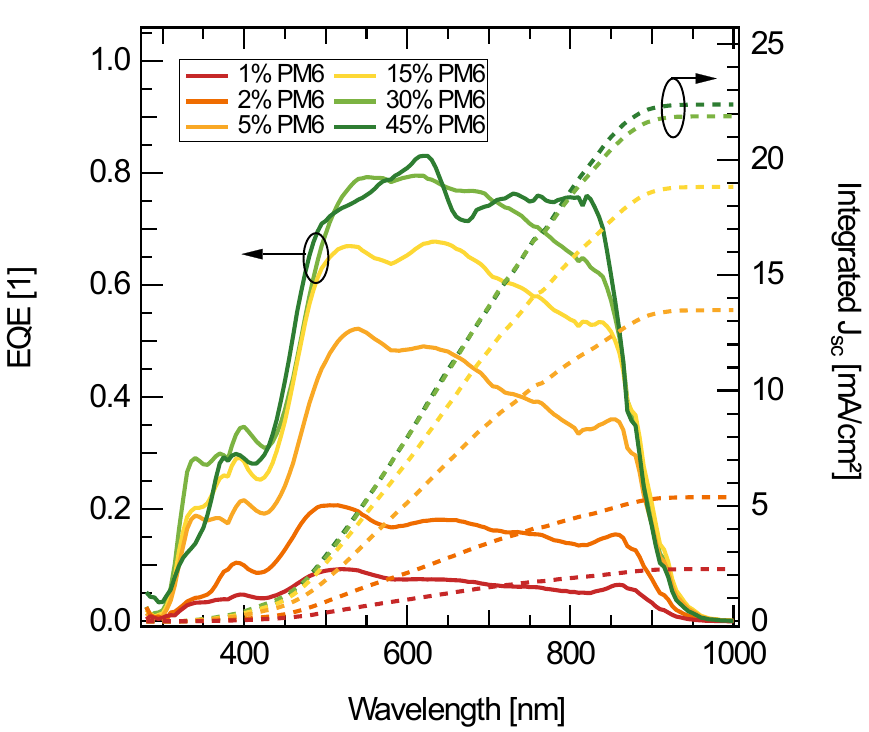}
    \caption{EQE of devices based on different PM6 content and the corresponding integrated $\jsc$ using the AM1.5G spectrum.}
    \label{SI_fig:EQE_intJsc}
\end{figure}

\begin{table}
\caption{Solar cell parameters based on different PM6 content}
\centering
\small
\begin{tabular}{cccccc}
\hline
PM6 (\%) & $\Voc$ (V) & $\jsc$ (mA\,cm$^{-2}$) & $\jsc^\mathrm{*}$ (mA\,cm$^{-2}$) & $FF$ & PCE$^\mathrm{*}$ (\%) \\
\hline
1  & 0.742 & 0.95 & 2.52 & 0.29 & 0.5 \\
2  & 0.824 & 3.29 & 5.78 & 0.42 & 2.0 \\
5  & 0.816 & 12.64 & 14.28 & 0.51 & 5.9 \\
15 & 0.824 & 18.33 & 19.85 & 0.61 & 10.0 \\
30 & 0.831 & 20.62 & 23.10 & 0.63 & 12.1 \\
45 & 0.838 & 21.18 & 24.60 & 0.72 & 14.9 \\
\hline
\multicolumn{6}{l}{\footnotesize $^\ast$ AM1.5G-calibrated $\jsc$ and PCE values.}
\end{tabular}\label{SI_table:solar_cell_parameters}
\end{table}

\section{Light-intensity-dependent JV (Suns-$\Voc$) analysis}\label{SI_sec:sunsvoc}

\begin{figure}[h]
    \centering
    \includegraphics[width=0.9\linewidth]{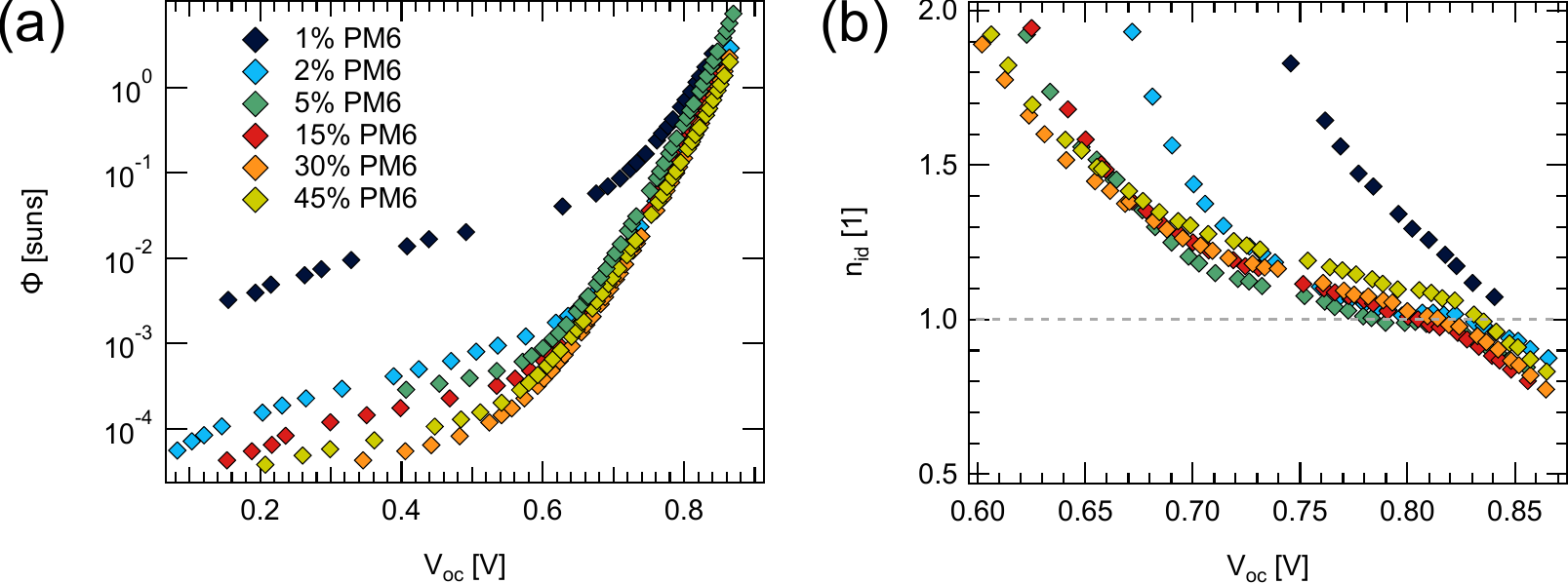}
    \caption{(a) Suns-$\Voc$ plot and (b) $\nid$ versus $\Voc$ based on different PM6 content. The ideality factor is calculated according to Section~\ref{SI_sec:nid}}
    \label{SI_fig:sunsvoc_nid}
\end{figure}

\section{Determination of the energy gap and the open-circuit voltage losses}\label{SI_sec:Eg_Voc_loss}

\begin{figure}[h]
    \centering
    \includegraphics[width=0.9\linewidth]{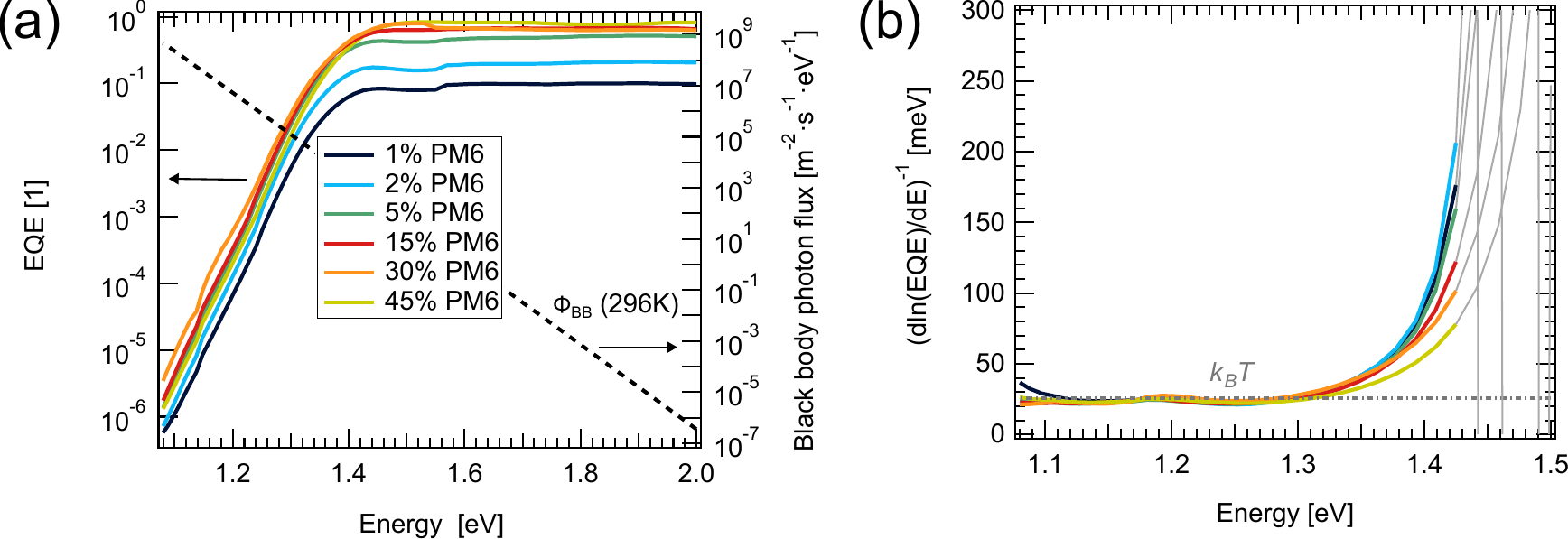}
    \caption{(a) sensitive EQE spectra and (b) the calculated $(\der \ln(EQE)/\der E)^{\mathrm{-1}}$ over photon energy based on different PM6 content.}
    \label{SI_fig:sensEQE}
\end{figure}

Figure~\ref{SI_fig:sensEQE} (a) shows the measured sensitive EQE response of devices based on different PM6 content. The Boltzmann statistics indicating the occupation probability is plotted together. The calculated $\Eu$ from $(\der \ln(EQE)/\der E)^{\mathrm{-1}}$ is shown in Figure~\ref{SI_fig:sensEQE} (b), and roughly equals the thermal energy $\kT$, indicating that it is dominated by thermal broadening instead of static energetic disorder.\cite{kaiser2021universal}

\begin{figure}[tb]
    \centering
    \includegraphics[width=0.4\linewidth]{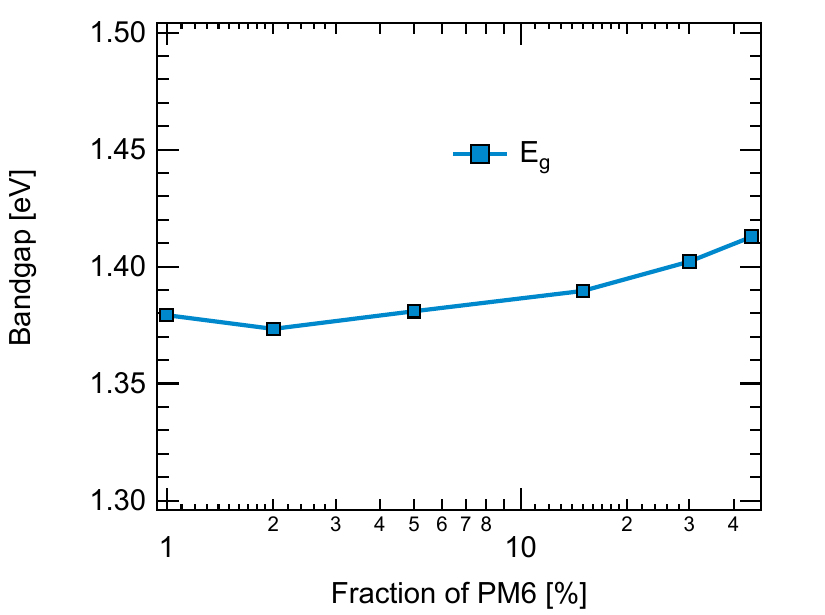}
    \caption{$\Eg$ plot as a function of PM6 fraction}
    \label{SI_fig:energy_gap}
\end{figure}

We determine the energy gap $\Eg$ based on the method by Rau et al.,\cite{rau2017efficiency} according to which the gap energy is the mean peak energy at the absorption edge of the distribution $P(E)$:
\begin{equation}
    P(E)=\frac{\der EQE}{\der E},\qquad
    E_\mathrm{g} = \frac{\displaystyle \int_a^b E P(E)\, \mathrm{d}E}{\displaystyle \int_a^b P(E)\, \mathrm{d}E} . 
\end{equation}
Here, the integration limits $a$ and $b$ are chosen as the energy where $P(E)$ is equal to $50\%$ of its maximum, i.e., $P(a)=P(b)=\frac{1}{2}\max[P(E)]$. The determined $\Eg$ is shown in Figure~\ref{SI_fig:energy_gap}

We further analyze the $\Voc$ losses of solar cells based on the reciprocity relation between absorption and emission:\cite{rau2007reciprocity,mueller2023charge}
\begin{equation}\label{eq:Voc losses SQ}
V_{\mathrm{oc,SQ}} = \frac{k_{\mathrm{B}} T}{e} \ln\!\left[
\frac{e \displaystyle\int_{E_g}^{\infty}\phi_{\mathrm{AM1.5G}}(E)\,\mathrm{d}E}{e \displaystyle\int_{E_g}^{\infty}\phi_{\mathrm{BB}}(E)\,\mathrm{d}E}+ 1\right].
\end{equation}
\begin{equation}\label{eq:Voc losses rad}
V_{\mathrm{oc,rad}} = \frac{k_{\mathrm{B}} T}{e} \ln\!\left[
\frac{e \displaystyle\int_0^{\infty}EQE_{\mathrm{PV}}(E)\phi_{\mathrm{AM1.5G}}(E)\,\mathrm{d}E}{e \displaystyle\int_0^{\infty}EQE_{\mathrm{PV}}(E)\phi_{\mathrm{BB}}(E)\,\mathrm{d}E}+ 1\right].
\end{equation}
Here, $V_{\mathrm{oc,SQ}}$ is the maximal achievable voltage, assuming no absorption below the optical bandgap of the absorber. $V_{\mathrm{oc,rad}}$ is the radiative limit, where there is only a radiative recombination channel for charge carriers. $\phi_{\mathrm{BB}}$ and $\phi_{\mathrm{AM1.5G}}$ are the blackbody and AM1.5G photon flux spectra, respectively. $EQE_{\mathrm{PV}}$ is the measured sensitive EQE response, $e$ is the elementary charge, and $\kT$ is the thermal energy. The calculated $\Delta V_{\mathrm{oc,SQ}}=\Eg/e-V_{\mathrm{oc,SQ}}$, $\Delta V_{\mathrm{oc,abs}}=V_{\mathrm{oc,SQ}}-V_{\mathrm{oc,rad}}$, and $\Delta V_{\mathrm{oc,nonrad}}=V_{\mathrm{oc,rad}}-V_{\mathrm{oc,measured}}$, are listed in Table~\ref{SI_table:voc_losses}.

\begin{table}[h]
    \centering
    \caption{Calculated $\Voc$ losses for PM6:Y12 with different donor fractions.}
    \begin{tabular}{c c c c c c c c}
        \hline
        \diagbox{ratio}{loss} & $\Eg [eV]$ & $V_{\mathrm{oc,SQ}}[V]$ & $V_{\mathrm{oc,rad}}[V]$ & $V_{\mathrm{oc,measured}}[V]$ & $\Delta V_{\mathrm{oc,SQ}}[V]$ & $\Delta V_{\mathrm{oc,abs}}[V]$ & $\Delta V_{\mathrm{oc,nonrad}}[V]$ \\
        \hline
        2\%PM6 & 1.37 & 1.11 & 1.06 & 0.83 & 0.26 & 0.05 & 0.23 \\
        5\%PM6 & 1.38 & 1.12 & 1.06 & 0.83 & 0.26 & 0.05 & 0.24 \\
        15\%PM6 & 1.39 & 1.12 & 1.07 & 0.83 & 0.27 & 0.06 & 0.24 \\
        30\%PM6 & 1.40 & 1.14 & 1.06 & 0.84 & 0.27 & 0.08 & 0.22 \\
        45\%PM6 & 1.41 & 1.15 & 1.08 & 0.84 & 0.27 & 0.06 & 0.25 \\
        \hline\label{SI_table:voc_losses}
    \end{tabular}
\end{table}

\clearpage
\section{GIWAXS and RSoXS}\label{SI_sec:GIWAXS}

\begin{figure}[h]
    \centering
    \includegraphics[width=0.9\linewidth]{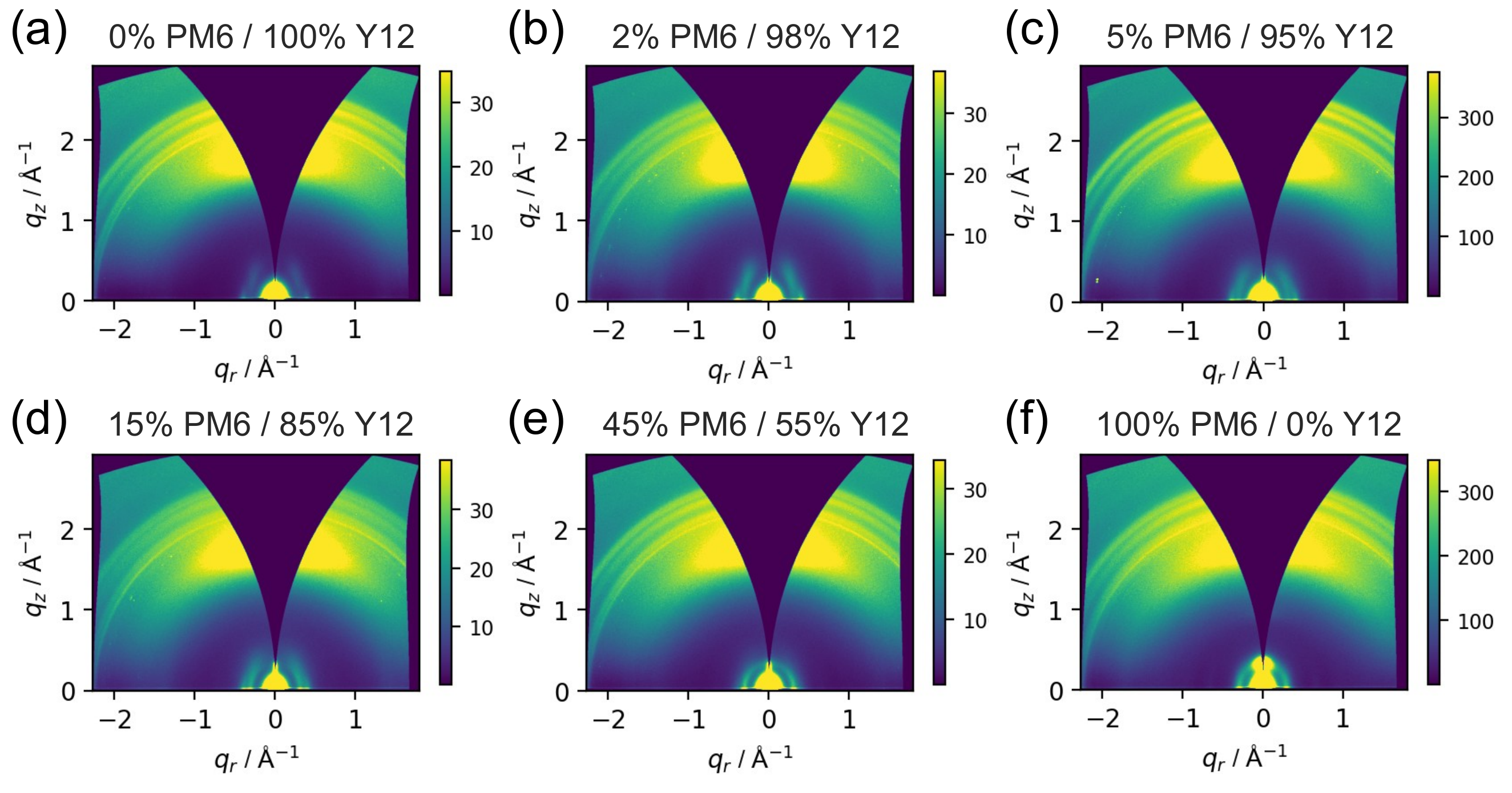}
    \caption{2D GIWAXS images of samples with varying PM6/Y12 content on ZnO/ITO substrates measured with $\alpha = 0.18^{\circ}$. Material content is indicated on top of the images. The signal below 0.2 \AA$^{-1}$ is direct beam dominated. Colour shows intensity on a linear scale.}
    \label{SI_fig:GIWAXS:2D}
\end{figure}

In this study, grazing-incidence wide-angle X-ray scattering (GIWAXS) was employed to investigate the molecular ordering on the nano- and sub-nanoscale in PM6/Y12 blend films with varying donor--acceptor ratios. The samples were prepared in a device-relevant geometry identical to that used for the solar cells. Accordingly, the PM6/Y12 active layers were measured on ZnO/ITO substrates. The GIWAXS patterns of the blends are additionally compared to those of neat Y12 and neat PM6 films (see Figure~\ref{SI_fig:GIWAXS:2D}).

To analyse the PM6 and Y12 contributions in more detail, we fit each line cut with the fixed FWHM of the neat PM6 and the fixed FWHM and peak amplitude ratio of the double peak of the neat Y12 to each of the compositions. This restriction during the fitting allows us to determine how much of each ordered contribution is present within the sample. The corresponding results are presented in Figure~\ref{fig:nano}(c). As expected, the rising PM6 content also leads to an increased ordered fraction of PM6 within the film.

\begin{figure}[h]
    \centering
    \includegraphics[width=0.9\linewidth]{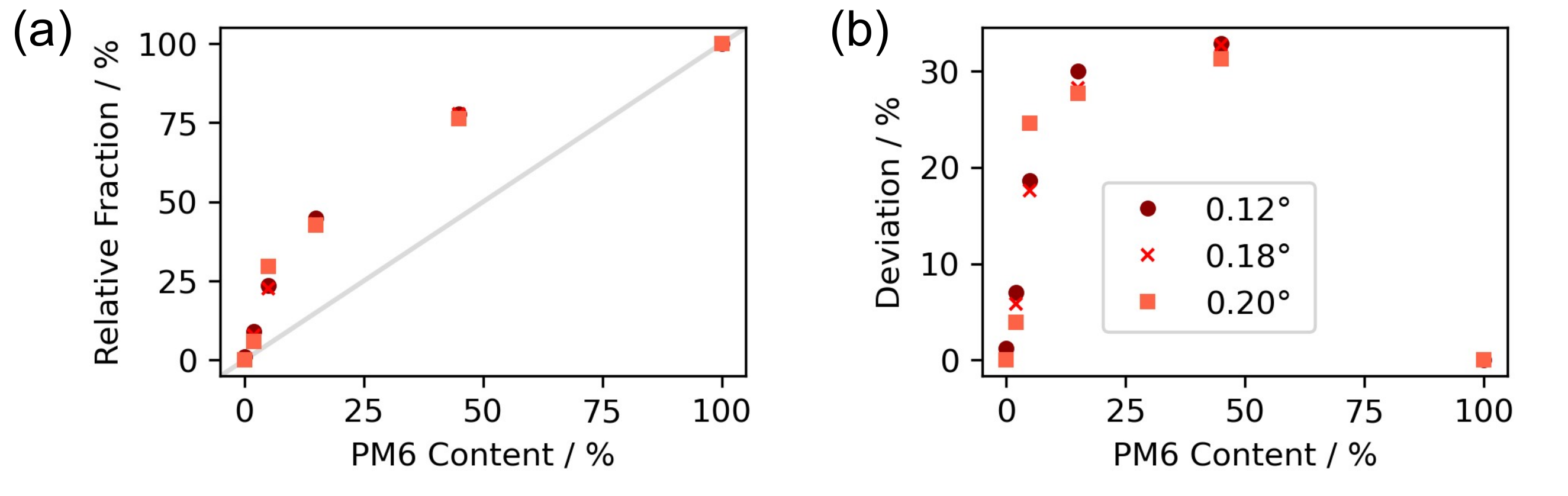}
    \caption{Relative ordered material fraction obtained from the fit results of the horizontal GIWAXS line cuts as a function of the input PM6 content. (a) Relative Fraction of ordered PM6. The light grey line indicates a 1:1 relationship between measured ordered material and input material. The evaluation was performed for different angles of incidence, which are represented by different symbols. (b) Deviation of the PM6 ordered material content from the direct 1:1 relationship with the input material (light grey line in (a)).}
    \label{SI_fig:GIWAXS:fraction}
\end{figure}

If both materials are unaffected by each other upon mixing, we would expect a 1:1 correspondence of the \textit{input material volume ratio} and the \textit{measured ordered material ratio} (light grey curve in Figure~\ref{SI_fig:GIWAXS:fraction}(a)). Hence, any observed deviation from the direct relationship will be a stronger relative reduction of ordering in one of the materials as a consequence of the blending. We observe, that the measured ordered material ratio is higher than the direct relationship, hence PM6 more dominantly orders than Y12. Already at low PM6 contents, the measured ratio deviates strongly from the direct relationship (Figure~\ref{SI_fig:GIWAXS:fraction}(b)), while the deviation does saturates above 15\% PM6 content, indicating disruption of the Y12 ordering already with low PM6 contents. To probe the depth dependence of this structural behavior, GIWAXS measurements were performed with varying incident angles. Firstly, below the critical angle ($\alpha_c$=0.16°) at 0.12° to probe the surface and above $\alpha_c$ (at 0.18° and  0.20°) to probe the bulk of the sample. Both the increase in the PM6 content and the deviation from the linear 1:1 relationship are observed consistently for all measured incident angles, indicating that no significant differences exist between the surface-sensitive and bulk-sensitive measurements (see Figure~\ref{SI_fig:GIWAXS:fraction}).

\begin{figure}[t]
    \centering
    \includegraphics[width=0.75\linewidth]{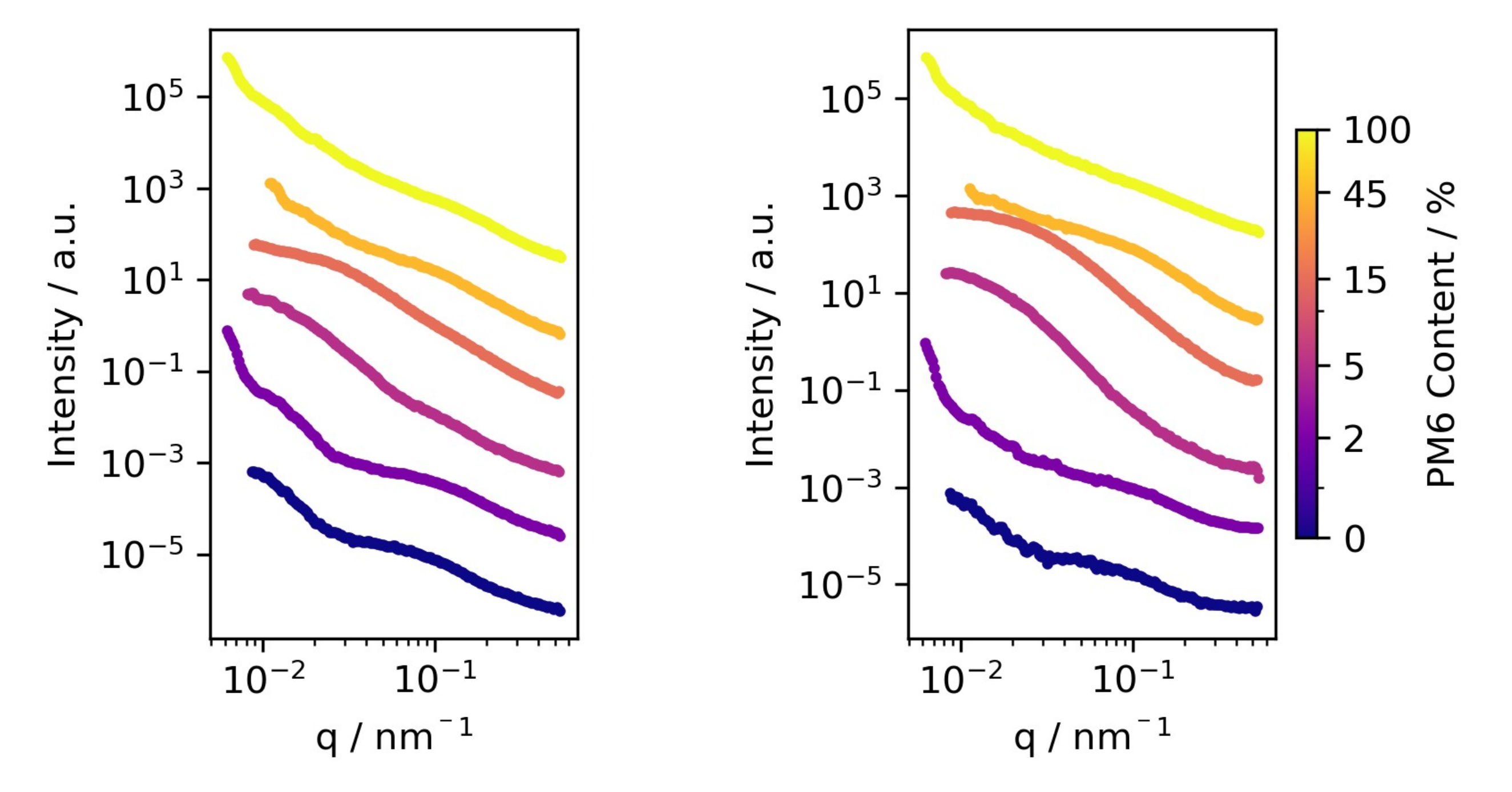}
    \caption{Radially averaged RSoXS line cuts of PM6/Y12 blend films with varying PM6 content, measured at 283.2~eV and 284.8~eV.}
    \label{SI_fig:GIWAXS:rsoxs}
\end{figure}

Resonant soft X-ray scattering (RSoXS) was used to probe structural features on longer length scales (5-100\,nm). Although the continued variation of material composition does not allow to use intensity changes to extract phase compositions, qualitative differences in scattering intensity and characteristic length scales are observable in Figure~\ref{SI_fig:GIWAXS:rsoxs}.

\clearpage
\section{VASE: results and observations}\label{SI_sec:ellipsometry}

\begin{figure}[h]
    \centering 
    \includegraphics[width=0.9\linewidth]{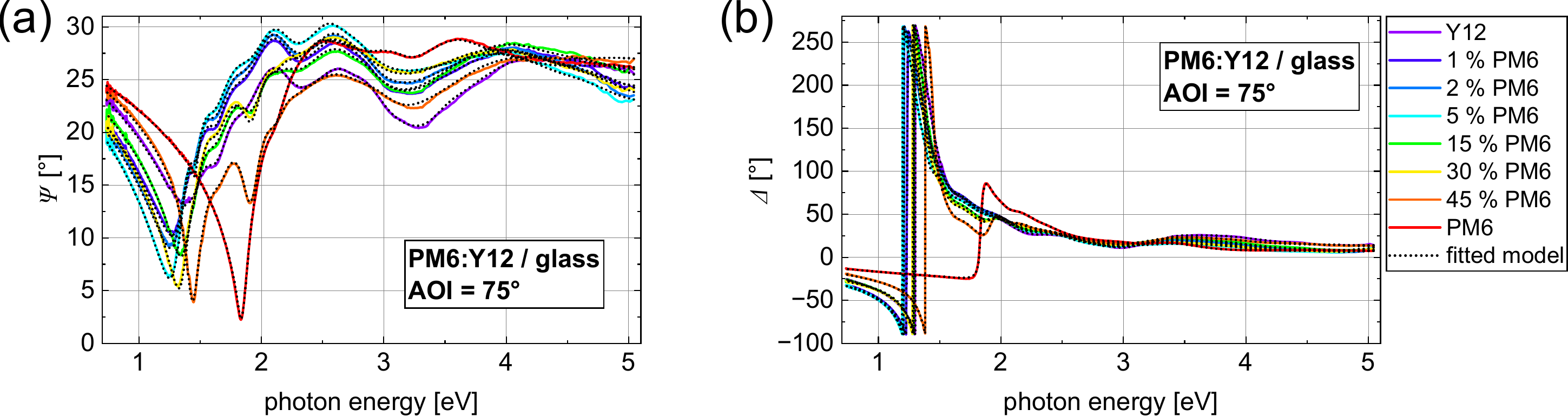}     
    \caption{Representative plots of the ellipsometric angles a) $\Psi$ and b) $\Delta$ at a 75$^\circ$ angle of incidence for neat Y12 and PM6, and PM6:Y12 blend films with varying concentrations of PM6 on glass. Solid lines correspond to measured data, dotted lines are the fitted curves generated from the optical model.}
    \label{SI_fig:ellips:PsiDelta}
\end{figure}

Figure~\ref{SI_fig:ellips:PsiDelta} shows representative measured and fitted ellipsometric angles $\Psi$ and $\Delta$ for neat PM6, Y12, and PM6:Y12 blend films, illustrating the overall quality of the optical model fits. Extracted film thickness $d$, surface roughness $r$, mean square error (MSE), and optical bandgaps are summarized in Table~\ref{SI_tab:ellips:VASE}. Error bars represent the standard deviation obtained from measurements at different sample positions.

Film thickness varied by up to $\sim$30~nm between samples and accounts for most visible differences in the measured spectra. A PM6-related feature between 1.8-2.0~eV appears in $\Psi$ and $\Delta$ only for PM6 concentrations $\ge$15\,\%, consistent with the spectral position observed for neat PM6. At low photon energies, spectra remain smooth up to $\sim$1.3~eV for Y12 and blend films and up to $\sim$1.8~eV for PM6, defining the transparent range used for Cauchy modelling. No clear trend in surface roughness with PM6 concentration is observed.

\begin{table}[h] 
    \centering
    \begin{tabular}{l|c|c|c|c|c}
       Sample  & $d$ [nm] & $r$ [nm] & MSE & $E_{g,CL}$ [eV] & $E_{g,Tauc}$ [eV]\\ \hline
       
       PM6  & $71.6 \pm 0.7$ & $5.6 \pm 0.1$ & $4.3$ & $1.80 \pm 0.01$ & $1.82 \pm 0.02$ \\
       Y12  & $48.44 \pm 2.9$ & $0.1 \pm 0.1$ & $7.4$ & $1.27 \pm 0.01$ & $1.30 \pm 0.03$ \\
       1\,\%  & $64.9 \pm 6.0$ & $0.3 \pm 0.3$ & $13.4$ & $1.31 \pm 0.01$ & $1.32 \pm 0.03$  \\
       2\,\%  & $65.5 \pm 5.2$ & $3.7 \pm 0.2$ & $13.6$ & $1.28 \pm 0.01$ & $1.32 \pm 0.03$ \\
       5\,\%  & $77.1 \pm 4.7$ & $2.6 \pm 0.6$ & $13.5$ & $1.29 \pm 0.01$ & $1.325 \pm 0.03$\\
       15\,\%  & $66.8 \pm 1.2$ & $0.25 \pm 0.2$ & $6.5$ & $1.26 \pm 0.01$ & $1.31 \pm 0.03$ \\
       30\,\%  & $74.6 \pm 4.2$ & $1.3 \pm 0.5$ & $9.4$ & $1.29 \pm 0.01$ & $1.32 \pm 0.03$  \\
       45\,\%  & $55.6 \pm 1.3$ & $0.2 \pm 0.2$ & $5.9$  & $1.28 \pm 0.01$ & $1.32 \pm 0.03$  \\
    \end{tabular}
    \caption{Summary of thickness $d$, surface roughness $r$, MSE, and gap energy $E_{g,CL}$ extracted directly from the optical models, as well as the band gap $E_{g,Tauc}$ estimated from a Tauc plot of the absorption coefficient extracted from the optical models.}
    \label{SI_tab:ellips:VASE}
\end{table}

\begin{figure}[tb]
    \centering
    \includegraphics[width=0.9\linewidth]{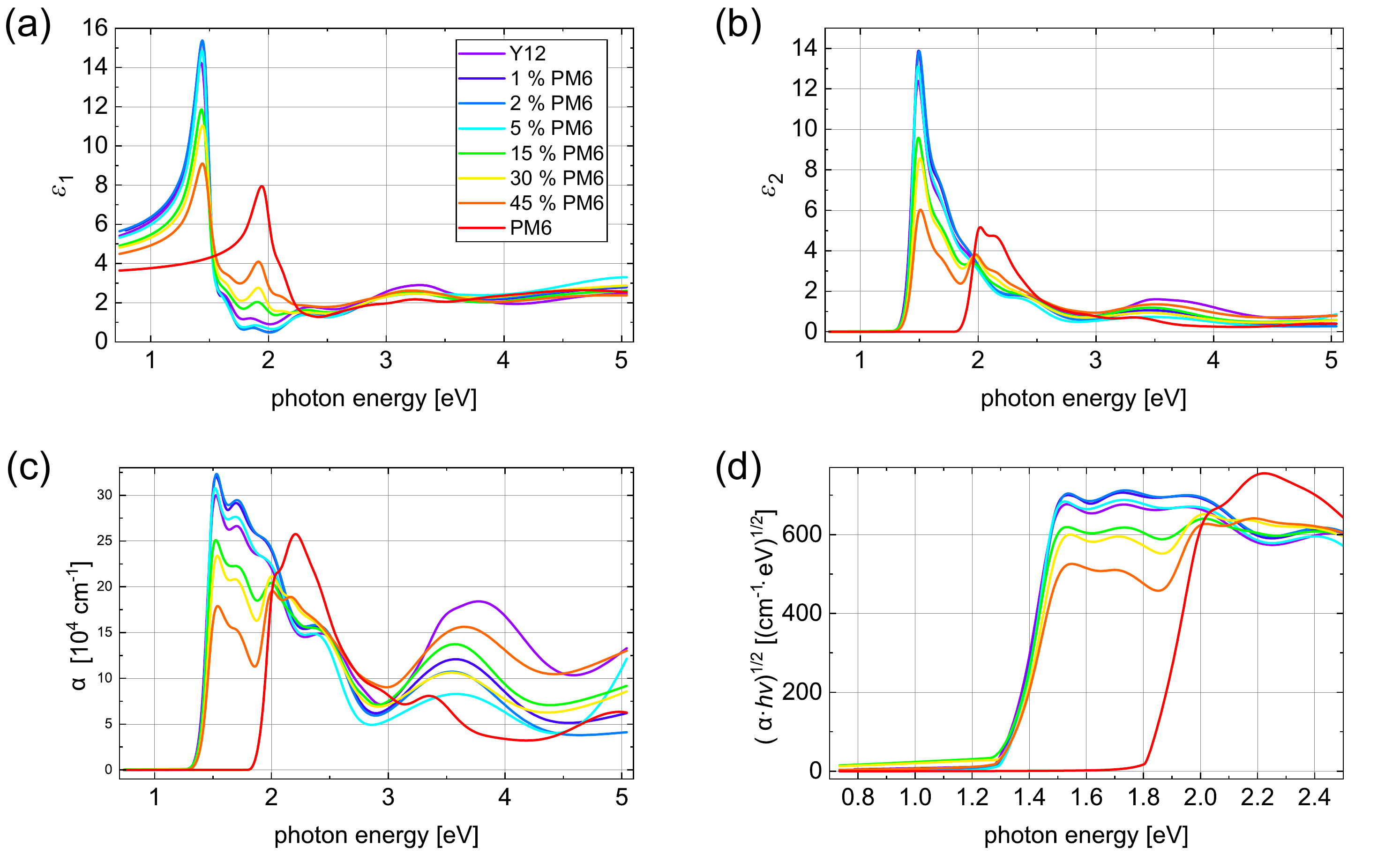}
    \caption{(a) Real part $\varepsilon_1$ and (b) imaginary part $\varepsilon_2$ of the dielectric function of neat Y12 and PM6, and PM6:Y12 blend films with varying concentration of PM6, extracted from the optical models. (c) Absorption coefficient $\alpha$ of neat Y12 and PM6, and PM6:Y12 blend films with varying concentration of PM6, as calculated from the optical models. (d) Tauc plot of $\alpha$ from (c) near the absorption onset.}
    \label{SI_fig:ellips:e1e2alpha}
\end{figure}

The real and imaginary parts of the dielectric function, $\varepsilon_1$ and $\varepsilon_2$, extracted from the optical models are shown in Figure~\ref{SI_fig:ellips:e1e2alpha}(a) and (b). Pronounced line-shape changes are observed only for PM6 concentrations $\ge$15\,\%. With increasing PM6 content, the main $\varepsilon_2$ peak of Y12 exhibits a small blueshift of up to $\sim$20~meV, while the PM6-related peak is redshifted by up to $\sim$60~meV relative to neat PM6.

Figure~\ref{SI_fig:ellips:e1e2alpha}(c) shows the absorption coefficient $\alpha$ calculated from the extracted dielectric functions. The trends observed in $\varepsilon_2$ are preserved in $\alpha$. For PM6 concentrations $\ge$15\,\%, the Y12-related absorption below 2~eV decreases and the PM6 absorption peak emerges at slightly shifted energies. At low PM6 concentrations (1-2\,\%), a weak increase in absorption below 2~eV is observed.


Optical bandgaps were determined from the Cody--Lorentz gap parameter $E_{g,CL}$ and from Tauc plots of the absorption onset, as shown in Figure~\ref{SI_fig:ellips:e1e2alpha}(d). The resulting values are listed in Table~\ref{SI_tab:ellips:VASE}. As expected, $E_{g,Tauc}$ is systematically slightly larger than $E_{g,CL}$ by up to $\sim$40~meV. No significant dependence of the bandgap on PM6 concentration is observed for the blend films, with values dominated by the Y12 component.

\section{XPS and UPS depth profiling}\label{SI_sec:XPSUPS}

\begin{figure}[b]
    \centering
    \includegraphics[width=0.85\linewidth]{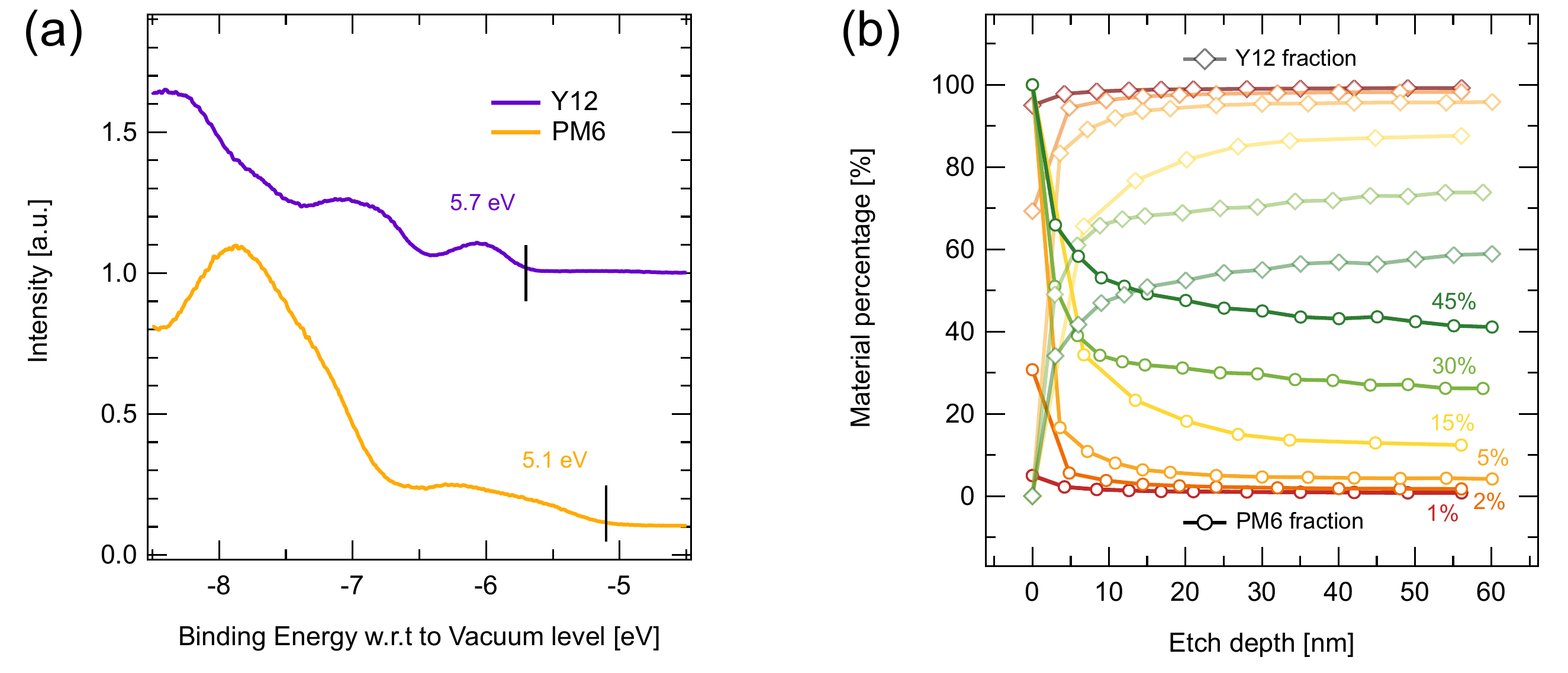}
    \caption{(a) UPS spectra of the HOMO onset and its corresponding ionization potential, (b) Vertical composition profiles of various donor: acceptor ratios derived from UPS depth profiles.}
    \label{SI_fig:UPSXPS}
\end{figure}

Since this study mainly focuses on different donor to acceptor ratios, confirming that the expected ratios are preserved after fabrication is essential. Hence, we performed X-ray photoemission spectroscopy (XPS) and ultraviolet photoemission spectroscopy (UPS) depth profiling to investigate the vertical compositional gradient throughout the blend with different donor--acceptor ratios. Previous studies have shown that UPS depth profile is a valuable technique for understanding the energetic landscape of the device and its compositional profiles.\cite{Lami2019} Argon gas cluster etching is utilized along with UPS to collect UPS spectra as a function of depth. The technique has been proven to reduce etching damage in organic materials and maintain their electronic and compositional structure. The method was explained in detail elsewhere.\cite{Lami2019} The results showed that the UPS depth profile outperforms the XPS depth profile as it can distinguish between similar chemical structures. 

In this study, PM6 and Y12 have identical elements, making it difficult to distinguish their composition via XPS depth profiling. However, as the donor and acceptor have different electronic structures (see Figure~\ref{SI_fig:UPSXPS}(a)), their compositional profiles were accurately determined by UPS depth profiling.

\section{Exciton quenching efficiency and exciton lifetime}\label{SI_sec:quenching}

We determined the PL quenching ratio $\etaexc$ from the steady-state PL. The total PL intensity ($I_{\mathrm{PL}}$) used for PL quenching calculation was obtained by integrating the PL signal over the emission range of PM6 (600--800~nm) and Y12 (830--1200~nm). Quenching ratio is determined as 
$$1-\frac{I_{\mathrm{PL,blend}}}{I_{\mathrm{PL,neat}}} . $$ 
Weighted quenching ratio was determined as
$$\etaexc = \frac{f_{\mathrm{\scriptsize PM6}} \cdot A_{\mathrm{\scriptsize PM6}}}{f_{\mathrm{\scriptsize PM6}} \cdot A_{\mathrm{\scriptsize PM6}}+f_{\mathrm{\scriptsize Y12}} \cdot A_{\mathrm{\scriptsize Y12}}} \cdot \eta_{\mathrm{exc},\mathrm{\scriptsize PM6}} +\frac{f_{\mathrm{\scriptsize Y12}} \cdot A_{\mathrm{\scriptsize Y12}}}{f_{\mathrm{\scriptsize PM6}} \cdot A_{\mathrm{\scriptsize PM6}}+f_{\mathrm{\scriptsize Y12}} \cdot A_{\mathrm{\scriptsize Y12}}} \cdot \eta_{\mathrm{exc},\mathrm{\scriptsize Y12}} , $$
where $f$ and $A$ represent the fraction and relative absorptance of corresponding component, respectively. PL decay spectrum excited at 556~nm and 820~nm, 930~nm emission was detected in both cases. PL lifetimes were extracted by exponential reconvolution fit of the decay traces with the IRF using the built-in analysis software and listed in Table~\ref{tab:TRPL exciton lifetime}.

\begin{table}[h]
    \centering
    \begin{tabular}{c|c|c}
    \hline
    \diagbox{ratio}{lifetime} & $\tau_{\mathrm{ex,556}}$[ps] & $\tau_{\mathrm{ex,820}}$[ps]\\
    \hline
         Y12 & 1168 & 1105 \\
         1\%PM6 & 965 & 932 \\
         2\%PM6 & 788 & 742 \\
         5\%PM6 & 464 & 396 \\
         15\%PM6 & 263 & 163 \\
         30\%PM6 & 226 & 32 \\
         45\%PM6 & 256 & 89 \\
    \hline
    \end{tabular}
    \caption{exciton lifetime excited at 556~nm and 820~nm of different samples.}
    \label{tab:TRPL exciton lifetime}
\end{table}

\section{Charge-transfer exciton dissociation efficiency}\label{SI_sec:etadiss}

\begin{figure}[h]
    \centering
    \includegraphics[width=0.9\linewidth]{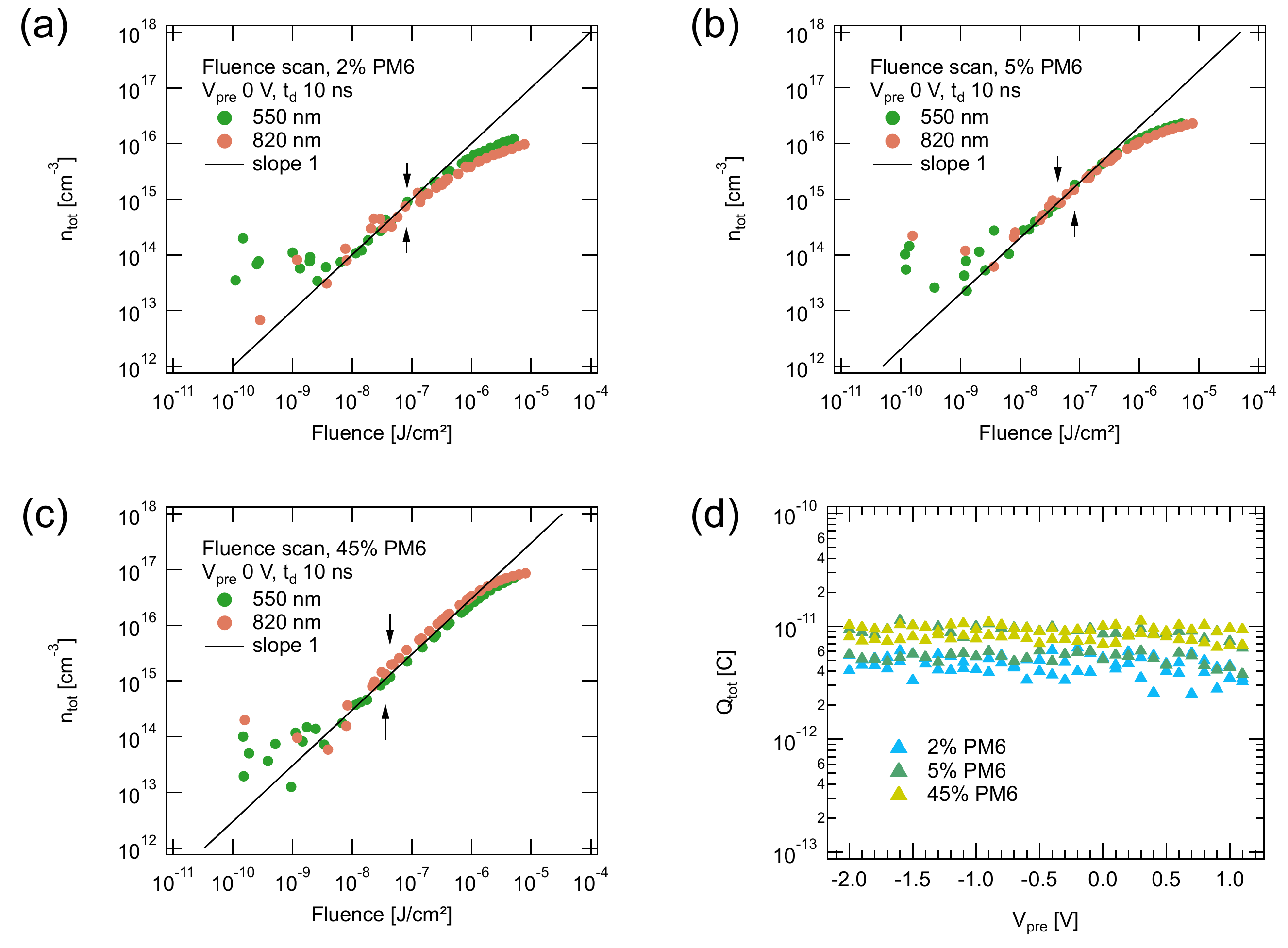}
    \caption{The total extracted charge carrier density, $n_\mathrm{tot}$, plotted as a function of laser fluence for devices with (a) 2\% PM6, (b) 5\% PM6, and (c) 45\% PM6 content. A slope of unity indicates that first-order recombination losses dominate. (d) The total extracted charge, $Q_\mathrm{tot}$, as a function of pre-bias voltage $V_\mathrm{pre}$. The laser fluence was adjusted to ensure similar charge carrier densities across different TDCF experiments.}
    \label{SI_fig:TDCF:Fluence}
\end{figure}

Time-delayed collection field (TDCF) was employed to study geminate recombination. Two parameters were varied between experiments: the pre-bias voltage $V_\mathrm{pre}$ and the excitation wavelength $\lambda$. Excitation wavelengths of 820~nm and 550~nm were used to preferentially excite the acceptor and mainly the donor, respectively. Excitons generated in the donor and acceptor phases dissociate into unbound charge carriers at the donor--acceptor interface. By changing $V_\mathrm{pre}$, the internal electric field of the device is modified, which allows the field dependence of charge-transfer (CT) exciton dissociation to be investigated. 

The fluence reaching the sample was kept sufficiently low to ensure that nongeminate recombination losses did not dominate. To determine the appropriate fluence range, the experiment was first performed at different illumination intensities. A log--log plot of the total extracted charge as a function of fluence was used to identify the recombination order (see Figure~\ref{SI_fig:TDCF:Fluence}(a)--(c)). At low illumination intensities, geminate recombination losses were dominant, as indicated by a slope of one, and this regime was therefore selected for the measurements. At higher illumination intensities, the slope decreased below one, indicating that nongeminate recombination became dominant. To maintain comparable experimental conditions for different excitation wavelengths, similar charge carrier densities within the device were required. This was achieved by adjusting the laser fluence, as shown in Figure~\ref{SI_fig:TDCF:Fluence}(d). The time delay was set to 10~ns for all measurements and the voltage applied during the charge carrier collection was set to -4~V. 

The dissociation efficiency of CT excitons was calculated as
\begin{equation}
    \etadiss\bl V_\mathrm{pre} \br = \frac{EGE\bl V_\mathrm{pre} \br}{EGE\bl -2.0\mathrm{ V} \br} ,
\end{equation}
where $EGE$, the external generation efficiency, is defined as the ratio of the extracted charge carrier density to the incident photon density,
$EGE = n_\mathrm{tot}/n_\mathrm{ph}$. The dissociation yield is assumed to be unity at $V_\mathrm{pre} = -2.0$~V.

The charge carrier density and the photon density are given, respectively, by
\begin{equation}
    n_\mathrm{tot} = \frac{Q_\mathrm{tot}}{eLA} , \qquad 
    n_\mathrm{ph} = \frac{F\cdot\lambda}{h c L} . 
\end{equation}
Here, $Q_\mathrm{tot}$ is the total extracted charge, $e$ is the elementary charge, $L$ is the photoactive layer thickness, $A$ is the device area, $F$ is the incident fluence, $\lambda$ is the excitation wavelength, $h$ is the Planck constant, and $c$ is the speed of light.

\begin{figure}[h]
    \centering
    \includegraphics[width=0.45\linewidth]{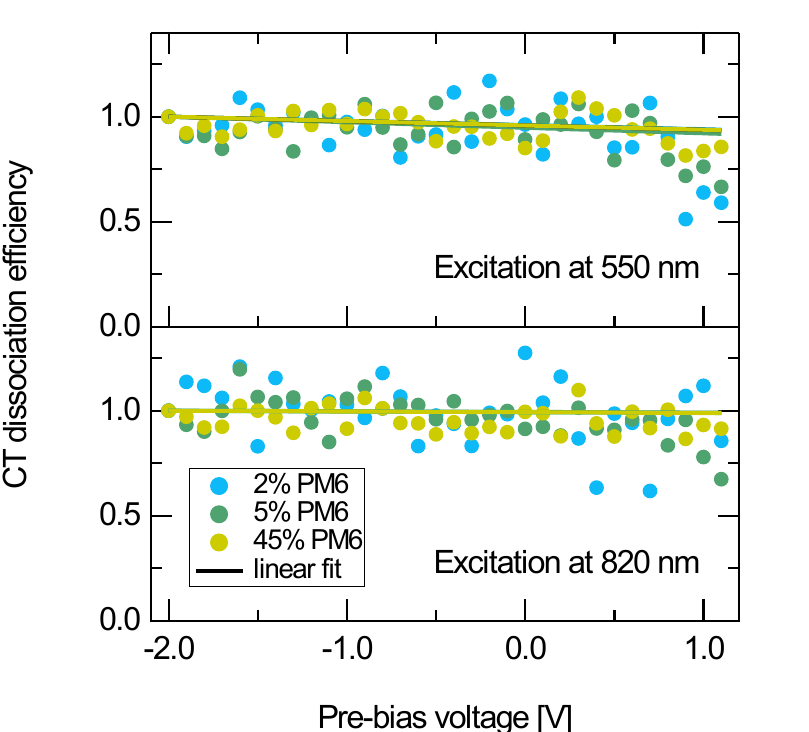}
    \caption{CT dissociation efficiency for solar cells with 2\%, 5\% and 45\% PM6 content using laser excitation at 550~nm (top) and 820~nm (bottom). To estimate the CT dissociation efficiency at specific device working conditions, the data were fitted using $\etadiss\bl V_\mathrm{pre}\br = m\cdot\bl V_\mathrm{pre} + 2\br + 1$.}
    \label{SI_fig:TDCF:etadiss}
\end{figure}

\clearpage
\section{Device simulations}\label{SI_sec:simulation}

\begin{figure}[h]
    \centering
    \includegraphics[width=0.85\linewidth]{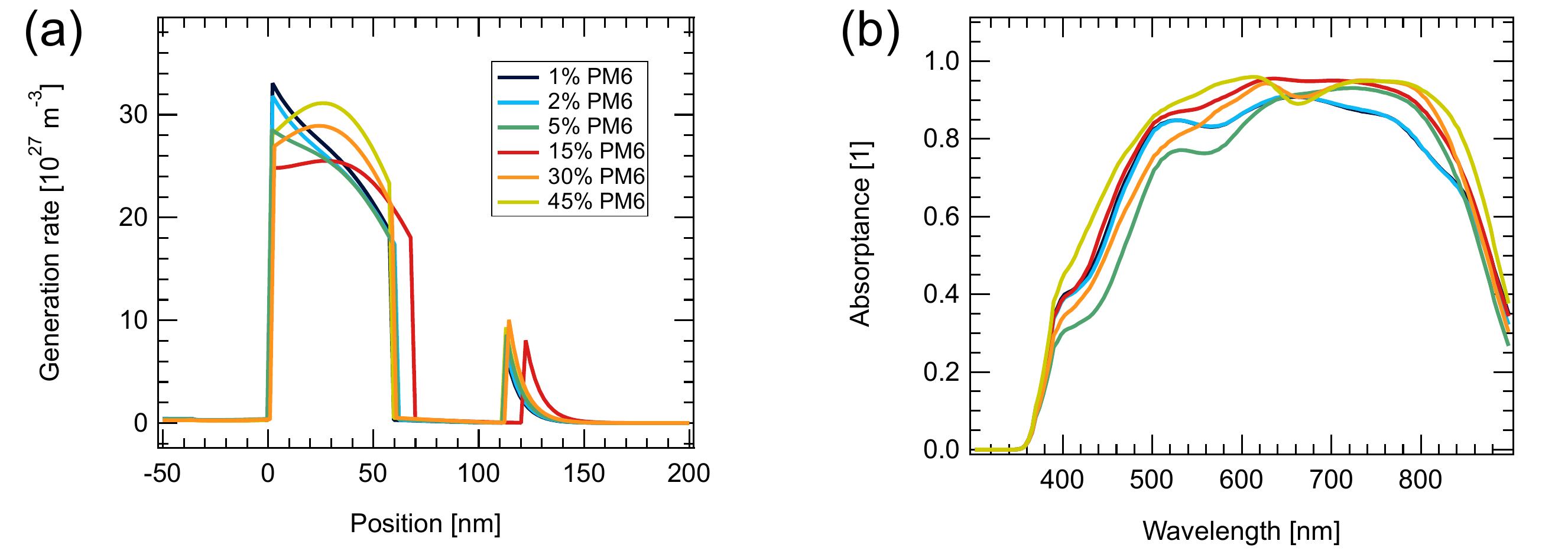}
    \caption{(a) Simulated spatially resolved optical generation rate within the device stack under AM1.5G illumination at 1000~W\,m$^{-2}$, calculated using OghmaNano. (b) Corresponding total device absorptance, obtained by integrating the absorbed photon profile over the active layer and normalizing to the incident photon flux.}
    \label{SI_fig:oghmanano}
\end{figure}

Optical generation profiles and total absorptance were obtained using OghmaNano simulations. The layer stack, thickness, and optical constants (refractive index and extinction coefficient obtained from VASE measurements) of each layer were implemented as input parameters, and the software was used to calculate the spatially resolved optical generation rate under AM1.5G 1000 W\,m$^{-2}$ illumination (Figure~\ref{SI_fig:oghmanano}(a)). The total absorptance of the device shown in Figure~\ref{SI_fig:oghmanano}(b) was determined by integrating the simulated absorbed photon profile over the active layer and dividing it by the integral of the incident photon flux at the corresponding wavelength.

\section{Internal quantum efficiency loss}\label{SI_sec:IQE}

The total IQE loss is defined as

\begin{equation}
L_{\mathrm{tot}} = 1 - \etaexc \cdot\etadiss \cdot\etacol.
\end{equation}

To attribute the total IQE loss to individual photoelectric conversion process, we employed a Shapley value decomposition.\cite{roth1988introduction} In the Shapley framework, the loss is treated as the value of a cooperative game with characteristic function
\begin{equation}
v(S) = 1 - \prod_{i\in S} \eta_i ,
\end{equation}
where $S$ denotes a subset of subprocesses. and $v(\varnothing)=0$. The Shapley value $\phi_i$ associated with subprocess $i$ is then given by
\begin{equation}
\phi_i =
\sum_{S \subseteq N \setminus \{i\}}
\frac{|S|!\,(N-|S|-1)!}{N!}
\left[ v(S \cup \{i\}) - v(S) \right] .
\end{equation}

where $i$ denotes exciton harvesting, CT state dissociation, or charge collection subprocesses.

\section{Donor fraction- and temperature-dependent effective conductivity}\label{SI_sec:sigmaoc}

\begin{figure}[h]
    \centering
    \includegraphics[width=0.9\linewidth]{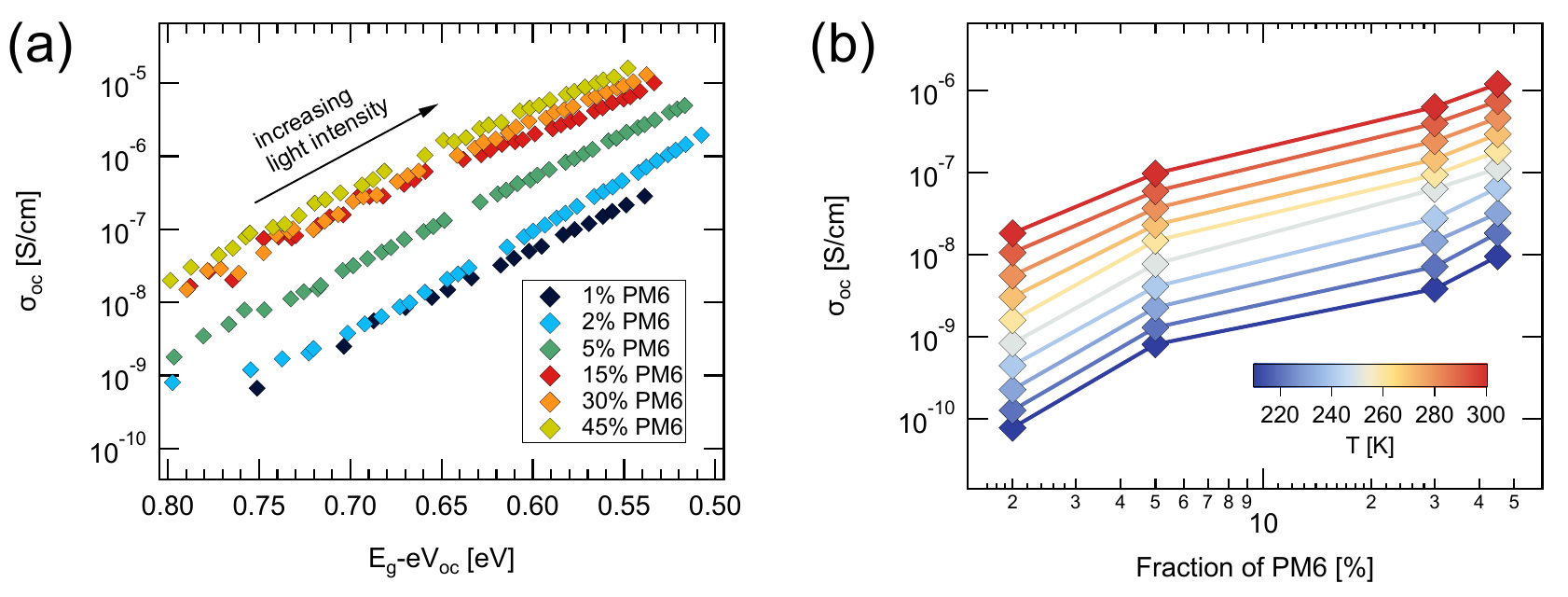}
    \caption{(a) $\sigma_{\mathrm{oc}}$ as a function of the energetic position, for the solar cells with different PM6 content. (b) $\sigma_{\mathrm{oc}}$ taken from the energetic cross-section of 0.66~eV as a function of PM6 fraction over different temperatures.}
    \label{SI_fig:sigma}
\end{figure}

Figure~\ref{SI_fig:sigma} shows the effective conductivity $\sigma_{\mathrm{oc}}$ determined from light intensity-dependent JV measurements. We plotted $\sigma_{\mathrm{oc}}$ as a function of $\Eg-e\Voc$ in order to compare charge transport at an equivalent energetic position within the bandgap, as indicated in Figure~\ref{SI_fig:sigma}(a).

\section{Charge carrier mobility from space-charge limited current}\label{SI_sec:SCLC}

\begin{figure}[t]
    \centering
    \includegraphics[width=0.9\linewidth]{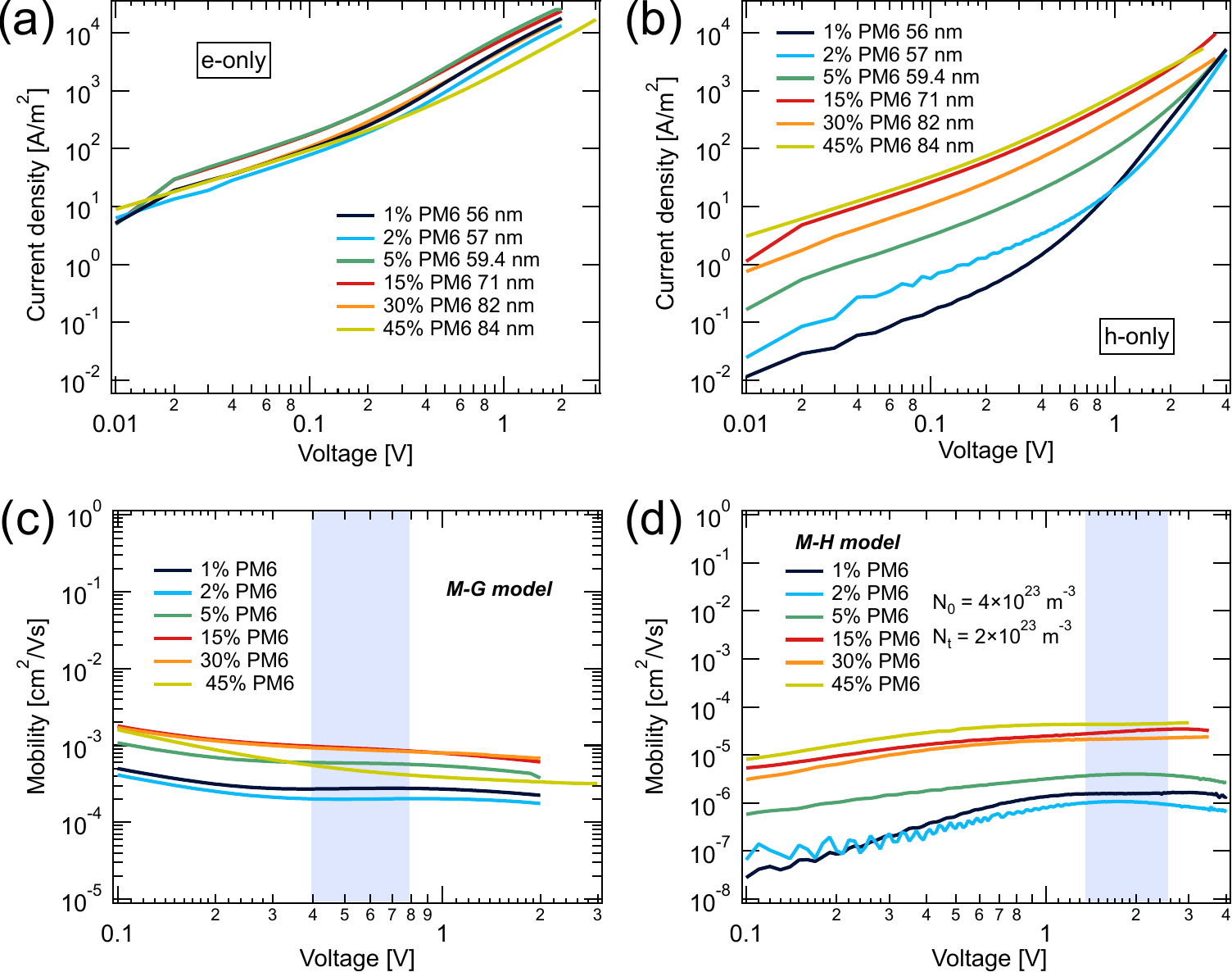}
    \caption{SCLC analysis of electron- and hole-only devices with varying PM6 content. JV characteristics of (a) electron-only, and (b) hole-only devices. 
    Voltage-dependent mobilities of (c) electrons and (d) holes, extracted using Eqs.~\eqref{eq:MG_model} and \eqref{eq:MH_model}, respectively. The mobilities exhibit voltage dependence at low bias and reach a plateau at higher voltages, from which the field-independent $\mu_{\mathrm{n}}$ and $\mu_{\mathrm{p}}$ values were extracted.}
    \label{SI_fig:SCLC}
\end{figure}

We measured the JV curves of electron-only and hole-only devices based on different PM6 fractions to determine the $\mu_{\mathrm{n}}$ and $\mu_{\mathrm{p}}$, as shown in Figure~\ref{SI_fig:SCLC}(a) and (b). For electron-only devices, the Mott-Gurney model was applied as the JV characteristics exhibit an approximately quadratic dependence on voltage.\cite{mott1948electronic} However, for the hole-only devices, the slopes ($\der\ln J/\der\ln V$) deviate from 2, particularly for devices with low PM6 contents. We therefore applied the Mark-Helfrich model,\cite{mark1962space} which is commonly employed for systems with the DOS that is not Gaussian. Mark-Helfrich model was found to provide a consistent description of our measured JV curves in the target bias region. This is consistent with our reported mixed DOS in state-of-the-art OSCs as a combination of a Gaussian and a power-law distribution.\cite{saladina2023power}

The Mott–Gurney equation is described as:
\begin{equation}\label{eq:MG_model}
    J=\frac{9}{8}\varepsilon\varepsilon_{0}\mu_\mathrm{n}\frac{V^2}{L^3}.
\end{equation}

The Mark-Helfrich equation is described as:
\begin{equation}\label{eq:MH_model}
    J=N_0e\mu_\mathrm{p}\left(\frac{\varepsilon_0\varepsilon}{eN_t}\right)^r\left(\frac{2r+1}{r+1}\right)^{r+1}\left(\frac{r}{r+1}\right)^r\frac{V^{r+1}}{L^{2r+1}}.
\end{equation}

Here, $J$ is the current density, $\varepsilon$ is the relative permittivity, assumed to be 3.5, $\varepsilon_{0}$ is the vacuum permittivity, $\mu_\mathrm{n}$, $\mu_\mathrm{p}$ is the carrier mobility, $V$ is the applied voltage, and $L$ is the active layer thickness. $N_\mathrm{0}$ is the total density of localised states, $N_\mathrm{t}$ is the total density of trap states. $N_\mathrm{0}$ and $N_\mathrm{t}$ were chosen according to the reported values.\cite{nicolai2011electron}. The exponent $r=T_\mathrm{t}/T$, $T_\mathrm{t}$ is the characteristic temperature corresponding to the exponential DOS, $e$ is the elementary charge. 

The $\mu_{\mathrm{n}}$ and $\mu_{\mathrm{p}}$ as a function of applied voltage is calculated according to Equation~\eqref{eq:MG_model} and Equation~\eqref{eq:MH_model}. We used $\der\ln J/\der\ln V$ to calculate the power of $r+1$. From Figure~\ref{SI_fig:SCLC}(c) and (d), it can be seen that mobility shows voltage dependence in the low voltage regime, and reaches a plateau around 0.6~V for e-only devices and 2~V for h-only devices, where we extracted the field-independent $\mu_{\mathrm{n}}$ and $\mu_{\mathrm{p}}$.

The values of extracted mobilities for electrons and holes are reported in Table~\ref{tab:SCLC}.

\begin{table}[h]
    \centering
    \begin{tabular}{|c|c|c|c|c|c|c|}
    \hline
    Parameter & 1\% PM6 & 2\% PM6 & 5\% PM6 & 15\% PM6 & 30\% PM6 & 45\% PM6 \\\hline
    $\mu_\mathrm{n}$ [cm$^2$\,V$^{-1}$\,s$^{-1}$] & $2.76\,\cdot\,10^{-4}$ & $2.00\,\cdot\,10^{-4}$ & $5.83\,\cdot\,10^{-4}$ & $8.97\,\cdot\,10^{-4}$ & $8.65\,\cdot\,10^{-4}$ & $4.52\,\cdot\,10^{-4}$ \\
    $\mu_\mathrm{p}$ [cm$^2$\,V$^{-1}$\,s$^{-1}$] & $1.56\,\cdot\,10^{-6}$ & $1.05\,\cdot\,10^{-6}$ & $3.99\,\cdot\,10^{-6}$ & $3.21\,\cdot\,10^{-5}$ & $2.20\,\cdot\,10^{-5}$ & $4.38\,\cdot\,10^{-5}$ \\
    \hline
    \end{tabular}
    \caption{SCLC mobility values for electron- and hole-only PM6:Y12 devices with varying PM6 content.}
    \label{tab:SCLC}
\end{table}

\clearpage
\section{Temperature-dependence of the recombination ideality factor}\label{SI_sec:nid}

To determine the electronic density of states in PM6:Y12, we use the method introduced earlier.\cite{saladina2023power} 
The ideality factor was determined from illumination-intensity- and temperature-dependent JV measurements. The relationship between the illumination intensity $\Phi$ and the open-circuit voltage $\Voc$ follows
$$\nid = \frac{e}{\kT} \bl \frac{\der\,\ln \Phi}{\der\,\Voc} \br^{-1}$$
The ideality factor was extracted from the local slope of $\ln\Phi(\Voc)$, evaluated in regions unaffected by leakage current (at low intensities) and by contact limitations (at high intensities). This procedure yields one ideality factor per $(T, \Voc)$ point.

Following the multiple-trapping-and-release framework, the dominant recombination pathway determines how the recombination ideality factor depends on temperature. The functional form depends on the assumed DOS. For a mixed DOS (mobile charge carriers in a Gaussian, trapped charge carriers in an exponential), 
\begin{equation}\label{SI_Eq:mixDOS}
    \nid(T) = \frac{\Eu + \kT}{2\kT} , 
\end{equation}
where $\Eu$ is the Urbach energy describing the exponential tail. For a purely exponential DOS, 
\begin{equation}
    \nid(T) = \frac{2\Eu}{\Eu + \kT} . 
\end{equation}

\begin{figure}[h]
    \centering
    \includegraphics[width=0.5\linewidth]{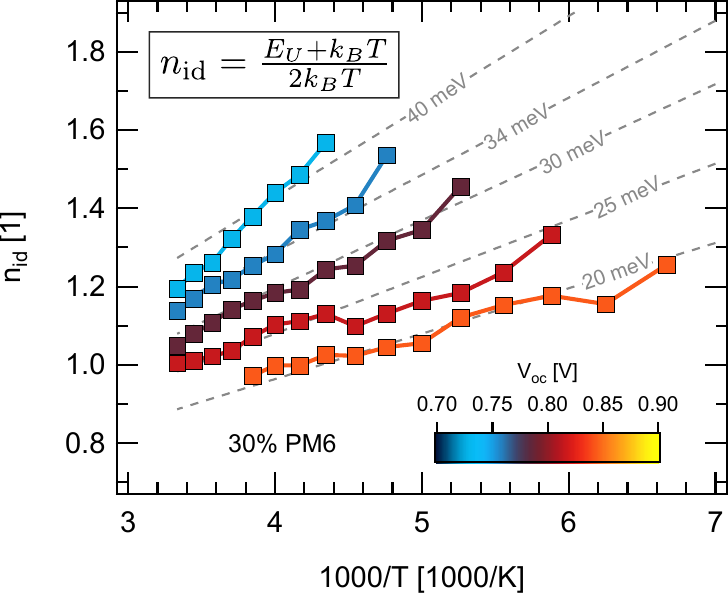}
    \caption{Recombination ideality factor for a PM6:Y12 solar cell with 30\% donor content, plotted as a function of inverse temperature. The data at specific $\Voc$ follows closely the mixed DOS model, Equation~\eqref{SI_Eq:mixDOS}. The associated Urbach energy increases at lower $\Voc$ values.}
    \label{SI_fig:DOS:nid_mixDOS}
\end{figure}

At each $\Voc$ (i.e., at a fixed quasi-Fermi level splitting), $\nid(T)$ data was related to the appropriate analytical expression. For PM6:Y12, mixed DOS was found to describe the data best, as shown in Figure~\ref{SI_fig:DOS:nid_mixDOS}. An effective Urbach energy $E_\mathrm{U}(V_\mathrm{oc})$ was then calculated from $\nid$ data using Equation~\eqref{SI_Eq:mixDOS}. This provides the disorder parameter of the trap distribution \textit{at that energy depth} in the DOS.

The resulting $\Eu$ values were then plotted as a function of energetic position below the gap edge, $\Eg - \Voc$. A constant $\Eu$ indicates a strictly exponential DOS, while a linear increase of $\Eu$ with the quasi-Fermi level splitting implies that the exponential function is only a local approximation to the real DOS. In this case, the DOS follows a power-law of the form $g(E) \propto (E_0 - E)^{\xi}$.\cite{saladina2023power,mazzolini2025discerning}
This method therefore allows reconstructing the shape of the DOS directly from experimental light-intensity-dependent JV data.

\section{Recombination parameters from long-time transient photoluminescence}\label{SI_sec:CTPL}

\begin{figure}[h]
    \centering
    \includegraphics[width=0.95\linewidth]{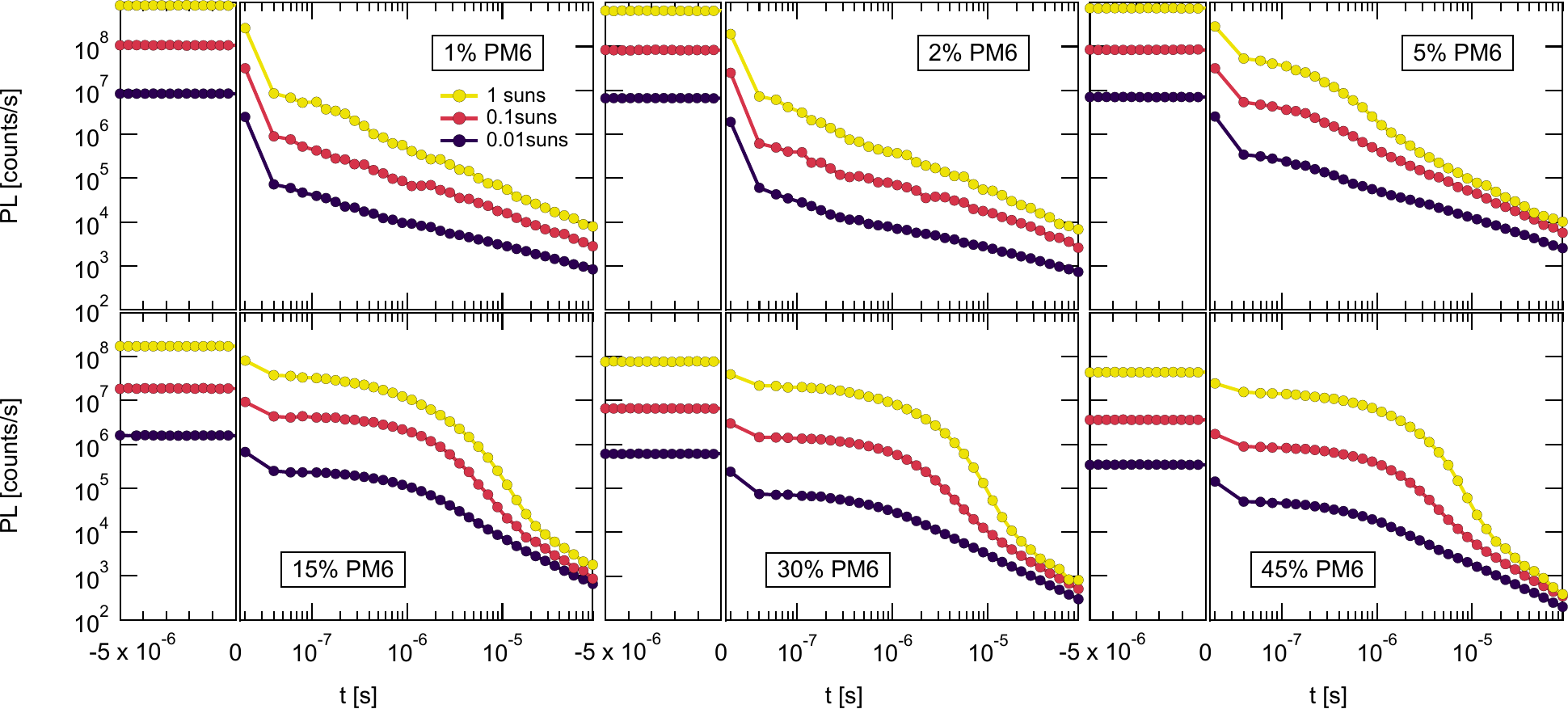}
    \caption{Transient PL decays of PM6:Y12 films with varying donor content measured at excitation intensities of 0.01, 0.1, and 1 sun.}
    \label{SI_fig:CTPL}
\end{figure}

The photoluminescence (PL) of organic solar cells can in general be described by a superposition of PL from non-dissociated photogenerated excitons ($\mathrm{PL}_\mathrm{ex}$) and that of separated charges recombining at the interface ($\mathrm{PL}_\mathrm{CS}$). To separate these contributors, transient PL measurements were conducted as outlined in previous works \cite{list2023determination,faisst2025implied}. Leveraging the much shorter lifetime of geminate exciton recombination ($\tau_\mathrm{ex}$<1\,ns) compared to separated charges ($\tau_\mathrm{CS} \approx \mathrm{\mu s}$), the PL of the latter dominates the transient PL decay for times exceeding the geminate exciton recombination lifetime ($\mathrm{PL}(t\gg\tau_\mathrm{ex})\approx\mathrm{PL}_\mathrm{CS}(t)$). The full dataset for each donor concentration and for intensities 0.01, 0.1 and 1 sun is depicted in Figure~\ref{SI_fig:CTPL}. 

\begin{figure}[h]
    \centering
    \includegraphics[width=0.54\linewidth]{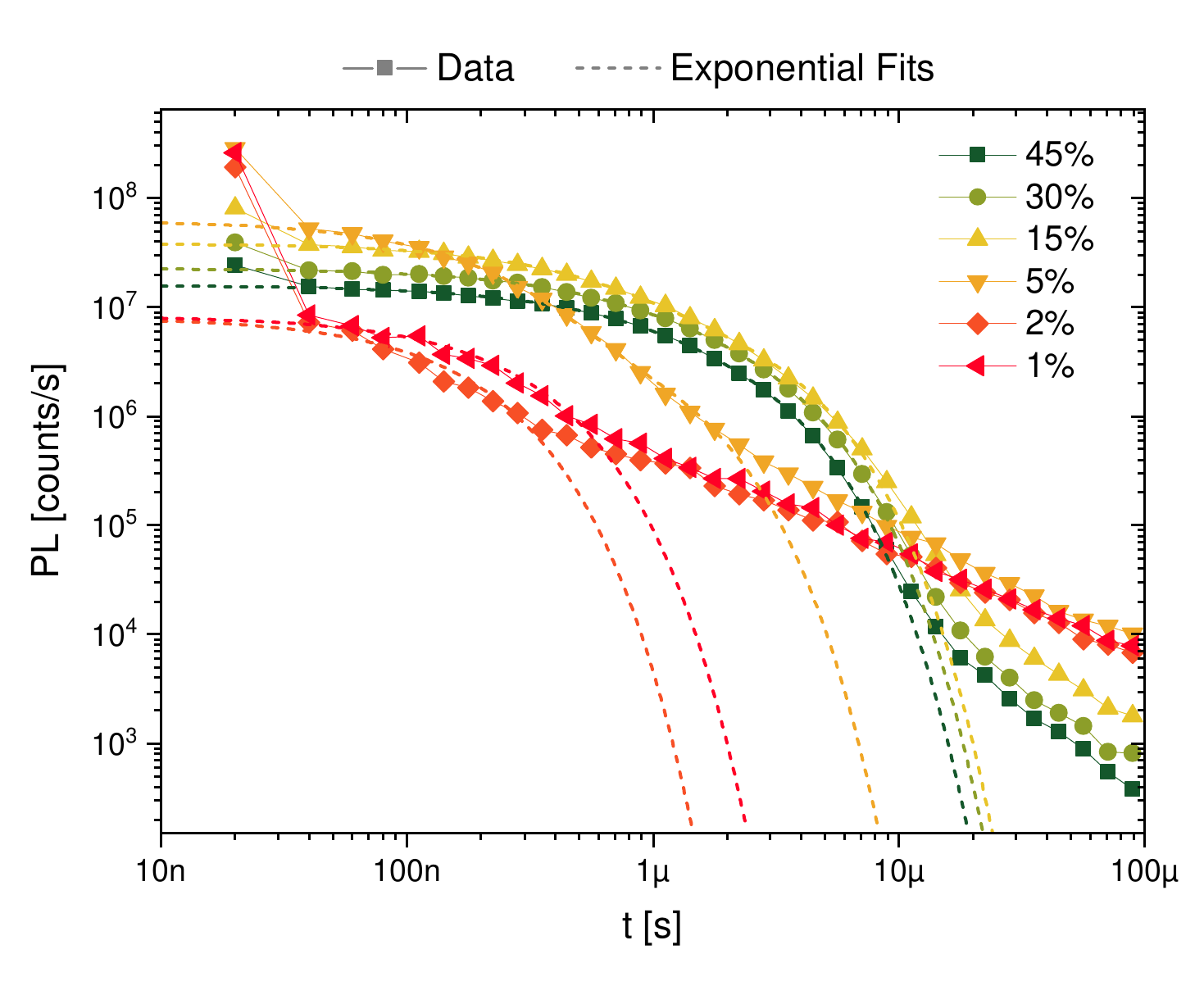}
    \caption{Log-smoothed transient photoluminescence decays measured for different donor contents under excitation equivalent to 1~sun (symbols), together with the corresponding (bi)exponential fits (dashed lines) applied in the regime dominated by Langevin-type recombination.}
    \label{SI_fig:TPL_ExpFits}
\end{figure}

For $t>10\,\mathrm{ns}$ and in the range where Langevin type recombination is dominant, the decays for 1 sun equivalent were fitted with a bi-exponential decay function
\begin{equation}
    \mathrm{PL}_\mathrm{CS}(t) = A_1 \exp\!\left(-\frac{t}{\tau_1}\right)
     + A_2 \exp\!\left(-\frac{t}{\tau_2}\right)
\end{equation}
The mean lifetime $\langle \tau \rangle$ can then be approximated by amplitude weighted averaging of both lifetimes\cite{thor2025shedding}, i.e. 
\begin{equation}
    \langle \tau \rangle=\frac{A_1 \tau_1 + A_2 \tau_2}{A_1 + A_2}
\end{equation}
The data with the corresponding fits are depicted in Figure \ref{SI_fig:TPL_ExpFits} and Table \ref{tab:lifetime_mean}. Note that the transient PL curves obtained for 2\% and 1\% donor content are fitted with a mono exponential fit, due to the small range (<500\,ns) of apparent Langevin type recombination. 

\begin{table}[h]
\centering
\renewcommand{\arraystretch}{1.2}
\begin{tabular}{c c c c c c c}
\hline
Donor (\%) 
& $A_1$ (cps) 
& $\tau_1$ (s) 
& $A_2$ (cps) 
& $\tau_2$ (s) 
& $\langle\tau\rangle$ (s) 
& $\mathrm{PL_{CS}}(t=0)$ (cps) \\
\hline
45 & $7.16\times 10^{6}$ & $5.53\times 10^{-7}$ & $8.56\times 10^{6}$ & $1.75\times 10^{-6}$ & $1.20\times 10^{-6}$ & $1.57\times 10^{7}$ \\
30 & $1.15\times 10^{7}$ & $5.54\times 10^{-7}$ & $1.11\times 10^{7}$ & $1.96\times 10^{-6}$ & $1.24\times 10^{-6}$ & $2.25\times 10^{7}$ \\
15 & $2.56\times 10^{7}$ & $4.54\times 10^{-7}$ & $1.29\times 10^{7}$ & $2.11\times 10^{-6}$ & $1.01\times 10^{-6}$ & $3.86\times 10^{7}$ \\
5  & $5.50\times 10^{7}$ & $1.71\times 10^{-7}$ & $7.44\times 10^{6}$ & $7.70\times 10^{-7}$ & $2.42\times 10^{-7}$ & $6.24\times 10^{7}$ \\
2  & $8.09\times 10^{6}$ & $1.34\times 10^{-7}$ & $0$              & $0$              & $1.34\times 10^{-7}$ & $8.09\times 10^{6}$ \\
1  & $8.30\times 10^{6}$ & $2.22\times 10^{-7}$ & $0$              & $0$              & $2.22\times 10^{-7}$ & $8.30\times 10^{6}$ \\
\hline
\end{tabular}
\caption{Transient PL fitting parameters}\label{tab:lifetime_mean}
\end{table}

\begin{figure}[h]
    \centering
    \includegraphics[width=0.5\linewidth]{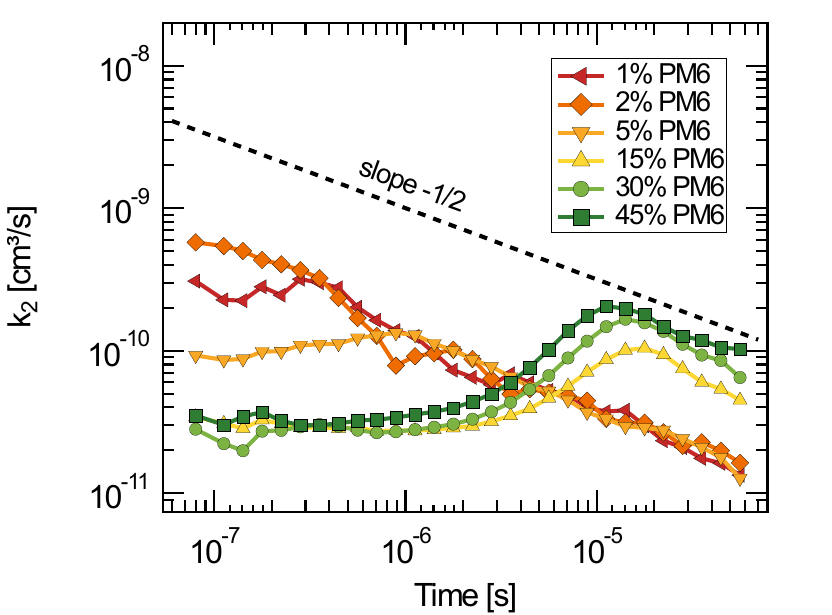}
    \caption{Time-dependent effective bimolecular recombination coefficient $k_2(t)$ derived from transient PL lifetime and charge carrier density. The latter was estimated from the signal amplitude. Slope $t^{-1/2}$ is indicative of Smoluchowski-type recombination.}
    \label{SI_fig:CTPL_k2}
\end{figure}

\subsection{Inferring the relative carrier concentration from the charge-transfer exciton PL measurements}

According to the ideality diode equation, we can write the open-circuit voltage as
\begin{equation}
    \Voc = \frac{\nid \kT}{e} \ln \left( \frac{\jgen}{J_0} \right),
\end{equation}
where $\nid$ is the recombination ideality factor, $\kT/e$ the thermal voltage, $\jgen$ and $J_0$ the generation and dark saturation current densities, respectively. Similarly valid, due to generation being equal to recombination under open-circuit, is
\begin{equation}
    \Voc = \frac{\nid \kT}{e} \ln \left( \frac{R}{R_0} \right).
\end{equation}
Assuming that the recombination rate in organic solar cells is dominated by $R \propto n_c n$ and $R_0 \propto n_i^2$, we can write
\begin{equation}
    \Voc = \frac{\nid \kT}{e} \ln \left( \frac{n_c n}{n_i^2} \right).
\end{equation}
From multiple-trapping-and-recombination, we get $n_c \propto n^{\delta-1}$, with the recombination order $\delta$. This allows us to write
\begin{equation}\label{eq:SI-Voc-classical}
    \Voc \approx \frac{\nid \kT}{e} \ln \left( \frac{n^\delta}{n_i^2} \right) = \frac{\frac{\delta}{2} \nid \kT}{e} \ln \left( \frac{n^2}{n_i^2} \right).
\end{equation}

In previous works by List \& Faisst \textit{et al.} \cite{list2023determination,faisst2025implied}, it was experimentally shown that $\Voc$ can be well described by the PL, which itself is assumed to be $\propto n_\mathrm{free}^2$:
\begin{equation}\label{eq:SI-Voc-PL}
    \Voc = \frac{kT}{e} \ln \left( \frac{n_\text{free}^2}{n_i^2} \right),
\end{equation}
without the recombination ideality $\nid$ in the prefactor.

The connection between ideality factor and recombination order depends on the details of recombination, which in turn depends on the density of states (DOS). The following comparison is exemplary and relevant this publication, but not general:

\textbf{Case 1}: In homogeneous systems, in particular (but not only) inorganic semiconductors, all charge carriers are available for recombination. There,
$$n_{id} \approx \frac{2}{\delta}.$$
which means that the term $\frac{\delta}{2}\nid$ in Eqn.~\ref{eq:SI-Voc-classical} becomes unity. Consequently, $n_\mathrm{free}$ corresponds to all charge carriers $n$ (that are controlled by the quasi-Fermi level splitting).

\textbf{Case 2}: In inhomogenous systems with trapping, such as phase-separated and energetically disorded donor--acceptor blends, not all charge carriers are available for recombination. If charge carriers are separated into mobile ($n_c$) and trapped ($n_c$) carriers that in sum determine the overall charge carrier concentration $n$, then the trapped electrons and holes cannot contribute to recombination as they can not meet one another. Therefore, the classically expected recombination rate $R\propto n^2$ is actually smaller (when the prefactors are equal): $R \propto n - n_t^2$. As far as we know, in many organic solar cells the recombination is dominated by mobile charge carriers from a Gaussian DOS and trapped ones from a power-law DOS, which can be locally approximated by an exponential DOS.\cite{saladina2023power} Therefore, we assume that our DOS can be approximated by a local exponential tail. The recombination order $\delta$ then depends on the characteristic tail energy (Urbach energy) $E_u$, so that $\delta-1 = E_u/kT$. The ideality factor is given by van Berkel et al.\cite{vanBerkel1993} as
\begin{equation}
    n_{id} = \frac{2 (\delta-1)}{\delta}.
\end{equation}
We rewrite this result,
\begin{equation}
    n_{id} = 2\left(1 - \frac{1}{\delta}\right),
\end{equation}
and again try to replace the above-mentioned factor $\frac{\delta}{2}\nid$,
\begin{equation}
    \frac{\delta}{2} n_{id} = \frac{\delta}{2} 2\left(1 - \frac{1}{\delta}\right) = \left(\delta - 1\right).
\end{equation}
While not as clear as the classical case~1, for $\delta \approx 2$ the ideality factor drops from Eqn.~\ref{eq:SI-Voc-classical}, and it becomes equivalent to Eqn.~\ref{eq:SI-Voc-PL},
\begin{equation}
    \Voc = (\delta-1) \frac{\kT}{e} \ln \left( \frac{n^2}{n_i^2} \right) \approx \frac{\kT}{e} \ln \left( \frac{n^2}{n_i^2} \right).
\end{equation}

Therefore, in both cases -- for case 2 under assumption of $\delta \approx 2$, $\Voc$ can be well described by the PL as
\begin{equation}\label{eq:SI-Voc-PL}
    \Voc = \frac{\kT}{e} \ln \left( \frac{n^2}{n_i^2} \right),
\end{equation}
\emph{without} the recombination ideality $\nid$ as prefactor.

\section{Nongeminate recombination dynamics using time-delayed collection field}\label{SI_sec:TDCF_rec}

TDCF measurements were performed to investigate nongeminate recombination dynamics. Charge carriers were photogenerated using a short laser pulse at an excitation wavelength of 550~nm while the device was held at a constant pre-bias voltage. After a variable delay time $t_\mathrm{delay}$, a strong reverse collection bias of $-4$~V was applied to extract the remaining mobile charge carriers. Delay times were varied from 10~ns to 10~$\text{\textmu}$s in order to probe recombination over the nanosecond to microsecond time scale.

\begin{figure}[h]
    \centering
    \includegraphics[width=1\linewidth]{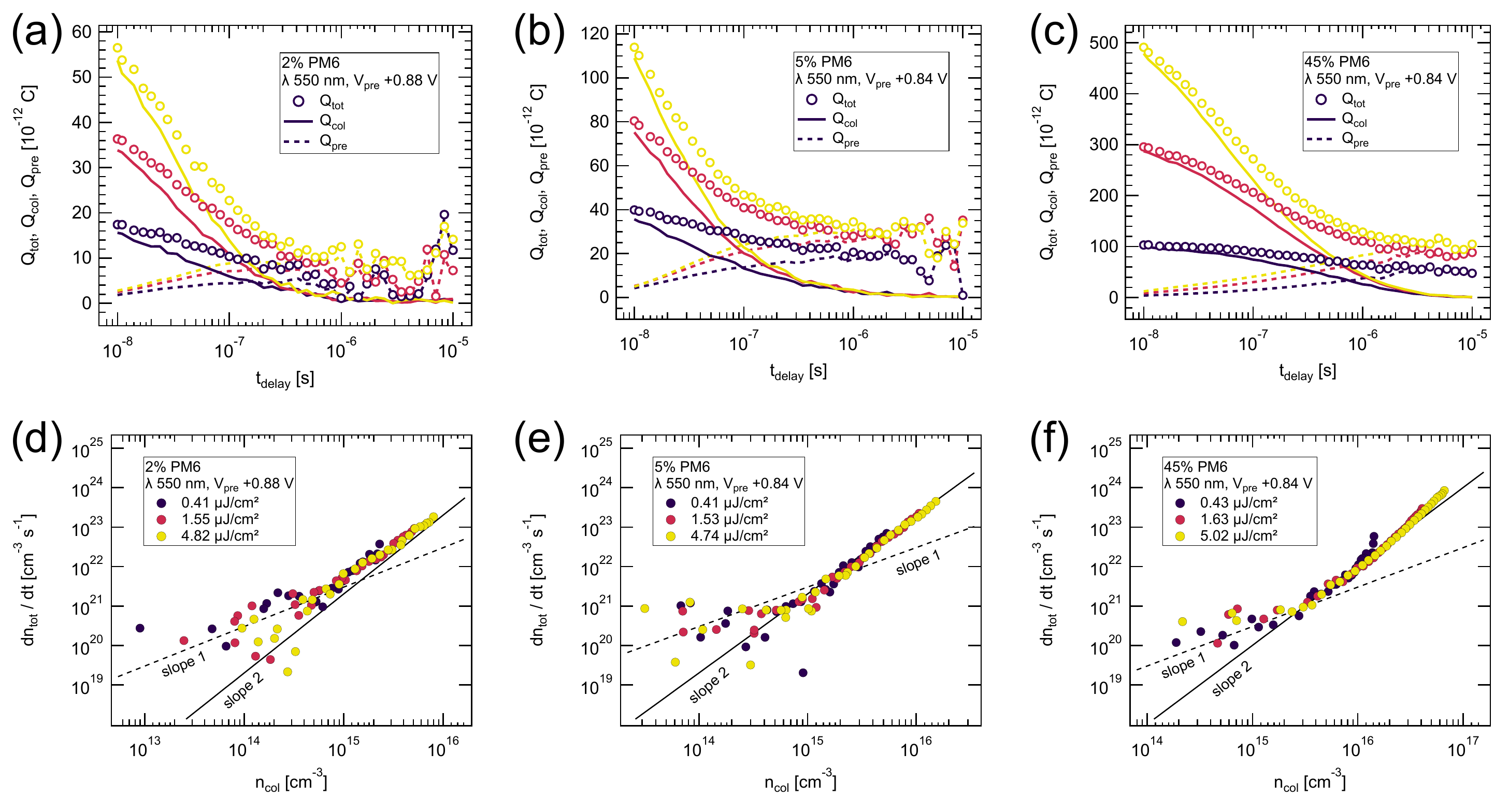}
    \caption{(a-c) TDCF photocurrent transients and (d-f) the corresponding recombination rate as a function of charge carrier density for devices with (a,d) 2\% PM6, (b,e) 5\% PM6, and (c,f) 45\% PM6 content.}
    \label{SI_fig:TDCF:delayscan}
\end{figure}

The transient photocurrent following photoexcitation was separated into pre-bias and collection-bias contributions. The total extracted charge $Q_\mathrm{tot}$ was obtained by summing the time integrals of the photocurrent measured during the pre-bias phase and during the collection-bias phase. The corresponding charge carrier density was calculated from $Q_\mathrm{tot}$ as described above. The obtained data is shown in Figure~\ref{SI_fig:TDCF:delayscan}(a)-(c).

Nongeminate recombination was analyzed following the procedure developed in the group of Dieter Neher.\cite{kniepert2014conclusive} The incremental loss of charge carriers due to recombination was evaluated from the delay-dependent total extracted charge, as shown in Figure~\ref{SI_fig:TDCF:delayscan}(d)-(f). A slope of unity indicates that pseudo-first-order recombination losses dominate, i.e.\ recombination of the photogenerated and dark injected charge carriers. At higher charge densities, where photogenerated charge carrier density is much higher than the injected one, recombination is close to the second order. 

Assuming bimolecular recombination, the recombination coefficient $k_2$ was directly calculated from the data using \cite{kniepert2014conclusive}
$$
\frac{\der n_\mathrm{tot}}{\der t} = k_2 \bl n_\mathrm{col}^2 + 2 n_\mathrm{col} n_0 \br , 
$$
where $n_0$ is the dark injected charge carrier density evaluated at a specific pre-bias voltage. 

\begin{figure}[h]
    \centering
    \includegraphics[width=0.45\linewidth]{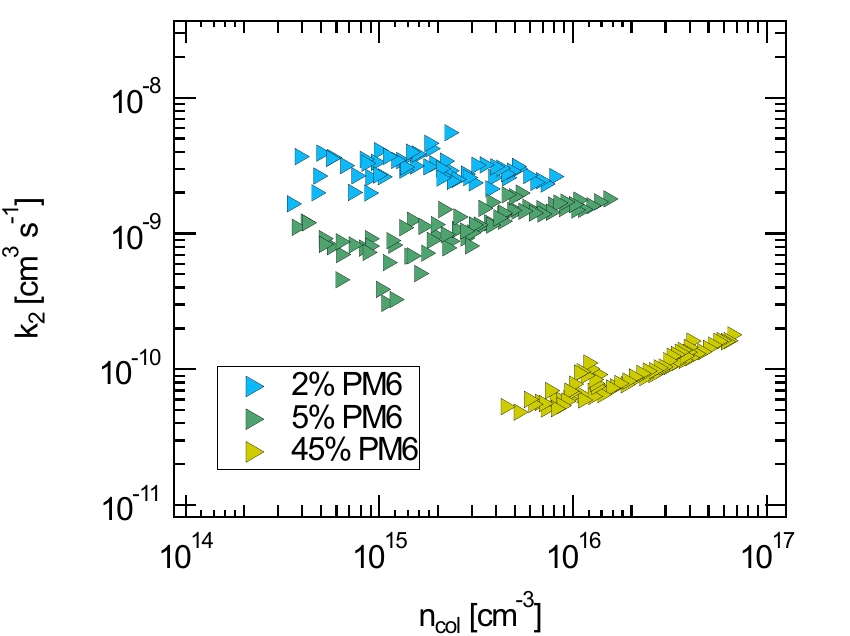}
    \qquad
    \includegraphics[width=0.45\linewidth]{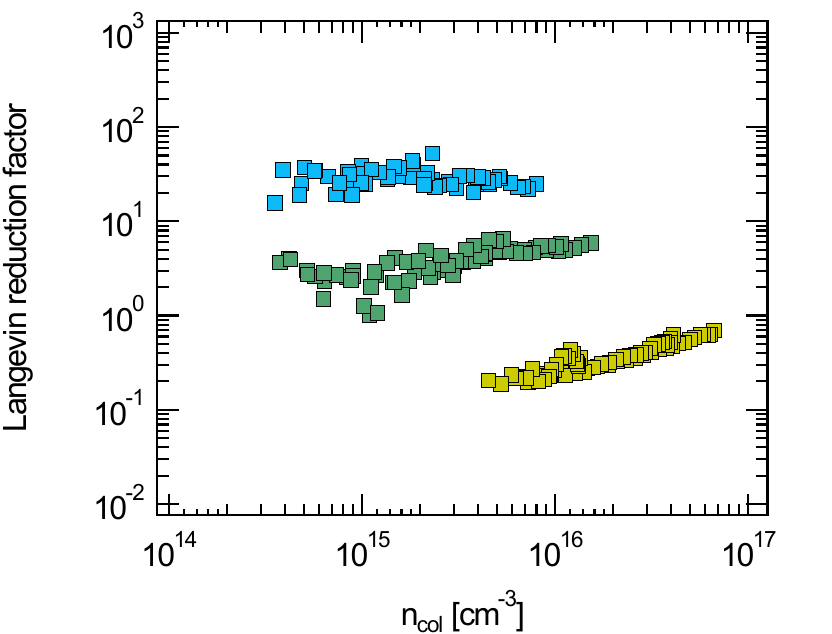}
    \caption{(a) The effective second order recombination coefficients $k_2$ and (b) the associated Langevin reduction factors for solar cells with 2\%, 5\% and 45\% PM6 content calculated from TDCF data in Figure~\ref{SI_fig:TDCF:delayscan}.}
    \label{SI_fig:TDCF:k2}
\end{figure}

The recombination coefficient $k_2$ was determined for each delay time from the above equation and is plotted in Figure~\ref{SI_fig:TDCF:k2}. Regions where $k_2$ could not be reliably extracted due to noise were excluded from the analysis. The Langevin reduction factor $\gamma$ was calculated from $k_2$ using SCLC mobility values according to 
$$
k_2 = \gamma\cdot \frac{e\bl\mu_n + \mu_p\br}{\varepsilon\varepsilon_0} , 
$$
where $e$ is the elementary charge, $\mu_n$ and $\mu_p$ the electron and hole mobilities, respectively, $\varepsilon$ the relative permittivity of the active layer, and $\varepsilon_0$ the vacuum permittivity. It can be seen that $\gamma>1$ under all the
investigated light intensities for 2\% and 5\% PM6 devices and the reduction factor depends only weakly on charge carrier density. We further determined the reduction factor at charge carrier densities of $\sim 1 \cdot 10^{16}$~cm$^{-3}$ for 2\% and 5\% PM6 devices and $\sim 2 \cdot 10^{16}$~cm$^{-3}$ for 45\% PM6 device, which are comparable to the carrier densities derived from CT PL under 1 sun-equivalent illumination.

\bibliographystyle{apsrev4-2}

\end{document}